\documentclass[letter,12pt, final]{article}
\usepackage{amsmath}
\usepackage{changes}
\definechangesauthor[color=orange]{cic}
\definechangesauthor[color=red]{jf}
\definechangesauthor[color=magenta]{kg}

\usepackage{bm}
\usepackage{amsfonts}
\usepackage{amssymb}
\usepackage{amsthm}
\usepackage{amsbsy}
\usepackage{mathrsfs}
\usepackage{float}
\usepackage{soul}
\usepackage{pdflscape}
\usepackage{bbm}
\usepackage{latexsym}
\usepackage{setspace}
\usepackage[top=2.5cm, bottom=2.5cm, right=2.5cm, left=2.5cm]{geometry}
\usepackage{graphicx}
\usepackage[utf8]{inputenc}			
\usepackage[english]{babel} 			
\usepackage[T1]{fontenc}				
\usepackage{verbatim}					
\usepackage{enumerate}					
\usepackage{array}						
\usepackage{multicol}					
\usepackage{paralist}					
\usepackage{multirow}
\usepackage{dsfont}
\usepackage{ctable}
\usepackage{longtable}
\usepackage{quoting}
\usepackage[]{caption}
\usepackage{sectsty}

\usepackage{subcaption}
\usepackage{lipsum}

\usepackage[all]{xy}					
\usepackage[colorlinks, citecolor = blue, linkcolor = blue, urlcolor = blue, bookmarksopen = true, pdfstartview = {FitH}]{hyperref}
\usepackage{url}
\usepackage[round]{natbib}			
\usepackage{epstopdf} 
\usepackage[normalem]{ulem}
\usepackage[hang,splitrule]{footmisc}


\usepackage{algorithm}
\usepackage{algorithmic}

\DeclareMathAlphabet\mathbfcal{OMS}{cmsy}{b}{n}

\theoremstyle{plain}

\theoremstyle{definition}

\newtheorem{hypothesis}{Hypothesis}

\usepackage{titling} 
\usepackage[round]{natbib}
\newcommand{\xfnm}[1][]{\ifx!#1!\else\unskip,\space#1\fi}

\usepackage[affil-it]{authblk}
\title{When David becomes Goliath: \\ Repo dealer-driven bond mispricing}

\author[a]{Carlos Ca\~{n}\'{o}n\thanks{\url{carlos.canonsalazar@bankofengland.co.uk}}}
\author[b]{Eddie Gerba\thanks{\url{eddie.gerba@bankofengland.co.uk}}}
\author[c]{Jozef Barun\'{i}k\thanks{\url{barunik@fsv.cuni.cz}}}
\affil[a]{Bank of England \& King's College London}
\affil[b]{Bank of England \& LSE}
\affil[c]{Charles University \& Czech Academy of Sciences}

\begin{document}
\maketitle
\vspace{-1cm}

\begin{abstract}
This paper studies the impact of funding market frictions on bond prices and market-wide liquidity. Using proprietary transaction‑level data on all gilt‑backed repo and reverse-repo trades, we demonstrate how the market power of individual dealers and their linkages generate frictions. Specifically, we show that frictions related to market power account for between 0.5 and 1.3 percentage points of bond yield deviation, while the transmission of heterogeneously persistent shocks between dealers accounts for between 2 and 4 percentage points of yield deviation.
\textbf{}
\end{abstract}
\begin{spacing}{1.0}\small

\end{spacing}
\vspace{-.2cm}
\begin{spacing}{1.0}\scriptsize
\noindent \textbf{Acknowledgments:}  We have benefited from detailed comments from Tobias Dieler (discussant), Hillary Stein (discussant), Robert Czech, Nick Vause, Patrick Coen, Yiming Ma.  We also thank Marco Bardoscia, Semih Uslu, Jorge Florez, Karoll Gomez, Rodrigo Guimaraes, Petros Katsoulis, Gabor Pinter, Andres Murcia, Matt Roberts-Sklar, Laura Silvestri, and seminar participants at Bank of England, Banco de la Republica, Charles University, Boston CEBRA 2025, 2025 Bristol Financial Markets Conference, Oxford IFABS 2025, Rome EARIE 2023, and Portsmouth MMF 2023 for valuable insights. The views expressed in this paper are those of the authors and do not reflect the views of the Bank of England, the PRA, or any of its committees.
\end{spacing}

\newpage

\section{Introduction}\label{Sect 1}

At what point does the structure of funding markets become apparent in government bond prices? Episodes such as the `dash for cash' in March 2020 suggest that stresses in repo markets can quickly leak into sovereign bond markets, causing price fluctuations and reduced liquidity. However, there is little systematic evidence on the mechanisms through which funding-market frictions lead to bond mispricing and impaired market liquidity. Most research focuses either on aggregate liquidity and funding conditions or on the microstructure of individual markets. Rarely does it combine micro-level measures of dealer behaviour with bond-level mispricing and a market-wide liquidity metric.
 
In this paper, we examine how market power and dealer linkages in the UK repo market may shape gilt prices and market-wide liquidity. Our focus is on the core, dealer-centric structure of the market, whereby a limited number of bank dealers facilitate funding between a wide range of non-bank financial institutions and central counterparties using UK government bonds as collateral. Using proprietary, transaction-level data on all repo and reverse repo trades involving gilts between 2016 and 2022, combined with a cross-section of gilt yields, we quantify the impact of dealer market power, its dispersion across dealers, and the transmission of shocks through the dealer network as the three channels impacting bond mispricing and market liquidity. Although our data is UK specific, the core mechanisms we identify are common for all large money markets. 
 
Our approach involves two steps. First, we measure market power specific to each dealer in both the repo and reverse-repo segments. Using a structural demand-and-supply framework that treats each dealer as supplying a differentiated funding `product' to non-dealers, we estimate dealer-level markups and markdowns and recover markups and markdowns from the implied Lerner indices. To address endogeneity in repo rates, quantities and relationship-trading variables, we exploit the granularity of the panel data to construct instrumental variables based on idiosyncratic shocks at the dealer–non-dealer level, as described by \cite{Garbaix.20}. Second, we embed dealers and non-dealers in a dynamic network. Using a time-varying VAR and persistence-based variance decompositions, we construct global dealer factors that capture the proportion of transitory and persistent shocks originating from a subset of major dealers and transmitted to the rest of the system. These factors quantify the extent to which the market as a whole is exposed to shocks affecting the most influential dealers.
 
We then link these funding-market primitives to gilt pricing and market liquidity. At the bond level, we measure mispricing as the absolute deviation of each gilt's yield from a smooth term-structure benchmark. At the market level, we use the “noise” index proposed by \cite{Huetal.13} as a proxy of overall market liquidity. By regressing bond-level mispricing and market liquidity on: i) dealer markups and markdowns, ii) their dispersion across dealers, and iii) the corresponding global dealer factors, we can decompose bond price distortions and liquidity  into components attributable to different funding-market frictions.\footnote{We control for bond characteristics, monetary policy and other macro-financial covariates.}
 
Our main findings are threefold. First, dealer-specific market power in the reverse-repo and repo segments leads to significant mispricing of gilts (channel A): conditional on controls, frictions related to markups and markdowns explain between 0.5 and 1.3 percentage points of bond yield deviations along the curve. Second, the heterogeneity of market power among dealers (channel B) and the transmission of shocks through the dealer network (channel C) are equally important. Our time-varying global dealer factors explain an additional 2–4 percentage points of mispricing. These are particularly strong effects for longer-maturity gilts. Taken together, these funding-market frictions account for between 2.5 and 5.3 percentage points of gilt yield deviation from the term-structure benchmark. Third, the same mechanisms shape market-wide liquidity. Higher markdowns and their dispersion, as well as stronger exposure to persistent dealer shocks in the reverse repo segment, are associated with significantly worse market liquidity. In contrast, the repo segment plays a more benign role. These effects are nonlinear and become especially pronounced during periods of stress, such as the 'dash for cash' period.
 

These results contribute to several areas of research. First, it makes a contribution to the literature on mispricing \citep{greenwood2014bond,LEWIS.21} that attributes persistent price deviations to frictions in core markets (such as the bond, repo and foreign exchange (FX) swap markets), to central bank decisions, or to changes in intermediation capacity. \cite{pelizzon2025central} show that Central Banks can induce bond-level mispricing through a collateral scarcity channel associated with the implementation of quantitative easing, while \cite{Kerseendischeretal.24} argue that outages in the futures market produce bond-level mispricing through hedging and price signalling channels.  \cite{BrauningStein.25,Barbeiroetal.24} and \cite{SteinWallen.25} discuss how the reduction in intermediation capacity affects asset prices in foreign exchange and Treasury markets.  From \cite{Brunnermeieretal.2008, Huhetal.21}, we are aware of the tight relationship between market and funding liquidity (\cite{baietal.17}), and we provide evidence that frictions in the core funding market arising from: i) dealer market power, ii) heterogeneous market power distribution across dealers, and iii) network-driven shocks translate directly into mispricing at the level of individual bonds and market-wide liquidity.

Second, we contribute to the expanding body of research on the microstructure of the repo market by demonstrating that the ramifications of dealer dominance extend beyond the transmission of policy rates to the pricing and liquidity of the underlying collateral  \citep{Duffieetal.2005,Afonso.14,mancini2016euro,DiMaggioetal.17}.  A close reference is \cite{eisenschmidt2024monetary}, who show how the market power of dealers in the European repo market impedes the pass-through of ECB's policy rate. Based on the notion that most repo trades are conducted over-the-counter by a very small number of dealers, these dealers have significant market power. The result is an imperfect and heterogeneous pass-through from interdealer repo rates to OTC repo rates (similar to the findings in \cite{huber2023market} but for the US), with the interdealer repo rate in turn determined by the central bank's deposit facility rate and the value of collateral. While undoubtedly innovative and insightful, our paper goes beyond that study in three ways. First, we take a much broader perspective by distinguishing market power in the repo to that of reserve repo segment, both in prices and volumes, and contrast individual market power with global dealer factors. Second, we link dealer-level frictions to market-level phenomenon through the lens of market-wide liquidity. Third, and perhaps most fundamentally, we examine market power and trading behaviour as inextricably linked to market pricing in the bond market.

Third, we contribute to the UK-specific literature on gilt and repo markets, which emphasises their role in providing liquidity and ensuring financial stability.  Our paper is particularly related to \cite{Huseretal.21, coental.24} as we study the UK repo market. \cite{Huseretal.21} provides a detailed description of what happened during the Dash-for-Cash period, and \cite{coental.24} follows a structural approach to highlight that collateral demand is an important driver in the repo market. Our paper contributes in two dimensions. One, we assess the impact of dealer-level market power in both repo segments, as well as transitory/persistent dealer factor shocks on repo prices and quantities. Two, we show that dealer-level frictions explain bond-level mispricing in the gilt market.

The remainder of the paper is structured as follows. Section 2 outlines our hypotheses and reasoning. In Section 3, we describe our data, present some stylized facts about the market, and outline our identification approach. Section 4 discusses our core results on the three frictions and their spillovers. Further details and results are also available in the Online Appendix. Concluding remarks wrap up the paper in Section 5.  

\section{Hypotheses}\label{Sect Method}

Dealers in the UK repo market may hold market power on non-dealers due to their access to central counterparty clearing services.  Thus, we expect the traditional deadweight loss to operate in the form of lower volumes and higher funding costs.\footnote{\cite{Huber.23,Eisenschmidt.22} document the impact of market power for the US and EUR repo markets.} We also expect to find evidence of losses associated with the dispersion of market power (across dealers), which is reflected in market-wide aggregate distortions (\cite{BoarMidrigan.2019,Brooketal.2021, DavidVenkareswaran.2019, Haltiwangeretal.2018, Liang.2023, Weinberger.2020}). Because the repo market is closely related to other core markets, \cite{Brunnermeieretal.2008}, markups and markdowns in the repo market could spill over to individual collateral acquired in the bond and reverse repo markets, and to market-wide measures of market liquidity. We will investigate these next.

The effect of repo market markups and markdowns is mediated by several factors. First, there is ample evidence of relationship lending in various OTC markets, e.g. \cite{Jukartisetal.22, DiMaggioetal.17,Hauetal.21,Afonso.14}, yet it is unclear whether stronger relationships counterbalance dealers' market power. Second, market power could operate through quantities and prices, and it is empirically interesting to measure this. Third, market power could operate differently across financial sector segments, i.e. hedge funds, asset managers, insurance companies, and pension funds. Finally, since we know the identity of non-dealers, we can test whether non-dealers active in both the repo and reverse repo segments receive different terms of trade vis-a-vis non-dealers active in only one segment.  

Let us, next, outline our reasoning for channel A. The first hypothesis focuses on the impact of dealer-specific market power on repo market quantities and prices. Concretely, we want to understand how market power operates at individual repo transaction level, and the role of other factors, e.g.  relationship lending, sector specific differences, and common shock to main dealers, in attenuating the impact on quantities or prices.

\begin{hypothesis} \label{hyp: 1}
   Dealer-specific market power (markup and markdown) creates frictions that affect \textbf{both} repo prices and volumes.
\end{hypothesis}

The second dimension focuses on dealer-specific market power spillovers. From the repo segment, non-dealers could face difficulties in accessing arbitrage capital, which should increase mispricing, e.g. \cite{Huetal.13}, and deteriorate market liquidity. From the reverse repo segment, non-dealers could face difficulties in acquiring gilts in exchange for funding, contributing to search frictions in the gilt market and ultimately increasing mispricing, e.g. \cite{Duffieetal.15}, also deteriorating market liquidity. 

\begin{hypothesis} \label{hyp: 2}
    Dealer-specific market power (markup and markdown) impacts individual gilts, bond-market liquidity, and (financial) market-wide liquidity.  
\end{hypothesis}

The literature on misallocation due to market power emphasises that this specific friction does not depend on repo dealers holding positive markups and markdowns, but is also related to how market power is distributed across dealers, \cite{Syverson.2024}.\footnote{Another way of framing it is that the welfare loss is proportional to the wedge between marginal revenue and marginal cost. We evaluate this interpretation empirically and the results hold.} The next hypothesis focuses on this channel B. 

\begin{hypothesis} \label{hyp: 3}
   There is a friction due to the distribution/dispersion of market power across dealers (markups and markdowns), which impacts bond-level mispricing and market-wide liquidity.
\end{hypothesis}

Next, we argue that common repo dealer shocks may generate similar spillovers.\footnote{The OTC segmentation literature is closely related as it predicts how dealers will segment clients, e.g. based on their characteristics, trading patterns, etc. The literature remains inconclusive on the determinants of dealer heterogeneity.  \cite{Sambalaibat.23} argues that the largest dealers in an OTC market will specialise in the most active non-dealers, and \cite{LEWIS.21} shows empirically that this intuition can be extended to other asset classes. There are authors who assume away any dealer heterogeneity, e.g. \cite{Farboodietal.2022,Wang.2021}, or who explain it on the basis of ex ante factors, e.g. \cite{Uslu.2019,Eisfeldtetal.2022}. In any case, assuming dealer heterogeneity, quantity (prices) will increase (decline) as core traders become more interconnected.} The next hypothesis focuses on channel C: How do common persistent and transitory shocks to repo dealers shape individual bonds, bond market-makers, and market-wide liquidity? If collateralised funding markets are short-term in nature, we should expect common persistent shocks to have powerful spillovers to the bond market, and to market-wide aggregates.  

\begin{hypothesis} \label{hyp: 4}
   Persistent shocks to repo dealers matter for bond-level mispricing and market-wide liquidity. 
\end{hypothesis}

So far, the stated hypotheses describe three mechanisms that relate repo dealers to individual gilt-, bond market-, and (financial) market-wide liquidity. The final hypothesis focuses on their interaction (channels A+B+C). The first two mechanisms, associated with dealer-specific market power, are known to be interrelated but operate under very specific conditions, i.e. if all dealers have the same level of market power, then the second mechanism disappears. The last mechanism is more general, as it zooms in on heterogeneously persistent shocks, that are not uniquely associated with market power, and which can potentially be the main driver of spillovers.

\begin{hypothesis} \label{hyp: 5}
   The combined effects of the three frictions on bond yield deviations and market liquidity are sizeable and larger than any of those individually.
\end{hypothesis}

\section{Empirics}
\label{Sect Data Identification}

\subsection{Data}
We use proprietary transaction-level data for the UK repo market between January 2016 and January 2022. As shown below, we focus on the subsample of transactions utilising gilts as collateral to avoid dealing with asset classes that are illiquid relative to gilts and to avoid biasing the formation of relationship trading based on ownership of illiquid assets.  Another reason for focusing on gilt repo is that our analysis on mispricing only uses these securities. The dataset includes all repo and reverse repo transactions between dealers and non-dealers, and between dealers and CCPs. A detailed description of the structural features of the UK repo market can be found in Online Appendix \ref{Sect Repo Description}.

\begin{table}[htbp]
\footnotesize
  \centering
  \caption{Summary statistics for the repo market by type of transaction (repo vs. reverse repo) between dealers and non-dealers. Our sample only uses gilts as collateral.  Volumes are expressed in $10^7$ of sterling and the interest rate spread (in percentage points) is relative to the BoE reference rate.}
    \begin{tabular}{lcccccc}
    \toprule
          & Median   & Mean  & SD    & Min   & Max   & N \\
    \midrule
          & \multicolumn{6}{c}{Repo} \\
          \cmidrule{2-7}
    Volume   & 3.12 & 4.59 & 4.41 & 0.11 & 27.40 & 335601 \\
    Rate spread & -0.01 & -0.01 & 0.09  & -0.75 & 0.77  & 335601 \\
          &       &       &       &       &       &  \\
    \cmidrule{2-7}
          & \multicolumn{6}{c}{Reverse} \\
          \cmidrule{2-7}
    Volume   & 4.50 & 6.51 & 7.26 & 0.11 & 75.00 & 350609 \\
    Rate spread & -0.08 & -0.10 & 0.08  & -0.79 & 0.64  & 350609 \\
    \bottomrule
    \end{tabular}%
  \label{SumStats: Repo OTC}%
\end{table}%

Table \ref{SumStats: Repo OTC} presents summary statistics on repo market data by type of transaction between dealers and non-dealers. The median transaction is £31 million at a repo rate of 0.01 pp below the BoE reference rate when dealers provide liquidity in exchange for collateral, and £45 million at a repo rate of 0.08 pp below the BoE reference rate when dealers receive liquidity in exchange for collateral. Further, summary statistics by type of transaction between dealers and CCPs (Table \ref{SumStats: Repo CCP} in Online Appendix \ref{appendix: tables}) shows the median transaction is £23 million with a repo rate of 6 pp below the BoE reference rate when dealers provide liquidity in exchange for collateral and £22 million with a repo rate of 6 pp below the BoE reference rate when dealers receive liquidity in exchange for collateral. 

To examine the impact of repo dealer market power on mispricing of gilts, we use ISIN-level metrics of mispricing, volatility, liquidity, residual maturity, and a set of bond-level characteristics. The mispricing metric is the difference between the bond yield and a daily (spline) benchmark yield constructed using all gilts available in the market at the time.\footnote{Later, we will construct the market liquidity proxy, following \cite{Huetal.13}, by adding the square of the mispricing of the individual bonds.}

\begin{table}[htbp]
\footnotesize
  \centering
  \caption{Summary statistics of the gilts market by residual maturity. Mispricing is the difference between the yield and the benchmark yield, liquidity is the ask/bid spread, volatility is the high/low spread. Market liquidity is constructed according to \cite{Huetal.13}.}
    \begin{tabular}{llcccccc}
    \toprule
     &       & Median   & Mean  & SD    & Min   & Max   & N \\
    \midrule
    3yr - 7yr & Mispricing & -0.11 & 0.10  & 2.18  & -4.92 & 6.61  & 7566 \\
          & Duration  & 5.28 & 5.32  & 1.22  & 2.86  & 7.66  & 7566 \\
          & Liquidity & 0.03 & 0.04  & 0.02  & 0.01  & 0.26  & 7566 \\
          & Volatility & 0.24 & 0.28  & 0.23  & 0.05  & 6.87  & 7566 \\
    \cmidrule{2-8}
    8yr - 19yr & Mispricing & 0.20 & 0.12  & 1.48  & -5.24 & 4.74  & 10810 \\
          & Duration  & 9.94 & 10.69 & 2.60  & 6.73  & 18.55 & 10810 \\
          & Liquidity & 0.05 & 0.06  & 0.04  & 0.01  & 0.54  & 10810 \\
          & Volatility & 0.66 & 0.77  & 0.48  & 0.15  & 11.00 & 10810 \\
    \cmidrule{2-8}
    20yr or more & Mispricing & 0.03 & -0.16 & 1.27  & -5.19 & 3.46  & 14309 \\
          & Duration  & 22.56 & 22.90 & 4.95  & 14.64 & 37.31 & 14309 \\
          & Liquidity & 0.11 & 0.12  & 0.06  & 0.03  & 0.57  & 14309 \\
          & Volatility & 1.57 & 1.87  & 1.18  & 0.33  & 37.94 & 14309 \\
    \cmidrule{2-8}
    Total & Mispricing & 0.066 & 0.00  & 1.60  & -5.24 & 6.61  & 32685 \\
          & Duration  & 13.40 & 14.79 & 8.27  & 2.86  & 37.31 & 32685 \\
          & Liquidity & 0.07 & 0.08  & 0.06  & 0.01  & 0.57  & 32685 \\
          & Volatility & 0.86 & 1.14  & 1.07  & 0.05  & 37.94 & 32685 \\
    \cmidrule{2-8}
    Market Liquidity &  & 1.73 & 1.92  & 0.55  & 1.23 & 3.62  & 1470 \\      
    \bottomrule
    \end{tabular}%
  \label{SumStats: Bond Market}
\end{table}%

Table \ref{SumStats: Bond Market} shows summary statistics for the gilt market. The median gilt has a positive mispricing of 0.07 pp. However, when zooming in on the three different residual maturity buckets, only short-end gilts have a negative mispricing of 0.11 pp. The measures of all other bond characteristics (duration, liquidity and volatility) increase as residual maturity increases. 

\subsection{Stylized facts}

Table \ref{fact: numberDealers, Families, counterparts} shows the number of potential counterparts that could form a trading relationship. While we observe a fixed number of dealers, the number of non-dealers can change depending on the segment. For example, in the repo segment we have 3380 different non-dealers belonging to 727 different families, and in the reverse segment these numbers are 2004 and 682 respectively.\footnote{By family we mean a financial institution that could offer different investment vehicles to end-investors. For example, imagine a large asset management firm offering many individual funds to retail investors. In our data, individual non-dealers are each of these funds and the asset management firm is the family. As we explain in the paper, our analysis is done at the family level.} Our data is granular enough to identify non-dealers that operate in both segments simultaneously with the same dealer.  

\begin{table}[htbp]
\footnotesize
  \centering
  \caption{Dealer (D), Non-Dealers (D), Family of Non-Dealers, average number of counterparts and average counterparty dependence in our sample.}
    \begin{tabular}{lcccccccc}
    \toprule
    &&&& \multicolumn{2}{c}{No Counter.}  &  & \multicolumn{2}{c}{ \% Funding} \\
     \cmidrule{2-3} \cmidrule{5-6} \cmidrule{8-9}
          & \multicolumn{1}{c}{Repo} & \multicolumn{1}{c}{Reverse Repo} & & \multicolumn{1}{c}{Repo} & \multicolumn{1}{c}{Reverse Repo} & & \multicolumn{1}{c}{Repo} & \multicolumn{1}{c}{Reverse Repo} \\
    \cmidrule{2-3} \cmidrule{5-6} \cmidrule{8-9}
    Dealers & 15    & 15   & & 9.46 & 6.77 && 41.95\% & 46.13\% \\
    Non-Dealers & 3380    & 2004  & & 2.73 & 2.93 && 52.39\% & 60.04\% \\
    Family of ND & 727      & 682 &   \\
    \cmidrule{2-3} 
    Dyads D/ND & 5803    & 3484  & \\
    Dyads D/Family & 2112    & 1767  & \\
    \bottomrule
    \end{tabular}%
  \label{fact: numberDealers, Families, counterparts}%
\end{table}%

Table \ref{fact: numberDealers, Families, counterparts} shows the number of dyads in our sample, which is the unit of analysis for our panel estimation. For the repo segment, we observe 2112 dealer/family non-dealer dyads, and for the reverse repo segment, this number decreases to 1767. If we zoom in on non-dealers that are active in both segments simultaneously, the number of dyads falls to 1048. Our analysis is at the non-dealer family level, so we drop the word "family" and use non-dealers. 

\paragraph{Dealers hold more counterparts than non-dealers} 

Table \ref{fact: numberDealers, Families, counterparts} further shows the number of counterparts for dealers and non-dealers. On average, a non-dealer has fewer counterparts than a dealer. While for dealers this number fluctuates around ten counterparts, for non-dealers it does not reach three counterparts. This feature is not unexpected as it's also observed in other OTC markets.  

\paragraph{OTC vs inter-dealer segments}

Dealers are active in both the OTC and inter-dealer markets on a daily basis. As emphasized above, the inter-dealer market accounts for a large proportion of daily volume, e.g. around  $70\%$. There are also more fundamental differences. While virtually all transactions in the inter-dealer segment are overnight, in the OTC segment this share does not reach 50\% and the remaining share is made up of transactions with maturities of less than one week or one month  (Table {\ref{fact: repo maturities, ISIN}}  in Appendix \ref{appendix: tables}). Dealers' use of the inter-dealer market is also closely related to their holdings. While dealers use, on average, three different gilts in the OTC segment, they use ten times as many in the inter-dealer segment (Table {\ref{fact: repo maturities, ISIN}}  in Appendix \ref{appendix: tables}).  

In addition, Figure \ref{fig:ISIN vs Vol} shows a negative relationship between the number of unique gilts and the size and duration of repo (as well as reverse repo) transactions. Repo transactions are, on average, larger and longer when dealers operate with fewer gilts, and this holds for reverse repo transactions as well. We interpret this as evidence of inventory constraints.
\begin{figure}[h!]
    \begin{minipage}[h!]{\linewidth}
        \centering
        \caption{Unique gilts vs Mean Volume \& Mean ResMaturity at OTC segment for both repo and reverse repo. Note the volumes are divided by $10^6$.}
        \includegraphics[width=0.9\textwidth]{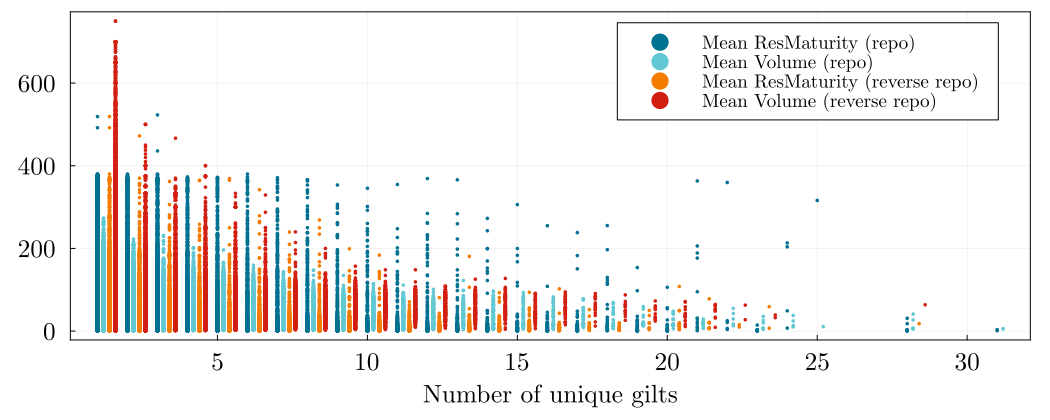}
        \label{fig:ISIN vs Vol}
    \end{minipage}
\end{figure}

\paragraph{Sector heterogeneity in OTC segment}

\begin{figure*}
\caption{Log volume and repo rate spread using the non-dealer dataset with daily frequency.  The spread is the difference between the repo rate and the reference rate. }
        \begin{subfigure}{0.48\textwidth}
        \caption{Volume}
            \label{network fig:CCP Volume}
            \includegraphics[width=\linewidth]{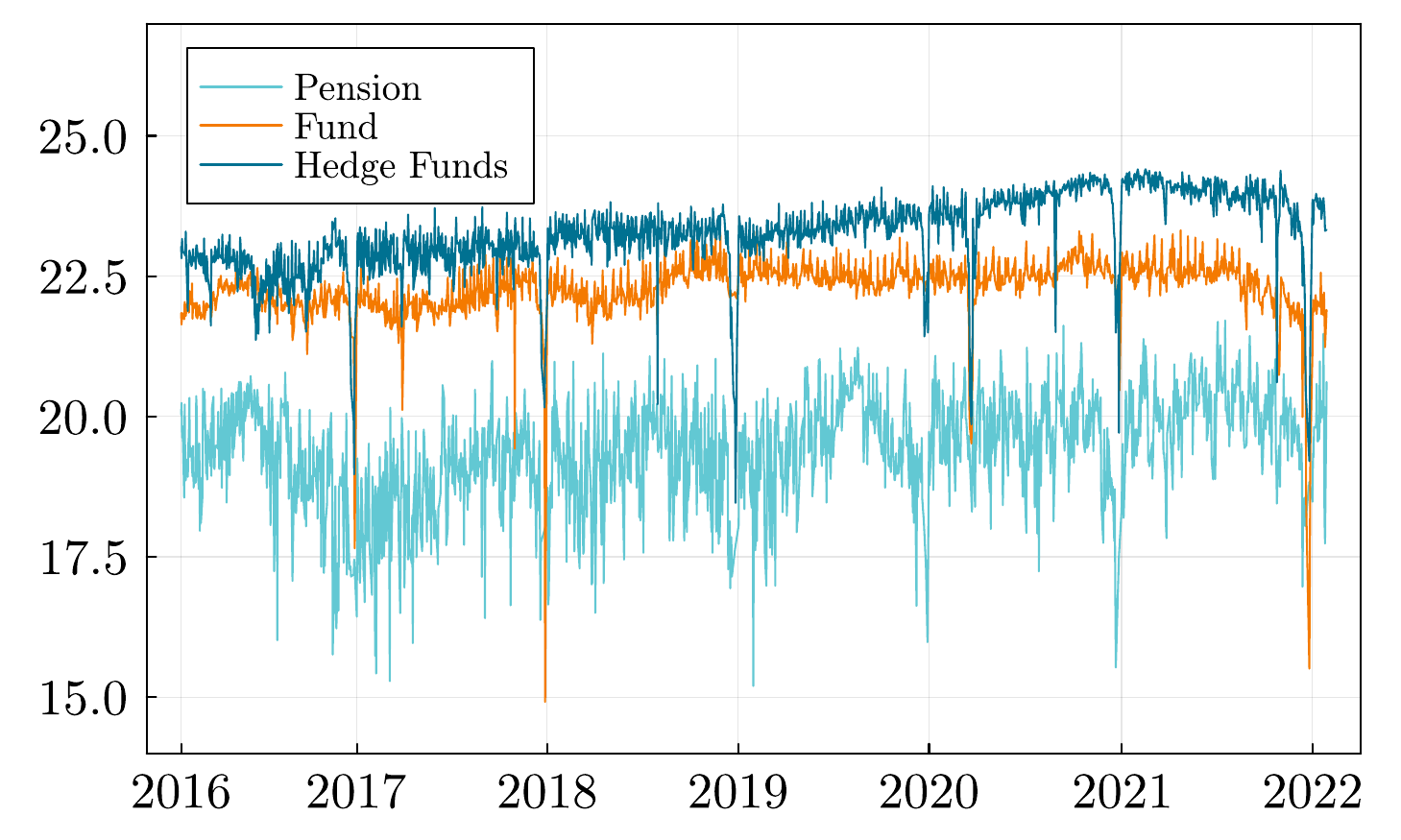}
        \end{subfigure}
        \hspace*{\fill}
        \begin{subfigure}{0.48\textwidth}
        \caption{Rate Spread}
            \label{network fig:CCP Spread}
            \includegraphics[width=\linewidth]{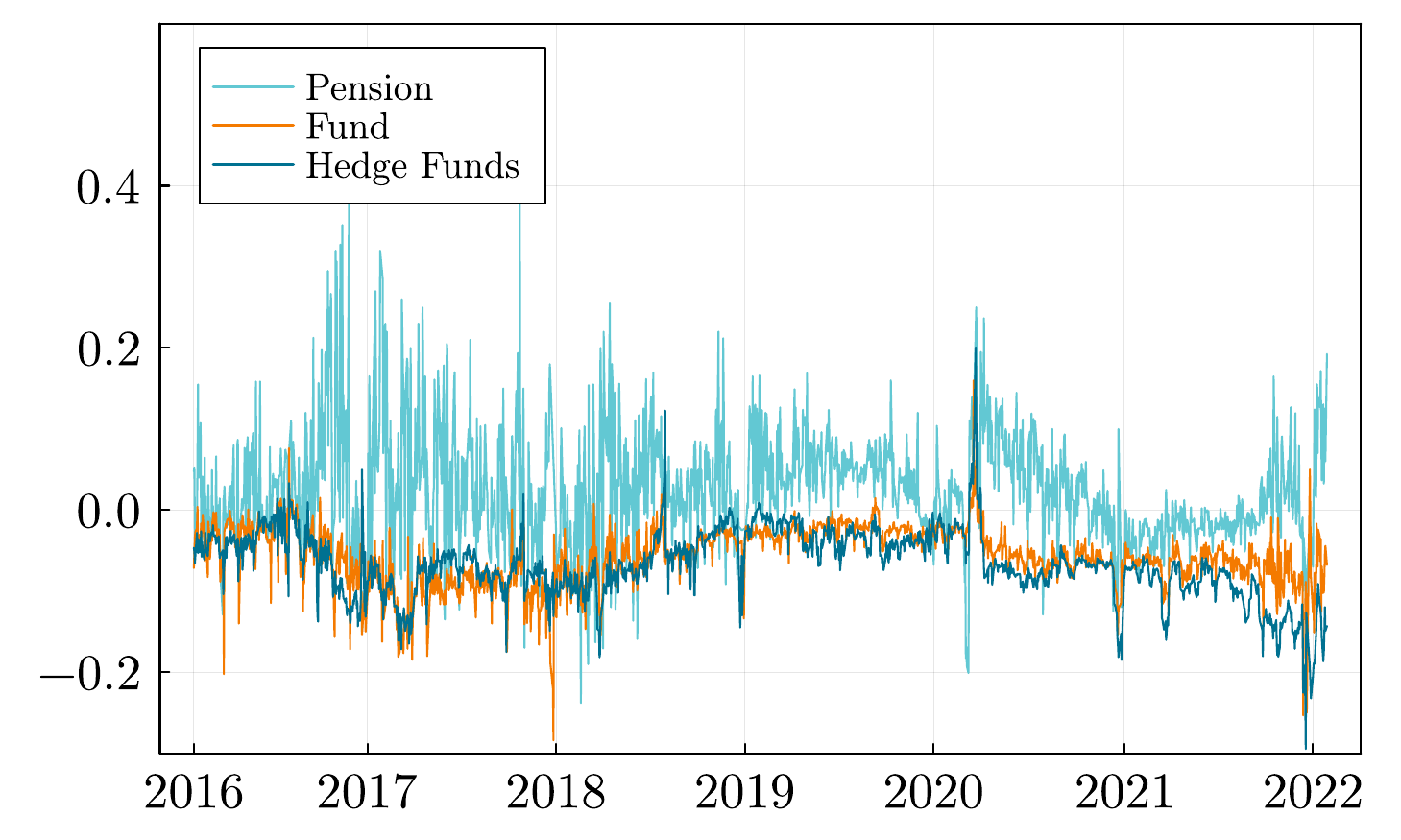}
        \end{subfigure}
\end{figure*}

Figures \ref{network fig:CCP Volume} - \ref{network fig:CCP Spread} show the historical volume and the repo spread. Hedge funds and other funds are the two most important sectors in the OTC segment, and the spread to the bank rate is usually negative. In contrast, pension funds have a smaller but still relevant presence, yet their spread is positive.   

\paragraph{Existence of global dealer factors}

\begin{figure*}
\caption{Global dealer factors derived from both transitory and persistent shocks to repo and reverse repo volumes and spreads.}
        \begin{subfigure}{0.48\textwidth}
            \caption{Volume}
            \label{fig:globalfactor_volume}
            \includegraphics[width=\linewidth]{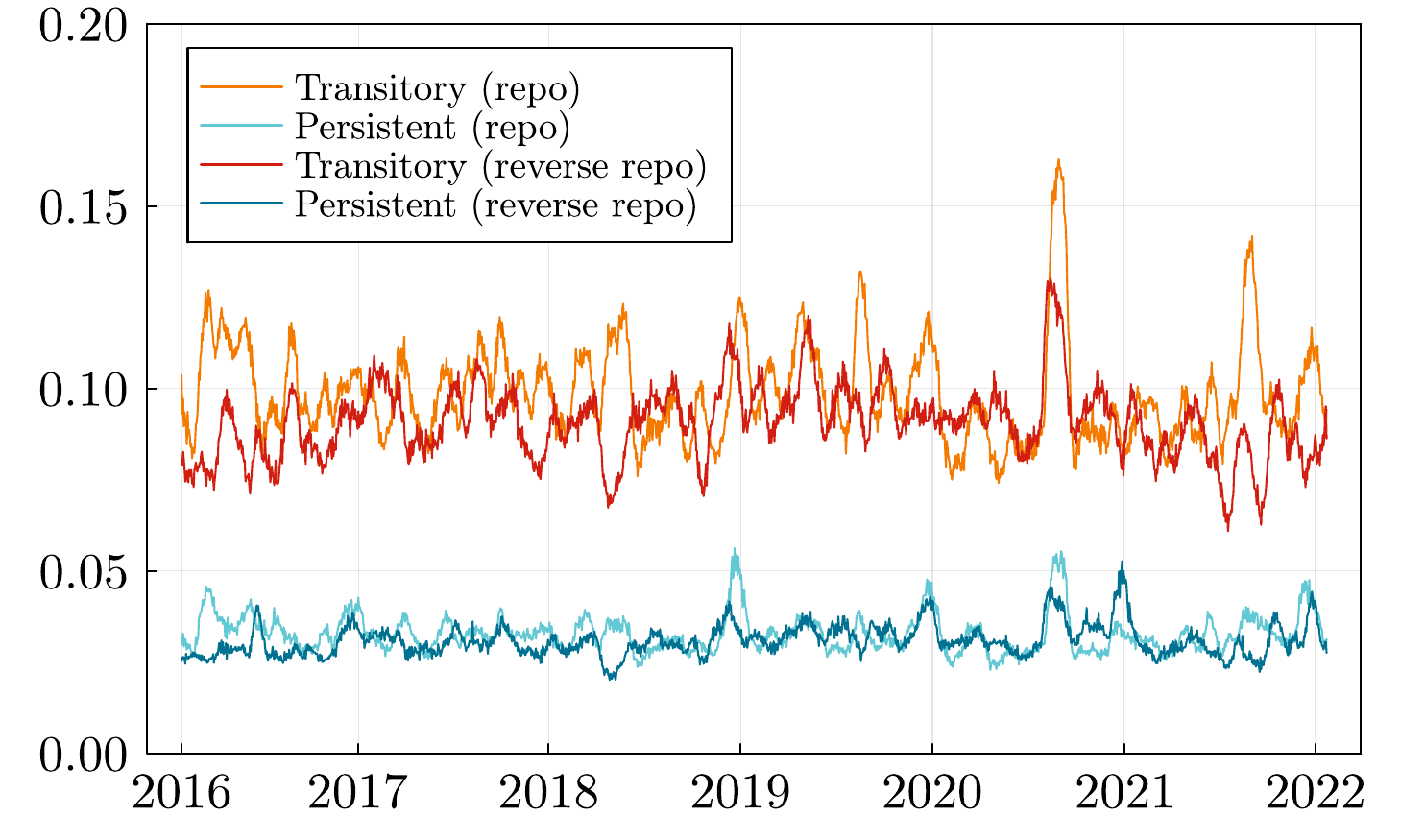}
        \end{subfigure}
        \hspace*{\fill}
        \begin{subfigure}{0.48\textwidth}
        \caption{Rate Spread}
            \label{fig:globalfactor_spread}
            \includegraphics[width=\linewidth]{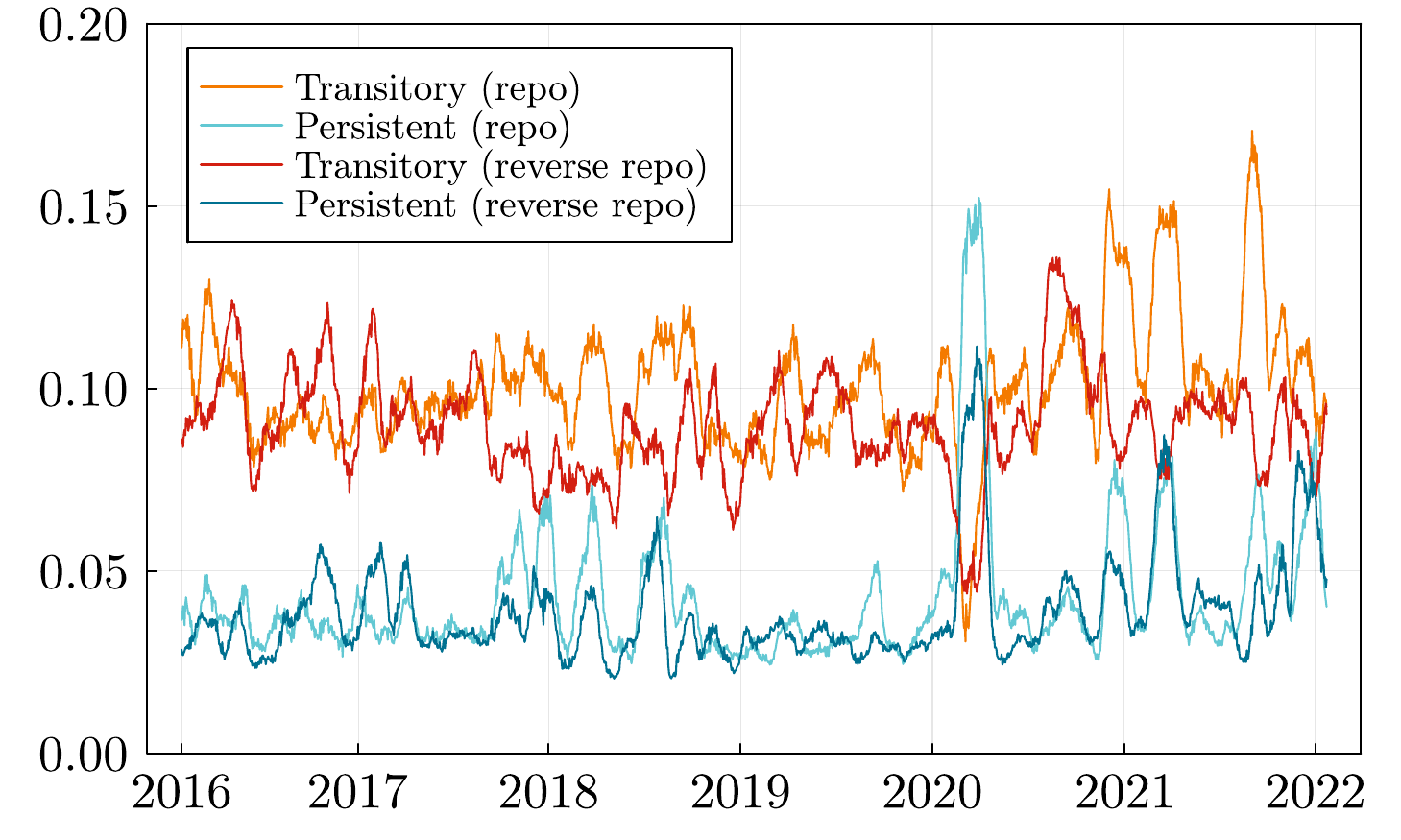}
        \end{subfigure}
\end{figure*}

Figures \ref{fig:globalfactor_volume} and \ref{fig:globalfactor_spread} depict the global dealer factors for both transaction volume and spread on both the repo and reverse repo segments, formed in response to transitory and persistent shocks. Note that we select 5 dealers that explain 40\% of all funding in both the repo and reverse repo markets (note that these may be different), and we can see that the total contribution of this subset of dealers is between 8\% and 18\% (as the sum of transitory and persistent factors). Note that the global factor derived from spread shocks is more volatile.

\subsection{Main Variables}

\paragraph{Market Power}

We use markups and markdowns as a proxy for market power. Dealers operate in the repo segment providing liquidity in exchange for collateral, and in the reverse repo segment, absorbing liquidity also in exchange for collateral. The repo dealers we focus on also have an important role as market makers in the gilt market, and have access to the repo market inter-dealer segment. As non-dealers operate with a small number of dealers, the latter can exercise their bargaining power in both repo segments in their favour.   

The markup in the repo segment is a function of the marginal cost of production, which is not observed in our data.\footnote{A common practice in reduced-form approaches, see \cite{Canonetal2022}, is to use dealers' balance sheet data and estimate their cost function, and thus obtain the marginal cost. This approach is not feasible in our setup due to data availability.} In the context of price competition with differentiated products, a firm's optimal pricing rule in a symmetric equilibrium equates the markup to the inverse of the elasticity of demand for liquidity. Our strategy is to back out the dealer-level markup using a simple structural model of supply and demand, following the standard demand estimation literature in empirical industrial organisation.

For the reverse repo segment, e.g. when dealers absorb liquidity in exchange for collateral, the market power proxy (markdown) is instead a function of the marginal revenue of liquidity, see \cite{Manning2003}, but again unobservable. Analogous to equation \ref{LI}, it's possible to use a one-to-one relationship between the markdown and the inverse elasticity of liquidity supply, see \cite{Yehetal2022}.

The estimation is done in two steps.\footnote{See Online Appendix \ref{appendix: Lerner} for details.} First, we construct supply and demand models for short-term funding and estimate the demand (supply) coefficients using our dataset of dealer/non-dealer transactions. We then use the coefficients to compute own-price elasticities and then apply the price-cost markup (or marginal revenue-price markdown) formula derived from the model to back out product-level markups (markdowns) and compute a proxy for dealer-level market power.  

The optimal pricing rule allows us to derive a markup and a markdown for each dealer in each period $t$. In Online Appendix \ref{appendix: Lerner} we explain the procedure in detail. 
\begin{align}
&Markup_{dt}\equiv \frac{r_{dt}-c_{dt}}{r_{dt}} = - \frac{1}{\eta_{dt}}, & 
Markdown_{dt}\equiv \frac{mrev_{dt}-r_{dt}}{r_{dt}} = \frac{1}{\nu_{dt}} 
\label{LI}
\end{align}

\noindent where $r_{dt}$ is the average repo rate of dealer $d$ at time $t$, $c_{dt}$ is the marginal cost of dealer $d$ at time $t$, $mrev_{dt}$ is the marginal revenue of liquidity of dealer $d$ at time $t$.  Finally, $\eta_{dt}$ ($\nu_{dt}$) is the elasticity of demand (supply).  

The supply and demand framework for short-term funding, though stylized, incorporates several key features observed in the data. On the supply side, a small number of dealers provide liquidity and hold inventories of gilts with varying characteristics, while competing aggressively to attract non-dealer. On the demand side, non-dealers typically operate with a limited set of dealers and are assumed to transact with only one dealer per day. Their choice of dealer depends on transaction characteristics (the repo rate and maturity), collateral characteristics (the market value of the gilt), and dealer-specific liquidity management, proxied by the number of transactions conducted with each dealer in the past month. Our measures of market power capture dealers’ ability to set prices that deviate from the benchmark of \textit{perfect competition}, without conflating this with search, switching, or transaction costs.

\paragraph{Dealer Factors}

Transaction volumes and spreads between dealers and non-dealers in the UK repo market (for both repo and reverse repo) can be altered by shocks between these market participants, which may form an asymmetric network of trading relationships between dealers. We are interested in measuring such a network, in particular, the extent to which dealers that drive most of the volume matter for the rest of the market. In addition, we want to identify such relationships formed by shocks of heterogeneous persistence, namely transitory and persistent shocks transmitted between dealers and non-dealers. More specifically, our global dealer factor will measure the percentage share of shocks from selected dealers that contribute to the overall market variation. In other words, we will measure the extent to which all dealers are influenced by selected dealers relative to the shock transmission across all dealers and non-dealers. Later, we will be interested in exploring how increasing the intensity of the global dealer factor affects volumes, spreads, gilt mispricing, and market-wide liquidity.

Formally, we consider a $N\times N$ dimensional system with $N=N^D+N^{ND}$ number of dealers ($N^D$) and non-dealers ($N^{ND}$) who can transmit shocks to each other throughout the system. To quantify the influence of shocks, we draw on the literature \citep{diebold:2012}, which notes that variance decompositions are useful for tracking how shocks affect the future variation of variables within a system, and are therefore a natural choice for inferring shock transmission. Following \cite{barunik:2018,barunik2024}, who further work with time dynamics as well as heterogeneous persistence of shocks, we will define the global dealer factors that vary smoothly over time and are inferred from shocks of heterogeneous persistence. Specifically, these shocks will be of transitory ($tr$) or persistent ($per$) nature, and we will collect them in a time-varying variance decomposition matrix $\theta_{t,d}$ identified from the underlying TVP-VAR model at a given level of persistence $d \in \{tr,per\}$ as detailed in the appendix \ref{appendix: global_factor}.

Specifically, $\theta_{t,d}$ holds information about how much of the future variation of a variable $j$ is due to a shock with persistence $d$ to the variable $k$ at a given time. This is important because the influence of shocks can evolve smoothly over time.

The impact of a selected set of dealers $Dsel$ on the rest of the dealers in the system, relative to all other shocks that may be transmitted between dealers and non-dealers, then forms our global dealer factor with a given persistence $d$ as
\begin{equation}
\mathcal{G}lobal\mathcal{F}_{t,d}^{Dsel} = \displaystyle\sum_{\substack{j=1\\ j\ne k \\ k\in Dsel}}^{N^D}\Big[\widehat \theta_{t,d}\Big]_{j,k} \Bigg/\displaystyle \sum_{j,k=1}^{N^D+N^{ND}} \Big[ \widehat \theta_t\Big]_{j,k},
\end{equation}
where $DSel$ is the set of selected dealers whose importance we want to measure. Note that $\widehat \theta_{t} = \widehat \theta_{t,tr}+ \widehat \theta_{t,per}$ and therefore $\mathcal{G}lobal\mathcal{F}_{t,d}^{Dsel} \in [0,1]$ measures the contribution of shocks of a specific persistence of selected dealers relative to the total shocks in the system, and $\Big[\widehat \theta_{t}\Big]_{j,k}$ is a $j,k$th element of the time-varying decomposition matrix.

In the estimation, we construct a global factor containing 15 dealers and 7 non-dealer sectors (banks, central banks, asset managers, hedge funds, pension funds, etc.), working with a $22\times 22$ matrix $\widehat \theta_{t}$ measuring the impact of dealers on dealers, dealers on non-dealers, non-dealers on dealers, and non-dealers on non-dealers over the whole period, estimated on volumes and spreads. We will select a number of dealers that account for a large proportion of all funding and we will be interested to see if there is a significant global factor that influences the whole market. The global factor for both repo and reverse repo markets, measured independently of volume and spread shocks, will measure the intensity with which a few selected dealers influence the rest of the market.

\paragraph{Other relevant covariates}

Dealers and non-dealers in the UK repo market are non-randomly matched. In our sample, dealers have an average of nine non-dealer counterparties, while non-dealers have an average of less than three dealers. Moreover, these trading relationships tend to be stable. These features are observed in other OTC markets, e.g. \cite{YueshenZou.23}, and we need to include them in the analysis.

We measure relationship trading in two dimensions, e.g. depth and intensity. Our measure of depth, denoted $Depth_{ND,t}^{D}$ and in line with \cite{Afonso.14,LI.21,Hanetal.22}, is calculated as the share of financing received/provided by dealer D to non-dealer ND out of all financing received/provided by all non-dealers in the previous month.\footnote{Analogously, we could calculate the share of financing received/provided by non-dealer ND from dealer D out of all financing received/provided by all dealers in the previous month. The results are qualitatively the same, and are available on request.}

\begin{equation}
Depth_{ND,t}^D= \dfrac{\sum_{t}^{t-20} volume_{ND,t}^D}{\sum_{t}^{t-20} \sum_{ND} volume_{ND,t}^D}
\label{depth_metrics}
\end{equation}

The equation \ref{depth_metrics} is a proxy for intensity of the trading relationship. A valuable feature of our dataset is that we are able to calculate each metric individually accounting for different segments. In principle, there is nothing to prevent a dealer from having differentiated relationships in the repo and reverse repo segments. Indeed, we observe that some non-dealers only interact with dealers in certain market segments. 

Trading relationships aren't just characterised by their depth.  From the point of view of dealers, two non-dealers with the same depth may have a different impact on inventory constraints if one executes frequently small trades and the other rare albeit large trades.\footnote{\cite{Sambalaibat.23} shows that OTC dealers segment based on the trading needs of their clients. Dealers at the core of the dealer network would serve the most active non-dealers.}  Recurring trades could be associated with proactive risk management strategies, while end-of-month/quarter trades could be associated with window-dressing strategies, see \cite{AnbilSenyuz.18,Gerba2024repo,Bassietal.23}.  

Our measure of intensity, $Intensity_{DND,t}$, is computed as the number of transactions in the last 20 working days between dealer D and non-dealer ND. There are variations of this metric in the relationship trading literature for OTC markets, e.g.\cite{Brauning.17}, but all use the frequency of transactions as the main input source. Higher intensity also implies more opportunities for dealers to learn private information from non-dealers, so while dealers may prefer to increase intensity, non-dealers may not want to, if the trading relationship is sufficiently beneficial to them.
\begin{equation}
Intensity_{DND,t}= \sum_{t}^{t-20} (\mathbf{1}_{volume_{DND,t} > 0})
\label{intensity}
\end{equation}

Our dataset is granular enough to construct the intensity metric individually for repo and reverse repo segments. This feature is convenient as we can measure the relationship of any given dealer non-dealer dyad in each repo segment.

\subsection{Identification}

\paragraph{Market power and relationship-trading metrics}

The identification strategy for the analysis of the repo market poses several challenges. First, reverse causality is expected in relational trading.\footnote{We will properly define these variables in the next section.} Volumes and repo rates between dealers and non-dealers are determined at the dyadic level, as are the measures of relational trading. Thus, if the repo rate increases (relative to a risk-free rate), non-dealers would have incentives to change the frequency of trading and the funding dependence with that dealer.  Second, there could also be reverse causality with market power. Markups and markdowns are constructed at the dealer level. If a dealer has to change the repo rate with an important counterparty, this will change its overall market power.  

To address reverse causality on the proxies for market power and relationship trading metrics, we exploit the granularity of our data to construct granular instrumental variables (see \cite{Garbaix.20}). We recover idiosyncratic shocks at the level of individual dealers and non-dealers and use them to construct valid instruments for each variable. Essentially, we run a principal components factor model for each endogenous variable using the characteristics of the dealer/non-dealer dyad and extract the residuals. The constructed instruments are the size-weighted residuals of the regression of each endogenous variable on the factors, controlled for fixed effects.

For example, depth from the dealers' perspective is assumed to be driven by common and idiosyncratic shocks. The instrument is based on netting dealers' idiosyncratic variation in $Depth_{ND,t}^D$ from their common shock. We use variables that explain the formation of relationships with non-dealers but have little or no correlation with non-dealers' terms of trade, such as volumes and repo rates, to extract the common shock using principal component analysis.  We retain the principal components that explain 90\% of the total variation and use them to run the following $r$ factor model,
\begin{equation} \label{GIV}
    Depth_{ND,t}^D=  \sum^{R}_{r=1} \lambda^{r}_{dl} \eta_{t} + e_{dlt}
\end{equation}
where $\eta_{t}$ is the common shock and $e_{dlt}$ is an idiosyncratic shock. The residuals from equation \ref{GIV}, denoted by $\hat{e}_{dlt}$, are used to compute the GIV as the share-weighted average idiosyncratic shock:

\begin{equation}
     z^{Depth^D}_{t}= \sum_{idl} S_{dl} \hat{e}_{t}
\end{equation}

\noindent where $S_{dl}$ is the dealer/non-dealer $dl$ share of transactions. As an additional robustness check, we include additional terms in equation \ref{GIV}: $\lambda_d$, $\xi_l$ and $\phi_m$ which are dealer, non-dealer, and month fixed effects. To account for non-linearities, we include the square of the granular instrument and call it $z^{Depth^{D}_2}_{t}$. We do the same for the other endogenous variables, so that we end up with eight GIVs, two for each endogenous variable $z^{Depth^D_{t}}$, $z^{Depth^{ND}}_{t}$, $z^{Intensity^{DND}}_{t}$, $z^{LI}_{t}$ and $z^{Depth^{D}_{2}}_{t}$, $z^{Depth^{ND}_{2}}_{t}$, $z^{Intensity^{DND}_2}_{t}$, $z^{LI_{2}}_{t}$. 

\paragraph{Dealer factors}

To identify dealer factors, we use the generalised identification scheme of \cite{pesaran1998generalized} adapted by \cite{barunik2024} to approximate models with time-varying parameters, since a shock to one variable in the model does not necessarily occur in isolation. The use of time-varying frequency responses further identifies the persistence of shocks \citep{barunik2024}. As the global factors derived from the aggregate shock transmission of dealers and non-dealers are used to explain volumes and spreads in both markets at the dyadic level, there is no other type of endogeneity to be addressed in the regressions.


\section{Results}\label{Sect Results}

\subsection{Market Power}

We implement a simple structural model to identify dealer-level markups and markdowns using data from bilateral OTC repo market segments.  We follow a demand approach \citep{Berry.94} and use the equality between predicted shares, given by equation (\ref{choiceprob}) in the Online Appendix, and observed market shares $S_{dt}$ to transform the initial non-linear model into a linear one.\footnote{The Online Appendix section \ref{appendix: Lerner} presents the model in detail.}

\begin{equation}
   \ln S_{dt}-\ln S_{0t} =  \mathbf{x}_{d}\bm{\beta}-\alpha r_{dt}+\gamma I_{dt}+\phi_t+\xi_{d}+\Delta\xi_{dt}.
   \label{berry_corpus}
\end{equation}

\noindent where $\mathbf{x}_{d}$ is a (row) vector of observable product (dealer) characteristics that do not vary over time; $\ln S_{0t}$ is the market share of the outside option; $r_{dt}$ is the average repo rate between dealer $d$ and non-dealers at $t$; $I_{dt}$ is the average frequency, across non-dealers, of dealer $d$'s lending/borrowing relationships at $t$; $\phi_t$ accounts for time shocks common to all transactions observed in the market at $t$; $\xi_{d}$ captures the mean valuation of unobserved dealer characteristics that do not vary over time; $\Delta\xi_{dt}$ are unobserved dealer characteristics that vary over time.

We have three potentially endogenous variables, e.g. the repo rate, frequency and residual maturity, as they are determined by the individual dealers. Dealers $d$ may have incentives to adjust these variables in response to changes in non-dealers' need for funding or securities, or changes in preferences for time-varying dealer characteristics, $\Delta\xi_{ft}$, that are unobserved by the econometrician. To correct for potential bias in our estimates, we exploit both the granularity and the panel structure of our data to generate instrumental variables: two granular IVs (GIVs) for frequency, two difference IVs (DIVs) for repo rate, and three DIVs for residual maturity.\footnote{GIVs are computed as in \cite{Garbaix.20}, and DIVs are computed as the difference of the dealer-level variable from the mean across dealers divided by the standard deviation across dealers. Repo rate DIVs are the unique ISINs in the OTC reverse repo segment and the unique ISINs in the interdealer reverse repo segment. Maturity DIVs are the average maturity in the interdealer repo segment, the percentage of overnight in the OTC repo segment and the average funding in the OTC reverse repo segment.} Thus, the identification assumption is that our instrumental variables are not correlated with demand shocks after controlling for aggregate shocks at the market level and observed and unobserved characteristics at the dealer level. 

We estimate the model (\ref{berry_corpus}) using two-stage least squares. We report the results of the estimation in table \ref{tab:demand}. The top panel presents estimates with the reverse repo segment used to calculate the markdown, and the bottom panel presents estimates with the repo segment used to calculate the markup.  Both panels have two columns: the first shows the estimates without correcting for endogeneity, and the second correcting for it. As very few non-dealers may be large enough to flip bargaining power in their favour, we eliminate the top 5\% of non-dealers (according to their transaction volume) to ensure that dealers' bargaining power is not distorted. Estimates with all non-dealers are qualitatively similar and are available upon request.

\begin{table}[htbp]
\footnotesize
  \centering
  \caption{Demand Estimation:  Repo and Reverse Repo Segments}
  \scriptsize
    \begin{tabular}{lcc}
    \toprule
          & \multicolumn{2}{c}{Reverse Repo Data} \\
    \cmidrule{2-3}
    Repo Rate & -1.649*** & 1.25***  \\
          &  0.139 & 0.130  \\
    Residual Maturity &  -0.003*** & 0.014***  \\
          &  0.001 & 0.003  \\
    Frequency &  0.046*** & -0.227***  \\
          &  0.001 & 0.034  \\
    Collateral Market Value &  -0.000*** & -0.000***  \\
          &  0.000 & 0.000  \\
          &        &        \\
    Control function &  no    & yes    \\
    Month FE &  yes   & yes  \\
    Obs   &  7,822,333 & 7,822,333  \\
    R2    &  0.1266 & 0.1594  \\
          &        &       \\
    \cmidrule{2-3}
          & \multicolumn{2}{c}{Repo Data} \\
    \cmidrule{2-3}
    Repo Rate &  0.992*** & -0.641***  \\
          &  0.139 & 0.183  \\
    Residual Maturity &  0.012*** & 0.009***  \\
          &  0.000 & 0.003  \\
    Frequency &  0.068*** & -0.071  \\
          &  0.002 & 0.054  \\
    Collateral Market Value &  -0.000*** & -0.000***  \\
          &  0.000 & 0.000  \\
          &        &        \\
    Control function &  no    & yes   \\
    Month FE &  yes   & yes    \\
    Obs   &  7,281,915 & 7,281,915  \\
    R2    &  0.1187 & 0.1252  \\
    \bottomrule
    \multicolumn{3}{l}{$^{*}$p$<$0.1; $^{**}$p$<$0.05; $^{***}$p$<$0.01} \\
\multicolumn{3}{@{}l}{\parbox[t]{.75\linewidth}{Note: This table reports the results of the regressions for demand estimation, as discussed in Section \ref{Sect Data Identification}, for the period 2016:M1 to 2022:M1. Definitions, sources and frequency of all independent variables are presented in Section \ref{Sect Data Identification}. The top panel present results using the reverse repo data, and the bottom panel presents the results using repo data.}} \\
    \end{tabular}%
  \label{tab:demand}%
\end{table}%

The coefficients for the repo rate have the expected sign in each repo segment and are highly significant. In the top panel, the positive sign implies that the supply of funds from non-dealers is upward sloping; in the bottom panel, the negative sign implies that the demand for liquidity from non-dealers is downward sloping.  It is interesting to observe that non-dealers providing liquidity are twice as reactive to price changes than when they demand liquidity, and this is consistent with the assumption that dealers have more bargaining power than non-dealers.

The estimates for frequency provide insight into the drivers of market power. Table \ref{tab:demand} shows that market power of dealers decreases as the number of interactions, between dealers and non-dealers in the last month, increases. This characteristic is only statistically significant when we calculate the markdown, but the sign remains when we calculate the markup. This result supports the argument that market power of repo dealers is, at least, partly driven by adverse selection as better trading conditions endure following more frequent transactions.  

Market power of dealers may differ between the repo and reverse repo segments. To suggest otherwise is misleading, as we show in table \ref{tab: demand_both_segments} in the online appendix, where we estimate the demand model using data from both repo segments. In particular, the estimate for the repo rate (column (4)) is no longer statistically significant, and its level (-0.18) is a fraction of that obtained in Table \ref{tab:demand}.


\subsection{Repo Market Impact}

Market power of repo dealers should affect the terms of trade for non-dealers. Theoretically we have two competing views. According to first, we should expect dealers with higher market power to deteriorate terms of trade, i.e. to reduce volumes and increase prices. Alternatively, if the main friction between dealers and non-dealers is asymmetric information, then dealers with higher market power will improve the terms of trade of non-dealers as counterparties trade more frequently.\footnote{\cite{Crawford.18} argue that higher markups reduce the negative impact on prices associated with an increase in adverse selection.} Empirical work has been largely confined to the banking sector; see \cite{Altunbasetal.22, Carlsonetal.22, CruzGarciaetal.21, Canonetal2022}. Studies with transaction-level data, such as the data quality we have, tend to support the first view, i.e. higher levels of market power affect negatively the borrower's terms of trade. 

Our preferred specification is below, 
\begin{eqnarray}\label{estimation: non dealer}
\nonumber \Gamma_{Dlt} &=& \beta_0 + \beta_1 {Market \hspace{.1cm} Power}_{D,t} + \beta_3 {depth}_{ND,t}^D + \beta_2 {frequency}_{DND,t}    \\
\nonumber &+& \beta_3 \log \hspace{.1cm} CB \hspace{.1cm} reserves_t + \beta_4 DF \hspace{.1cm} tot \hspace{.1cm} repo_t + \beta_5 DF \hspace{.1cm} tot \hspace{.1cm} reverse_t   \\
&+& \beta_6 \mathbf{X}_{DND,t} + \beta_7 \mathbf{G}_{D,t} + \beta_8 \mathbf{H}_{t} + \lambda_{d} + \epsilon_{l} + \eta_t + \nu_{dlt}
\end{eqnarray}

\noindent where $\Gamma_{Dt}$ can be the logarithmic volume
between dealer D and non-dealer ND at time t ($lvol_{DND,t}$) or the repo rate spread over the reference rate for the same transaction ($rate_{DND,t}$). The explanatory variables are the market power proxy (i.e. markup or markdown) for dealer D at time t ($Market \hspace{.1cm} Power_{Dt}$), the depth metrics for dealer D (${depth}_{ND,t}^D$) and the monthly frequency of interactions of dealer D and non-dealer ND at time t (${frequency}_{DND,t}$), and the dealer factors for each repo segment ($DF \hspace{.1cm} tot \hspace{.1cm} repo_t$ and $DF \hspace{.1cm} tot \hspace{.1cm} reverse_t$).\footnote{This specification does not distinguish between transitory and persistent shocks to repo dealers.  We introduce the distinction later.}  We include dealer-specific controls ($\mathbf{G}_{D,t}$) and time-specific controls ($\mathbf{H}_{t}$). Finally, we include dealer (non-dealer) fixed effects, non-dealer (dealer) fixed effects interacted with month fixed effects for the situation where dealers (non-dealers) provide liquidity to non-dealers (dealers).\footnote{Following the banking literature (see \cite{Canonetal2022}), the interaction of dealer (non-dealer) fixed effects with monthly fixed effect in the reverse (repo) segment is to controls for unobserved liquidity demand by dealers (non-dealers) confounders.  Alternatively, we use dealer/non-dealer dyad fixed effects and the results are very similar, qualitatively and quantitatively.}

Some explanatory variables are potentially endogenous, i.e. the relationship trading proxies and market power, or might exhibit measurement error. Relationship trading proxies vary at the dyadic level and it may be that changes in the terms of trade affect depth or frequency. Market power varies only at the dealer level, but it could happen that changes in the terms of trade of large non-dealers affect the market power of dealers. Also, market power proxies could suffer from measurement errors attributed to model selection. To correct for these potential estimation biases, we exploit the granularity of our dataset and the panel structure of our data to generate a granular IV for each variable. The identifying assumption is that our instruments are not correlated with demand shocks after controlling for market-level aggregate shocks and dealer-level observed and unobserved characteristics.  

We estimate our preferred specification using two-stage least squares. In all tables, we distinguish estimates for repo transactions from those for reverse repo transactions. As our data is granular enough to identify pairs of dealers and non-dealers that engage in repo and reverse repo on the same day, in online appendix we show results using the subsample of non-dealers that simultaneously operate in both segments. 

Tables \ref{tab:OTC repo vol spread by frequency} - \ref{tab:volume and spread OTC by sector by frequency} show the impact of market power and dealer factors in the repo market. The structure of the tables is always the same. In the top panel, we present estimates for funding, measured as the logarithm of volume, and in the bottom panel we present estimates for the repo spread, in absolute values. In both panels we present estimates for both segments separately. We always distinguish dealer factors attributed to transitory shocks from persistent shocks.\footnote{Table \ref{tab:OTC repo simultaneous by frequency} in the Online Appendix \ref{appendix: tables} shows the results for the estimates on simultanoues dyads.} 

Dealers' interact with non-dealers in two ways, either by providing liquidity in exchange for gilts as collateral, or by receiving liquidity and providing gilts as collateral. In both situations, dealers are in an advantageous position, but we would expect market power to operate differently in each case. In repo transactions, when dealers are liquidity providers, we would expect dealers with higher market power to provide less funding. There is a large literature in banking showing that lenders with higher market power reduce lending, \cite{Altunbasetal.22,Carlsonetal.22,CruzGarciaetal.21,Canonetal2022}. In reverse repos, on the other hand, dealers have opposite incentives and we would expect qualitatively different results.

Table \ref{tab:OTC repo vol spread by frequency} shows the estimates of our benchmark model (\ref{estimation: non dealer}) for the log volume, and highlight the role of relationship metrics and repo dealer factors. As expected the estimates of market power at both repo segments are negative, but they are not statistically significant. On the other hand, the coefficients for relationship metrics (specially \textit{Depth}) are positive and highly significant.\footnote{We provide evidence of a non-monotonic relationship between the importance of non-dealers to dealers and the funding they receive. In tables  \ref{tab: impact on repo Client non-Client} - \ref{tab: impact on repo Client non-Client by frequency} of the Online Appendix \ref{appendix: tables}, we replace the relationship trading proxies with two dummies, e.g. `Client' and `Client Low'.  We find that Client is equal to one if the non-dealer in a given pair is simultaneously above quantile 70 in all relationship trading metrics; on the other hand, a non-dealer is a Low Client if it is simultaneously below quantile 30. Consistently, both dummies are statistically significant, but while Client has a positive sign, Client Low has a negative sign.}  Similarly, the estimates for dealer factors are positive, highly significant, and slightly higher in the reverse repo segment. In terms of magnitudes, the impact on volume is rather small. For example, as the average transaction of dealers in the reverse segment is £65 million, the impact of a persistent shock in the reverse segment, which has the largest magnitude of all estimates, is equivalent to $10^{4.3} = \pounds 19,952m$ .

The bottom panel of Table \ref{tab:OTC repo vol spread by frequency} shows the impact on the (absolute) spread between the repo rate and the bank rate. We consistently find that dealers with higher market power charge or pay higher spreads (which are negative on average, see Table \ref{SumStats: Repo OTC}). In the repo segment, this is consistent with dealers charging a repo rate below the reference rate to access gilts. In the reverse repo segment, dealers with higher market power also receive larger (negative) spreads. It's worth noting that the estimates for the relationship trading proxies are not statistically significant. In terms of magnitude, the story is different vis-a-vis the volume.  Dealers with ($1\%$) higher market power in the repo segment will increase the spread by 0.039 pp, which is roughly half of it's standard deviation (see Table \ref{SumStats: Repo OTC}). In the reverse repo segment, the impact is large as dealers with ($1\%$) higher market power will increase their spread by 0.53 pp, which greatly exceeds its' standard deviations of 0.08.

Dealer factors, particularly transitory shocks, always reduce the spread.  This effect is consistent with the presence of \textit{centrality discount} from core dealers, see \cite{Weill.2020, 
Feldhutter.2012,Sambalaibat.23}. In magnitude, their impact is larger than the impact of market power in the repo segment, e.g. 0.25 pp reduction versus 0.04 pp increase from market power, but smaller for the reverse repo segment, e.g. 0.134 pp decrease versus 0.53 pp increase from market power. 

Returning to Hypothesis \ref{hyp: 1} around channel A, Table \ref{tab:OTC repo vol spread by frequency} shows that dealer-specific market power produces the expected distortions in the repo market, but only through repo rates, as volumes are determined by relationship trading and dealer factors. Hence, we have sufficient evidence to reject the null hypothesis, as the transmission is only through prices.

\begin{table}[htbp]
  \centering
  \caption{OTC impact by persistence of shocks: volume and spread}
  \scriptsize
    \begin{tabular}{lccccc}
    \toprule
           & \multicolumn{5}{c}{Log volume} \\
           \cmidrule{2-6}
         & \multicolumn{2}{c}{Repo} && \multicolumn{2}{c}{Reverse} \\
         \cmidrule{2-3} \cmidrule{5-6}
    Market Power & -0.204 & -0.184 && -0.378 & -0.420 \\
          & [0.136] & [0.124] && [0.260] & [0.264] \\
    Depth & 0.719*** & 0.717*** && 0.927*** & 0.933*** \\
          & [0.238] & [0.238] & &[0.301] & [0.300] \\
    Frequency & 0.008*** & 0.008*** & &0.012*** & 0.012*** \\
          & [0.001] & [0.001] & &[0.002] & [0.002] \\
    DF tran repo & 1.837*** &       & &4.349*** &  \\
          & [0.634] &      & & [0.741] &  \\
    DF tran reverse & -1.037 &     &  & 2.227*** &  \\
          & [0.917] &      & & [0.651] &  \\
    DF per repo &       & 3.944** &&       & -1.031 \\
          &       & [1.562] &&       & [1.259] \\
    DF per reverse &       & -0.511 &&       & 4.290*** \\
          &       & [1.070] &&       & [1.193] \\
          &       &       &  &     &  \\
    Dealer FE & Yes   & Yes   && No    & No \\
    NonDealer FE & No    & No    && Yes   & Yes \\
    NonDealer*Week FE & Yes   & Yes  & & No    & No \\
    Dealer*Week FE & No    & No   & & Yes   & Yes \\
    Year FE & Yes   & Yes  & & Yes   & Yes  \\
    Controls & Yes   & Yes  & & Yes   & Yes \\
    Observations & 92,314 & 92,314 && 115,555 & 115,555 \\
    R-squared & 0.597 & 0.597 && 0.484 & 0.483 \\
    \cmidrule{2-6}
          & \multicolumn{5}{c}{Repo rate spread} \\
          \cmidrule{2-6}
         & \multicolumn{2}{c}{Repo} && \multicolumn{2}{c}{Reverse} \\
        \cmidrule{2-3} \cmidrule{5-6}
    Market Power & 0.034** & 0.039*** && 0.532*** & 0.528*** \\
          & [0.013] & [0.013] & &[0.020] & [0.020] \\
    Depth & -0.003 & -0.003 && 0.008 & 0.009 \\
          & [0.012] & [0.012] && [0.009] & [0.009] \\
    Frequency & -0.000 & -0.000 && 0.000 & 0.000 \\
          & [0.000] & [0.000] && [0.000] & [0.000] \\
    DF tran repo & -0.259*** &    &   & -0.134*** &  \\
          & [0.038] &      & & [0.025] &  \\
    DF tran reverse & -0.073** &  &     & -0.066*** &  \\
          & [0.037] &      & & [0.025] &  \\
    DF per repo &       & 0.021 &  &     & -0.030 \\
          &       & [0.064] & &      & [0.041] \\
    DF per reverse &       & 0.064 &    &   & -0.239*** \\
          &       & [0.079] &    &   & [0.047] \\
          &       &       &   &    &  \\
    Dealer FE & Yes   & Yes   && No    & No \\
    NonDealer FE & No    & No  &  & Yes   & Yes \\
    NonDealer*Week FE & Yes   & Yes &  & No    & No \\
    Dealer*Week FE & No    & No   & & Yes   & Yes \\
    Year FE & Yes   & Yes   && Yes   & Yes  \\
    Controls & Yes   & Yes  & & Yes   & Yes \\
    Observations & 92,314 & 92,314 && 115,555 & 115,555 \\
    R-squared & 0.090 & 0.087 && 0.129 & 0.129 \\
    \bottomrule
    \multicolumn{6}{l}{$^{*}$p$<$0.1; $^{**}$p$<$0.05; $^{***}$p$<$0.01} \\
\multicolumn{6}{@{}l}{\parbox[t]{.75\linewidth}{Note: This table reports the results of the regressions for repo market impact by persistence of shocks to dealer, as discussed in Section \ref{Sect Results}, for the period 2016:M1 to 2022:M1. Definitions, sources and frequency of all independent variables are presented in Section \ref{Sect Data Identification}. Panels report the results for the dependent variables as follows: the top panel uses log of volume, and the bottom panel uses the absolute value of repo spread to the reference rate determined by Bank of England. Clustered standard errors on dealer / non-dealer dyads.}} \\
    \end{tabular}%
  \label{tab:OTC repo vol spread by frequency}%
\end{table}%

Dealers tend to operate in both repo segments, as do a non-negligible fraction of non-dealers. As a robustness test, we repeat the same analysis, but only with dealer/non-dealer dyads operating simultaneously in both segments with the objective of testing if dealers provide different treatment to counterparts that enable them to remain active in both segments. Table \ref{tab:OTC repo simultaneous by frequency} in the Online Appendix \ref{appendix: tables} shows that the results are qualitatively unchanged. For market power, there are no major differences between estimates using all dyads and dyads active in both segments.  The only difference is for the dealer factors, as once the main dealers become more interconnected, only transitory shocks increase repo volumes.  

\subsection{Bond-Level Mispricing}\label{Sect Mispricing}

Market power of repo dealers generates two different frictions that could result in bond mispricing, see \cite{Syverson.2024}. Market power induces the traditional friction of reducing funding (access to gilts) through the repo (reverse repo) segment and distorting repo rates in favour of dealers. A second friction is related to the distribution in markups or markdowns across dealers, which can lead to misallocation of funds in the repo market, as more efficient dealers receive less funding in the reverse repo segment, which in turn can lead to arbitrage frictions in the bond market.\footnote{Relatedly, \cite{Eisfeldt.2024} show, for the corporate bond market, that the interdealer price dispersion explains changes in the yield spread.}  

Our benchmark specification includes both the individual and collective impact of repo dealers on bond market mispricing.

\begin{eqnarray} \label{spec: benchmark mispricing}
    \nonumber y_{bt} &=& \alpha_0 + \alpha_1 Markup_{bt} + \alpha_2 Markdown_{bt} + \alpha_3 DF^{Repo}_{t} + \alpha_4 DF^{Rev}_{t} \\
    & + & \beta_1 X_{bt} + \beta_2 \gamma_{t} + \beta_3 FE_b + \beta_4 FE_t + \epsilon_{bt} 
\end{eqnarray}

\noindent where $y_{bt}$ is the difference, in absolute value, between the yield on bond $b$ at time $t$ and a benchmark yield constructed using a spline-based method, see \cite{AndersonSleath.01},\footnote{\cite{Nymand-Andersen.18} compares well-known methods using ECB data and finds that spline-based methods perform slightly better}. $Markup_{bt}$ and $Markdown_{bt}$ are the (volume-)weighted average market power of dealers trading bond $b$ at time $t$ in each of the repo segments,\footnote{In other specifications, we replace these variables with the (volume-)weighted average standard deviation across dealers trading bond $b$ at time $t$ in each of the repo segments, i.e. $Markup \hspace{.1cm} Dispersion$ and $Markdown \hspace{.1cm} Dispersion$, or the interaction between them.} $DF^{Repo}_t$ and $DF^{Rev}_t$ are the repo dealer factors\footnote{At the implementation stage, we include both the transitory and persistent shocks.} at time $t$ in each repo segment, $X_{bt}$ are bond-level time-varying characteristics such as bid-ask spread, volatility, duration, repo premium,\footnote{Measured as the difference between repo trades using bond $b$ at time $t$ and repo trades using the general collateral category. } residual maturity and the Bank of England free-float share. $\gamma_t$ are market-level time-varying confounders such as the log of Bank of England reserves, a proxy for UK systemic risk, and individual period-specific dummy variables (March 2020 dash-for-cash in the UK, September 2019 US repo market stress period, and Brexit period). Finally, we include bond fixed effects, and month fixed effects.

Table \ref{tab: Dealer Factor} shows the estimates for different versions of equation \ref{spec: benchmark mispricing} using gilts at all maturities. The first two columns show the impact of markup and markdown controlling for endogeneity bias using GIVs, one column includes transitory shocks to dealer factors, and the other includes persistent shocks. Both columns show that, as the (weighted) average market power of dealers accepting/using a specific gilt in repo market increases, the mispricing of that same gilt increases in the bond market. In terms of magnitude, the impact of markup (markdown) is non-negligible, at 0.52 pp (0.64 pp). 

The dispersion of market power between repo dealers can also lead to mispricing of individual gilts.\footnote{Previous studies have raised awareness of the role of market power dispersion in explaining macroeconomic aggregates, see for example \cite{Syverson.2024}.} Columns three and four of Table \ref{tab: Dealer Factor} modify our benchmark specification of equation \ref{spec: benchmark mispricing} by replacing the (weighted) average markup (markdown) with the (weighted) average standard deviation of the markup (markdown). Columns five and six modify the benchmark specification with the interactions between both (weighted) averages, i.e. markup and markup dispersion, and markdown and markdown dispersion. 

Table \ref{tab: Dealer Factor} shows evidence that the (weighted) average standard deviation of market power between dealers is positively correlated with gilt mispricing. The evidence is particularly strong, in a statistical sense, when repo dealers receive funding from non-dealers in exchange for gilts. Then, an increase in markdown dispersion by $1\%$ increases bond mispricing (in absolute terms) by 1.3 pp. We also find that the interaction between market power proxies and their dispersion is not significant, at least when we include gilts of all maturities, but later we observe that this result does not hold for gilts with more than 20 years of residual maturity. 


\begin{table}[]
\centering
\scriptsize
  \caption{Mispricing:  All sample by persistence of shocks}
  \scriptsize
\begin{tabular}{lcccccc}
\toprule
 & \multicolumn{2}{c}{(1) - (2)} & \multicolumn{2}{c}{(3) - (4)} & \multicolumn{2}{c}{(5) - (6)}   \\
\cmidrule{2-7}
 &  &  &  &  &  &    \\
Markup & 0.523*** & 0.522*** & \multicolumn{1}{l}{ } & \multicolumn{1}{l}{ } & 0.189* & 0.197*  \\
 & {[}0.150{]} & {[}0.144{]} & \multicolumn{1}{l}{ } & \multicolumn{1}{l}{ } & {[}0.103{]} & {[}0.102{]}   \\
Markup Dispersion & \multicolumn{1}{l}{ } & \multicolumn{1}{l}{ } & 0.227 & 0.200 & 0.766 & 0.822    \\
 & \multicolumn{1}{l}{ } & \multicolumn{1}{l}{ } & {[}0.288{]} & {[}0.284{]} & {[}2.388{]} & {[}2.354{]}   \\
Markup*Markup Dispersion & \multicolumn{1}{l}{ } & \multicolumn{1}{l}{ } & \multicolumn{1}{l}{ } & \multicolumn{1}{l}{ } & -0.466 & -0.530  \\
 & \multicolumn{1}{l}{ } & \multicolumn{1}{l}{ } & \multicolumn{1}{l}{ } & \multicolumn{1}{l}{ } & {[}1.863{]} & {[}1.822{]}   \\
Markdown & 0.638*** & 0.653*** & \multicolumn{1}{l}{ } & \multicolumn{1}{l}{ } & 0.733*** & 0.737***    \\
 & {[}0.186{]} & {[}0.181{]} & \multicolumn{1}{l}{ } & \multicolumn{1}{l}{ } & {[}0.191{]} & {[}0.179{]}   \\
Markdown Dispersion & \multicolumn{1}{l}{ } & \multicolumn{1}{l}{ } & 1.362*** & 1.397*** & -4.155 & -4.043    \\
 & \multicolumn{1}{l}{ } & \multicolumn{1}{l}{ } & {[}0.258{]} & {[}0.256{]} & {[}3.596{]} & {[}3.605{]}      \\
Markdown*Markdown Dispersion & \multicolumn{1}{l}{ } & \multicolumn{1}{l}{ } & \multicolumn{1}{l}{ } & \multicolumn{1}{l}{ } & 4.362 & 4.295   \\
 & \multicolumn{1}{l}{ } & \multicolumn{1}{l}{ } & \multicolumn{1}{l}{ } & \multicolumn{1}{l}{ } & {[}2.872{]} & {[}2.880{]}   \\
Markdown - Markup & \multicolumn{1}{l}{ } & \multicolumn{1}{l}{ } & \multicolumn{1}{l}{ } & \multicolumn{1}{l}{ } & \multicolumn{1}{l}{ } & \multicolumn{1}{l}{ }  \\
 & \multicolumn{1}{l}{ } & \multicolumn{1}{l}{ } & \multicolumn{1}{l}{ } & \multicolumn{1}{l}{ } & \multicolumn{1}{l}{ } & \multicolumn{1}{l}{ }  \\
DF tran repo & -0.785 &  & -1.084 &  & -0.823 & \multicolumn{1}{l}{ }  \\
 & {[}0.643{]} &  & {[}0.681{]} &  & {[}0.637{]} & \multicolumn{1}{l}{ }    \\
DF tran reverse & 2.080** &  & 1.644* &  & 2.026** & \multicolumn{1}{l}{ }    \\
 & {[}0.895{]} &  & {[}0.950{]} &  & {[}0.897{]} & \multicolumn{1}{l}{ }   \\
DF per repo &  & -3.008** &  & -2.600* &  & -2.974**  \\
 &  & {[}1.235{]} &  & {[}1.377{]} &  & {[}1.237{]}   \\
DF per reverse &  & 4.151** &  & 3.971** &  & 4.124**   \\
 &  & {[}1.711{]} &  & {[}1.955{]} &  & {[}1.725{]}   \\
CB Market Share & -0.230*** & -0.228*** & -0.250*** & -0.249*** & -0.232*** & -0.230***   \\
 & {[}0.079{]} & {[}0.079{]} & {[}0.077{]} & {[}0.077{]} & {[}0.078{]} & {[}0.079{]}  \\
Log CB Reserves & -0.358*** & -0.389*** & -0.101 & -0.149 & -0.313** & -0.348***  \\
 & {[}0.134{]} & {[}0.128{]} & {[}0.148{]} & {[}0.149{]} & {[}0.133{]} & {[}0.128{]}  \\
 &  &  & \multicolumn{1}{l}{ } & \multicolumn{1}{l}{ } &  & \multicolumn{1}{l}{ } \\
Bond*Month FE & Yes & Yes & Yes & Yes & Yes & Yes  \\ 
Year FE & Yes & Yes & Yes & Yes & Yes & Yes  \\
 Controls &  Yes &  Yes &  Yes &  Yes &  Yes &  Yes  \\
 Mat Bucket FE &  Yes &  Yes &  Yes &  No &  Yes &  No   \\
 GIVs &  Yes &  Yes &  No &  No &  No &  No   \\
 Observations &  32,585 &  32,585 &  32,585 &  32,585 &  32,585 &  32,585  \\
 R-squared &  0.033 &  0.035 &  0.030 &  0.031 &  0.035 &  0.037    \\
\bottomrule
\multicolumn{7}{l}{$^{*}$p$<$0.1; $^{**}$p$<$0.05; $^{***}$p$<$0.01} \\
\multicolumn{7}{@{}l}{\parbox[t]{\linewidth}{Note: This table reports the results of the regressions for bond-level mispricing by frequency of shock on dealers, as discussed in Section \ref{Sect Results}, for the period 2016:M1 to 2022:M1.  Definitions, sources and frequency of all independent variables are presented in Section \ref{Sect Data Identification}. The dependent variable is the absolute value of the spread between the bond yield and the predicted yield based on a spline. We use Driscoll-Kraay standard errors with 20 working days lag.}} \\
\end{tabular} \label{tab: Dealer Factor}
\end{table}
  
Table \ref{tab: Dealer Factor by maturity} extends the previous table and reports estimates for gilts of different maturities.\footnote{Short-term gilts have a residual maturity of between 3 and 7 years, medium-term gilts have a residual maturity of between 8 and 19 years and long-term gilts have a residual maturity of 20 years, or more. We exclude gilts with less than three years to maturity because the model selection error is higher for the front segment of the yield curve. All results are robust to the inclusion of the latter and are available on request.} We observe that market power operates mainly through the reverse repo segment. In the first six columns, we observe a positive and significant impact of markdown for all maturities, equivalent to 1 pp at the short-end and 0.5 pp for the rest of the curve, but for markup it is only significant for medium-term gilts and equivalent to 1.2 pp.  

The dispersion of market power between dealers is relevant for the longer end of the curve. Columns seven to twelve show that for short- and medium-term gilts, a $1\%$ increase in markdown correlates with a 1.1 pp increase in bond mispricing. For markup we only observe a 0.6 pp increase for medium-term gilts. Moreover, long-term gilt mispricing is solely affected by the dispersion of markdown and markdown. In terms of magnitude, the effect of markdown is nonlinear. For lower dispersion values, the mispricing effect is close to zero. Yet, it rises rapidly as dispersion increases.  For example, when markdown disperson exceeds one standard deviation, long-term gilt mispricing easily exceeds several percentage points.  

\begin{landscape}
\begin{table}[htp]

\centering
  \caption{Mispricing:  All sample by persistence of shocks and maturity}
  \tiny
\begin{tabular}{lcccccccccccc }
\toprule
 & \multicolumn{6}{c}{(1) - (6)} & \multicolumn{6}{c}{(7) - (12)}  \\ 
 & \multicolumn{2}{c}{  Short} & \multicolumn{2}{c}{  Medium} & \multicolumn{2}{c}{  Long} & \multicolumn{2}{c}{  Short} & \multicolumn{2}{c}{  Medium} & \multicolumn{2}{c}{  Long} \\
\cmidrule{2-13}
& \multicolumn{2}{c}{  } & \multicolumn{2}{c}{  } & \multicolumn{2}{c}{  } & \multicolumn{2}{c}{  } & \multicolumn{2}{c}{  } & \multicolumn{2}{c}{  } \\
Markup & 0.043 & 0.066 & 1.223*** & 1.217*** & 0.126 & 0.109 & 0.064 & 0.079 & 0.642*** & 0.660*** & -0.015 & -0.017   \\
 & {[}0.218{]} & {[}0.214{]} & {[}0.261{]} & {[}0.250{]} & {[}0.179{]} & {[}0.177{]} & {[}0.172{]} & {[}0.175{]} & {[}0.233{]} & {[}0.224{]} & {[}0.132{]} & {[}0.133{]}  \\
Markup Dispersion &    &    &    &    &    &    & -8.775 & -8.262 & 3.852 & 4.116 & 0.756 & 0.787  \\
 &    &    &    &    &    &    & {[}5.720{]} & {[}5.775{]} & {[}4.624{]} & {[}4.948{]} & {[}3.003{]} & {[}2.968{]}  \\
Markup*Markup Dispersion &    &    &    &    &    &    & 7.212 & 6.804 & -3.058 & -3.297 & -0.375 & -0.408  \\
 &    &    &    &    &    &    & {[}4.463{]} & {[}4.498{]} & {[}3.685{]} & {[}3.976{]} & {[}2.381{]} & {[}2.345{]}  \\
Markdown & 1.056*** & 1.072*** & 0.550** & 0.623** & 0.517** & 0.524** & 1.115*** & 1.150*** & 1.072*** & 1.102*** & 0.320 & 0.317  \\
 & {[}0.248{]} & {[}0.244{]} & {[}0.277{]} & {[}0.265{]} & {[}0.234{]} & {[}0.238{]} & {[}0.181{]} & {[}0.174{]} & {[}0.251{]} & {[}0.242{]} & {[}0.230{]} & {[}0.226{]}   \\
Markdown Dispersion &    &    &    &    &    &    & 6.050 & 6.508 & -4.679 & -3.797 & -13.116*** & -12.994***  \\
 &    &    &    &    &    &    & {[}6.852{]} & {[}6.805{]} & {[}5.245{]} & {[}5.175{]} & {[}4.490{]} & {[}4.553{]}   \\
Markdown*Markdown Dispersion &    &    &    &    &    &    & -4.642 & -4.952 & 4.352 & 3.662 & 11.915*** & 11.842***   \\
 &    &    &    &    &    &    & {[}5.431{]} & {[}5.379{]} & {[}4.111{]} & {[}4.048{]} & {[}3.660{]} & {[}3.716{]}  \\
DF tran repo & -0.496 &    & -2.529** &    & -0.512 &    & -0.469 &    & -2.491** &    & -0.557 &     \\
 & {[}0.947{]} &    & {[}1.098{]} &    & {[}0.736{]} &    & {[}0.941{]} &    & {[}1.080{]} &    & {[}0.720{]} &     \\
DF tran reverse & 0.208 &    & 2.409 &    & 3.185*** &    & 0.265 &    & 2.392 &    & 3.117*** &    \\
 & {[}1.284{]} &    & {[}1.498{]} &    & {[}1.095{]} &    & {[}1.289{]} &    & {[}1.487{]} &    & {[}1.091{]} &     \\
DF per repo &  & -3.157** &  & -4.053** &  & -2.687* &  & -3.197** &  & -4.019** &  & -2.643**  \\
 &  & {[}1.366{]} &  & {[}1.880{]} &  & {[}1.371{]} &  & {[}1.351{]} &  & {[}1.869{]} &  & {[}1.347{]}   \\
DF per reverse &  & 4.022** &  & 7.608*** &  & 2.404 &  & 4.009** &  & 7.541*** &  & 2.361   \\
 &  & {[}1.819{]} &  & {[}2.329{]} &  & {[}1.985{]} &  & {[}1.800{]} &  & {[}2.308{]} &  & {[}1.976{]}     \\
CB Market Share & 0.734*** & 0.738*** & -0.677*** & -0.654*** & -0.537*** & -0.533*** & 0.744*** & 0.750*** & -0.659*** & -0.637*** & -0.539*** & -0.534***   \\
 & {[}0.172{]} & {[}0.173{]} & {[}0.107{]} & {[}0.109{]} & {[}0.182{]} & {[}0.181{]} & {[}0.174{]} & {[}0.174{]} & {[}0.108{]} & {[}0.109{]} & {[}0.181{]} & {[}0.181{]}    \\
Log CB Reserves & -0.721*** & -0.900*** & -0.564** & -0.670** & -0.060 & -0.026 & -0.740*** & -0.919*** & -0.557** & -0.658** & -0.007 & 0.021    \\
 & {[}0.240{]} & {[}0.230{]} & {[}0.269{]} & {[}0.275{]} & {[}0.198{]} & {[}0.175{]} & {[}0.240{]} & {[}0.230{]} & {[}0.267{]} & {[}0.274{]} & {[}0.189{]} & {[}0.167{]}    \\
 &  &    &  &    &  &    &  &    &  &    &  &     \\
Bond*Month FE & Yes & Yes & Yes & Yes & Yes & Yes & Yes & Yes & Yes & Yes & Yes & Yes    \\ 
Year FE & Yes & Yes & Yes & Yes & Yes & Yes & Yes & Yes & Yes & Yes & Yes & Yes   \\
Controls & Yes & Yes & Yes & Yes & Yes & Yes & Yes & Yes & Yes & Yes & Yes & Yes    \\
Mat Bucket FE & No & No & No & No & No & No & No & No & No & No & No & No  \\
Observations & 7,536 & 7,536 & 10,770 & 10,770 & 14,274 & 14,274 & 7,536 & 7,536 & 10,770 & 10,770 & 14,274 & 14,274   \\
R-squared & 0.102 & 0.106 & 0.115 & 0.120 & 0.015 & 0.013 & 0.102 & 0.107 & 0.119 & 0.123 & 0.017 & 0.016\\
\bottomrule
\multicolumn{13}{l}{$^{*}$p$<$0.1; $^{**}$p$<$0.05; $^{***}$p$<$0.01} \\
\multicolumn{13}{@{}l}{\parbox[t]{\linewidth}{Note: This table reports the results of the regressions for bond-level mispricing by bond residual maturity and frequency of dealer shocks, as discussed in Section \ref{Sect Results}, for the period 2016:M1 to 2022:M1. Definitions, sources and frequency of all independent variables are presented in Section \ref{Sect Data Identification}. The dependent variable is the absolute value of the spread between the bond yield and the predicted yield based on a spline. We use Driscoll-Kraay standard errors with 20 working days lag.}} \\
\end{tabular} \label{tab: Dealer Factor by maturity}
\end{table}
\end{landscape}

Repo dealers can collectively influence the functioning of the bond market, via the repo market. Returning to \cite{Sambalaibat.23}, if the largest dealers specialise in non-dealers based on their activity or importance, then as the main dealers become more interconnected, certain non-dealer sectors will have access to either more funding or more gilts. We argue that our dealer factors (DFs) capture new information relative to other market-level variables\footnote{For example, the correlation between the DF from the reverse segment and CB market share is 0.3, with log of CB reserves is 0.22, and with UK systemic risk is -0.01. Similar correlations are obtained with the DF from the repo segment}, have the advantage of distinguishing between transitory and persistent shocks and should therefore be included in the benchmark specification.

On the one hand, in Tables \ref{tab: Dealer Factor} and \ref{tab: Dealer Factor by maturity} we observe that transitory and persistent shocks in the reverse repo segment increase gilt mispricing. On the other hand, we observe the opposite effect, for persistent shocks only in the repo segment (same table). In terms of magnitudes, the increase in mispricing through the reverse repo segment is around 1 pp for transitory shocks, and around 4 pp for persistent shocks. The reduction on mispricing through the repo segment, and associated to persistent shocks, is around 3 pp. 

The qualitative difference between the repo and reverse repo segments corresponds to the way in which the main dealers segment the repo market. We argue that the main dealers in the repo segment specialise in pension funds and insurance companies, while in the reverse segment they specialise in hedge funds and asset managers. For example, as the main dealers in the reverse segment become more interconnected, hedge funds and asset managers will increase their funding to dealers and acquire gilts, and the pricing of gilts will deviate from the benchmark price. Conversely, as the repo segment's main dealers become more interconnected, on the margin pension funds and insurance companies will receive more funding in exchange for gilts, and the pricing of gilts will move closer to the benchmark price.

The ``dash-for-cash'' period permits to further test the role of repo dealer segmentation during periods of stress. According to \cite{Czechetal.21, Gerba2024repo, Huseretal.21}, in the dash-for-cash period, hedge funds and PFs \& ICs were net borrowers (in the repo segment), and other non-bank sector, mostly money market funds, decreased their lending to repo dealers (in the reverse repo segment). Finally, non-bank financial institutions significantly increased their activity in the gilt market.

Table \ref{tab:Dash for Cash by frequency} shows that dealer factors have different effects across different gilt maturities during periods of high liquidity stress and high bond market activity. We observe that for medium-term gilts, the impact of DFs was much stronger during the dash-for-cash compared to normal periods. Notably, for long-term gilts, as the main dealers in the reverse repo segment become more interconnected, mispricing during dash-for-cash does not increase, but rather decreases. The latter effect correlates with the fact that hedge funds tend to trade short- and medium-term gilts, whereas pension funds and insurance companies prefer long-term gilts.     

\begin{landscape}
\begin{table}[htbp]
  \centering
  \caption{Mispricing: Dash-for-Cash by persistence}
  \tiny
\begin{tabular}{l|cccccccccccc }
\toprule
 & \multicolumn{4}{c}{(1) - (4)} & \multicolumn{4}{c}{(5) - (8)} & \multicolumn{4}{c}{(9) - (12)} \\ 
 & \multicolumn{4}{c}{   Short} & \multicolumn{4}{c}{   Medium} & \multicolumn{4}{c}{   Long} \\ 
 \hline
 & \multicolumn{4}{c}{ } & \multicolumn{4}{c}{  } & \multicolumn{4}{c}{  } \\ 
Markup & 0.120 & 0.130 & 0.133 & 0.136 & 0.601*** & 0.607*** & 0.619*** & 0.628*** & -0.007 & -0.009 & -0.009 & -0.009 \\
 & {[}0.169{]} & {[}0.170{]} & {[}0.173{]} & {[}0.174{]} & {[}0.218{]} & {[}0.219{]} & {[}0.211{]} & {[}0.211{]} & {[}0.117{]} & {[}0.117{]} & {[}0.119{]} & {[}0.119{]} \\
Markdown & 1.076*** & 1.084*** & 1.109*** & 1.111*** & 1.139*** & 1.153*** & 1.161*** & 1.169*** & 0.415* & 0.409* & 0.410* & 0.402* \\
 & {[}0.178{]} & {[}0.178{]} & {[}0.173{]} & {[}0.172{]} & {[}0.248{]} & {[}0.247{]} & {[}0.239{]} & {[}0.238{]} & {[}0.226{]} & {[}0.227{]} & {[}0.223{]} & {[}0.224{]} \\
DF tran repo & -0.477 & -0.503 &  &  & -2.501** & -2.524** &  &  & -0.577 & -0.563 &  &  \\
 & {[}0.942{]} & {[}0.938{]} &  &  & {[}1.084{]} & {[}1.083{]} &  &  & {[}0.724{]} & {[}0.725{]} &  &  \\
Dum Dash * DF tran repo &  & 10.026* &  &  &  & 8.157 &  &  &  & -9.136*** &  &  \\
 &  & {[}6.034{]} &  &  &  & {[}11.498{]} &  &  &  & {[}2.336{]} &  &  \\
DF tran reverse & 0.251 & 0.228 &  &  & 2.385 & 2.364 &  &  & 3.088*** & 3.098*** &  &  \\
 & {[}1.288{]} & {[}1.286{]} &  &  & {[}1.495{]} & {[}1.496{]} &  &  & {[}1.094{]} & {[}1.094{]} &  &  \\
Dum Dash * DF tran reverse &  & 5.233 &  &  &  & 38.077*** &  &  &  & 1.524 &  &  \\
 &  & {[}13.890{]} &  &  &  & {[}9.466{]} &  &  &  & {[}4.681{]} &  &  \\
DF per repo &  &  & -3.189** & -3.180** &  &  & -4.006** & -3.977** &  &  & -2.574* & -2.612* \\
 &  &  & {[}1.352{]} & {[}1.354{]} &  &  & {[}1.879{]} & {[}1.884{]} &  &  & {[}1.374{]} & {[}1.376{]} \\
Dum Dash * DF per repo &  &  &  & -12.393 &  &  &  & -52.292*** &  &  &  & 22.775*** \\
 &  &  &  & {[}7.752{]} &  &  &  & {[}12.023{]} &  &  &  & {[}4.639{]} \\
DF per reverse &  &  & 4.029** & 4.024** &  &  & 7.556*** & 7.523*** &  &  & 2.316 & 2.351 \\
 &  &  & {[}1.799{]} & {[}1.803{]} &  &  & {[}2.321{]} & {[}2.328{]} &  &  & {[}2.004{]} & {[}2.005{]} \\
Dum Dash * DF per reverse &  &  &  & -15.434 &  &  &  & -58.753*** &  &  &  & 0.433 \\
 &  &  &  & {[}19.758{]} &  &  &  & {[}17.547{]} &  &  &  & {[}4.573{]} \\
CB Market Share & 0.742*** & 0.745*** & 0.748*** & 0.748*** & -0.665*** & -0.666*** & -0.643*** & -0.644*** & -0.536*** & -0.535*** & -0.531*** & -0.530*** \\
 & {[}0.173{]} & {[}0.173{]} & {[}0.174{]} & {[}0.174{]} & {[}0.108{]} & {[}0.108{]} & {[}0.109{]} & {[}0.109{]} & {[}0.182{]} & {[}0.182{]} & {[}0.181{]} & {[}0.181{]} \\
Log CB Reserves & -0.741*** & -0.741*** & -0.919*** & -0.920*** & -0.563** & -0.564** & -0.662** & -0.663** & -0.006 & -0.006 & 0.024 & 0.022 \\
 & {[}0.240{]} & {[}0.240{]} & {[}0.230{]} & {[}0.230{]} & {[}0.267{]} & {[}0.267{]} & {[}0.274{]} & {[}0.274{]} & {[}0.191{]} & {[}0.191{]} & {[}0.170{]} & {[}0.170{]} \\
Dum Dash & -0.285*** & -0.968* & -0.218** & 3.023 & -0.220* & -2.434*** & -0.101 & 12.982*** & -0.049 & 0.270 & 0.009 & -3.312*** \\
 & {[}0.096{]} & {[}0.534{]} & {[}0.090{]} & {[}2.377{]} & {[}0.130{]} & {[}0.403{]} & {[}0.110{]} & {[}2.095{]} & {[}0.070{]} & {[}0.240{]} & {[}0.079{]} & {[}0.869{]} \\
 &     &     &     &     &     &     &     &     &     &     &     &     \\
Bond*Month FE & Yes & Yes & Yes & Yes & Yes & Yes & Yes & Yes & Yes & Yes & Yes & Yes \\
Year FE & Yes & Yes & Yes & Yes & Yes & Yes & Yes & Yes & Yes & Yes & Yes & Yes \\
Controls & Yes & Yes & Yes & Yes & Yes & Yes & Yes & Yes & Yes & Yes & Yes & Yes \\
Observations & 7,536 & 7,536 & 7,536 & 7,536 & 10,770 & 10,770 & 10,770 & 10,770 & 14,274 & 14,274 & 14,274 & 14,274 \\
R-squared & 0.102 & 0.102 & 0.106 & 0.106 & 0.118 & 0.118 & 0.122 & 0.123 & 0.015 & 0.015 & 0.013 & 0.013\\
\bottomrule
\multicolumn{13}{l}{$^{*}$p$<$0.1; $^{**}$p$<$0.05; $^{***}$p$<$0.01} \\
\multicolumn{13}{@{}l}{\parbox[t]{\linewidth}{Note: This table reports the results of the regressions for bond-level mispricing, by frequency of shocks to dealers, including a dummy for the dash-for-cash period and interaction with dealer factors, as discussed in Section \ref{Sect Results}, for the period 2016:M1 to 2022:M1. Definitions, sources and frequency of all independent variables are presented in Section \ref{Sect Data Identification}. The dependent variable is the absolute value of the spread between the bond yield and the predicted yield based on a spline. We use Driscoll-Kraay standard errors with 20 working days lag.}} \\
\end{tabular}  \label{tab:Dash for Cash by frequency}
\end{table}%
\end{landscape}

\subsection{Impact on Market Liquidity}\label{Sect Mispricing Aggregate}

We now turn to the question of whether the individual repo dealer market power and the collective segmentation of the repo market have an impact on (financial) market-level aggregates.  

This paper explores the linkages between the funding and bond markets, see \cite{Brunnermeieretal.2008}, but so far we have focused the analysis at the bond (or gilt) level. We now move on to assess the impact on market liquidity, understood as in \cite{Huetal.13}. To this end, we aggregate equation \ref{spec: benchmark mispricing} at the daily level.

Table \ref{tab:aggregate effect market liquidity} shows evidence that both individual and collective repo dealer actions affect market liquidity. The first three columns incorporate transitory DF shocks, and the last three columns the persistent DF shocks.  We show that higher market power and dispersion of market power indeed deteriorate market liquidity, but only in the reverse repo segment. The latter is an interesting result, as the literature that includes market power generally does not distinguish between repo segments. Also, the relationship between both variables is non-linear, since markdown dispersion increases as market liquidity deteriorates, but holding dispersion constant, as average markdown increases, market liquidity improves. 

Market liquidity responds also to transitory and persistent shocks to dealer factors. Table \ref{tab:aggregate effect market liquidity} shows that persistent shocks in the reverse repo segment deteriorate market liquidity, but same shocks in the repo segment improve it. This result is consistent with previous evidence that shocks to dealers in the reverse segment improves HFs \& MAs activity, and shocks in the repo segment improves PFs \& ICs. Transitory shocks unequivocally deteriorates market liquidity.  

\begin{table}[htbp]
\footnotesize
  \centering
  \caption{Aggregate Effect on Market Liquidity}
  \scriptsize
    \begin{tabular}{lcccccc}
    \toprule
          & \multicolumn{3}{c}{(1) - (3)} & \multicolumn{3}{c}{(4) - (6)} \\
   \cmidrule{2-7}
    Markup &   5.519 &       & 0.947 & 4.911 &       & 0.844 \\
          &   [4.274] &       & [2.352] & [4.090] &       & [2.337] \\
    Markup Dispersion &          & -3.389 & -60.236 &       & -5.155 & -58.373 \\
          &          & [3.540] & [54.056] &       & [3.651] & [53.726] \\
    Markup*Markup Dispersion &          &       & 44.507 &       &       & 41.924 \\
          &          &       & [40.290] &       &       & [40.072] \\
    Markdown &   -3.776 &       & 1.173 & -3.262 &       & 1.091 \\
          &   [4.383] &       & [2.306] & [4.169] &       & [2.299] \\
    Markdown Dispersion &         & 12.458** & -77.307* &       & 11.351** & -82.390** \\
          &          & [4.903] & [39.858] &       & [4.874] & [36.748] \\
    Markdown*Markdown Dispersion &          &       & 64.114** &       &       & 67.709** \\
          &          &       & [30.885] &       &       & [28.206] \\
    DF tran repo &        4.483*** & 4.397*** & 4.342*** &       &       &  \\
          &      [1.042] & [1.099] & [0.962] &       &       &  \\
    DF tran reverse    &   5.729*** & 5.230*** & 5.345*** &       &       &  \\
           & [1.274] & [1.298] & [1.136] &       &       &  \\
    DF per repo &          &       &       & -3.985** & -4.309** & -3.815** \\
          &         &       &       & [1.733] & [1.793] & [1.545] \\
    DF per reverse &       &       &       & 7.616*** & 9.252*** & 6.727*** \\
          &     &       &       & [2.346] & [2.700] & [2.107] \\
    CB Market Share  & 2.725** & 3.048*** & 2.604*** & 2.910** & 3.200*** & 2.854*** \\
           & [1.065] & [1.128] & [0.985] & [1.136] & [1.176] & [1.052] \\
    Log CB Reserves  & -1.468*** & -1.308*** & -1.556*** & -1.414*** & -1.285*** & -1.503*** \\
           & [0.379] & [0.367] & [0.394] & [0.391] & [0.376] & [0.407] \\
    Month FE  & Yes   & Yes   & Yes   & Yes   & Yes   & Yes \\
    GIVs  &  Yes   & No    & No    & Yes   & No    & No \\
    Controls  & Yes   & Yes   & Yes   & Yes   & Yes   & Yes \\
    Observations  & 852   & 852   & 852   & 852   & 852   & 852 \\
    R-squared & 0.231 & 0.222 & 0.242 & 0.226 & 0.219 & 0.236 \\
    \bottomrule
    \multicolumn{7}{l}{$^{*}$p$<$0.1; $^{**}$p$<$0.05; $^{***}$p$<$0.01} \\
    \multicolumn{7}{@{}l}{\parbox[t]{0.8\linewidth}{Note: This table reports the results of the regressions of market liquitidy, as discussed in Section \ref{Sect Results}, for the period 2016 to 2022.  Definitions, sources and frequency of all independent variables are presented in Section \ref{Sect Data Identification}. The dependent variable is market liquidity metrics from \cite{Huetal.13} We use robust standard errors.}} \\
    \end{tabular}%
  \label{tab:aggregate effect market liquidity}%
\end{table}%

Returning to our hypotheses, the evidence from the last 2 subsections indicates that there is not enough evidence to reject the null hypotheses in Hypothesis \ref{hyp: 2} - Hypothesis \ref{hyp: 4}, noting that a larger share of the transmission is via the reverse repo segment. This implies we have sufficient evidence for the existence and quantitative significance of channels B and C. 

\subsection{Overarching discussion}

This paper pins down three channels that result in spillover of frictions from repo to bond market, and create dysfunctionalities at the bond level. Furthermore, we show that all three mechanisms have a market-wide impact by determining overall (financial) market liquidity.  We are \textit{ex ante} agnostic about how these mechanisms interact, as there currently is no theoretical literature to anchor to. Yet, we wish to test, as stated in Hypothesis \ref{hyp: 5}, whether these three mechanisms interact, either additively, or off-setting. 

Again, in Table \ref{tab:aggregate effect market liquidity} we provide evidence that once we include them all in our benchmark specification, they individually and additively contribute to market (il)liquidity.\footnote{The impact on other macro variables is certainly of interest but it's out of scope and we leave it for future research.} As expected, most qualitative results at the bond-level are also observed at the market-level, but noticeably we obtain two new insights.  First, at the aggregate level, the level of market power is less important than its dispersion between dealers. Second, the difference between transitory and persistent shocks to repo dealers is very clear. While transitory shocks undoubtedly worsen market liquidity, persistent shocks on the repo (reverse) segment actually improve (worsen) market liquidity. We argue that the latter qualitative difference between repo segments is explained by our evidence of OTC segmentation, e.g. main dealer from the repo segment specialize in PFs \& ICs, and those from the reverse segment specialize in HFs \& AMs.

Finally,  Table \ref{tab:aggregate effect market liquidity} shows that the interaction between the (weighted) average market power and its dispersion is statistically significant for the reverse segment. So why don't we observe it in both repo segments?  The reverse segment is used by hedge funds and asset managers to obtain gilts from dealers, most of which are on-the-run and might have high demand. Moreover, collateral scarcity observed during most of the sample period generated excess demand for it, which may explain why market power has a disproportionately large role to play in the reverse segment. We therefore have enough evidence to reject the null hypothesis \ref{hyp: 5} given that, although the combined effects are larger than any individual, the transmission of market power to bond mispricing and market liquidity from the two repo segments are, at times, off-setting.

Table \ref{tab: hypotheses} summarize the main takeaways.

\begin{table}[htbp]
\footnotesize
  \centering
  \caption{Hypotheses and Takeaways}
    \begin{tabular}{c|cl}
    \toprule
    \textbf{Hypothesis}	& \textbf{Accept/Reject} &  \textbf{Takeaway}  \\
    \hline
    & & \\
    1  (channel A) & REJECT & Dealer-specific market power (markup and markdown) \\ & & 
    creates inefficiencies
    that affect \textbf{both} repo prices and volumes. \\
    2 (channel A) & ACCEPT & Dealer-specific market power (markup and markdown) impacts \\ & & 
    individual gilts, bond-market liquidity, and (financial) market-wide liquidity. \\
    3 (channel B) & ACCEPT & There is a sizeable inefficiency effect due to the distribution/dispersion \\ & & 
    of market power across dealers (markups and markdowns), \\ & & which impacts bond-level mispricing and market-wide liquidity.\\
    4 (channel C) & ACCEPT & Persistent shocks to repo dealers matter for bond-level mispricing \\ & & 
    and market-wide liquidity.\\
    5 (A+B+C) & REJECT & The combined effects of the three inefficiencies on bond yield deviations \\ & & and market liquidity are sizeable and larger than any of those individually. \\
    \bottomrule
    \end{tabular}%
  \label{tab: hypotheses}%
\end{table}%

\section{Concluding remarks}\label{Sect Conclusion}

Using a proprietary gilt market dataset, we identify how microstructure inefficiencies impact gilt mispricing and market-wide liquidity, and we provide new perspective on how inefficiencies in the money market, measured by market power and global factors, affect the real economy through shocks to individual dealer capacity. 

We contribute to the growing literature on financial market frictions in three ways. First, we pin down the size of the friction using market liquidity metrics and suggest a way to conceptualise it in aggregate or system-wide terms. Second, we explicitly link industrial organisation features in financial markets to individual bond mispricing, which operates through the notion of market liquidity and collateral availability. Finally, we also show how transitory and persistent shocks to repo core dealers affect individual bond mispricing, and market-wide liquidity. The conceptual novelty, we believe, lies in linking micro and macro phenomena in financial markets, by providing empirical basis for financial market frictions in aggregate fluctuations.

\phantomsection
\addcontentsline{toc}{section}{References}
\bibliography{ref}

\newpage
\appendix
\begin{center}
{\huge\textbf{Online Appendix}}
\end{center}

\numberwithin{table}{section}
\setcounter{table}{0}
\numberwithin{equation}{section}
\setcounter{equation}{0}


\section{The UK Repo Market}\label{Sect Repo Description}

\subsection{Overview}

We wish to describe the UK repo market, both in terms of the institutional set-up and the participants.\footnote{It's important to point out the UK's institutional peculiarities in the repo market, as it differentiates significantly from that of the US or the Euro Area.} We follow the description laid out in \cite{Gerba2024repo}.

A repo transaction involves selling a security and agreeing to buy it back at a later date for a pre-agreed price. The security acts as collateral in the case the seller cannot buy it back and offers counterparty risk protection to the buyer who can sell it in order to limit losses. The difference between the sell price and the repurchase price determines the interest rate paid by the seller to borrow cash and is known as the \textbf{repo rate}, or price of the contract. The buyer can demand collateral haircuts to further reduce the risk of losses in case of a counterparty default by overcollateralising the loan. The transaction is called \textbf{repo} from the cash borrower's perspective and \textbf{reverse repo} from the cash lender's perspective. When a pre-specified security is exchanged as collateral, the transaction is known as \textbf{special repo}. On the other hand, when any security satisfying certain criteria, such as specific credit rating, is accepted the transaction is known as \textbf{general repo}. In special repo trades, the rate is typically \textit{lower} compared to general repo, or even negative as demand is driven by cash lenders who are willing to pay interest to obtain the security (e.g. for collateral management purposes). In general repo, the rate is typically higher and positive as demand is driven by cash borrowers who seek funding.

\subsection{Role of the repo market}

\begin{figure}{c}
\begin{center}
    \includegraphics[width=0.6\linewidth]{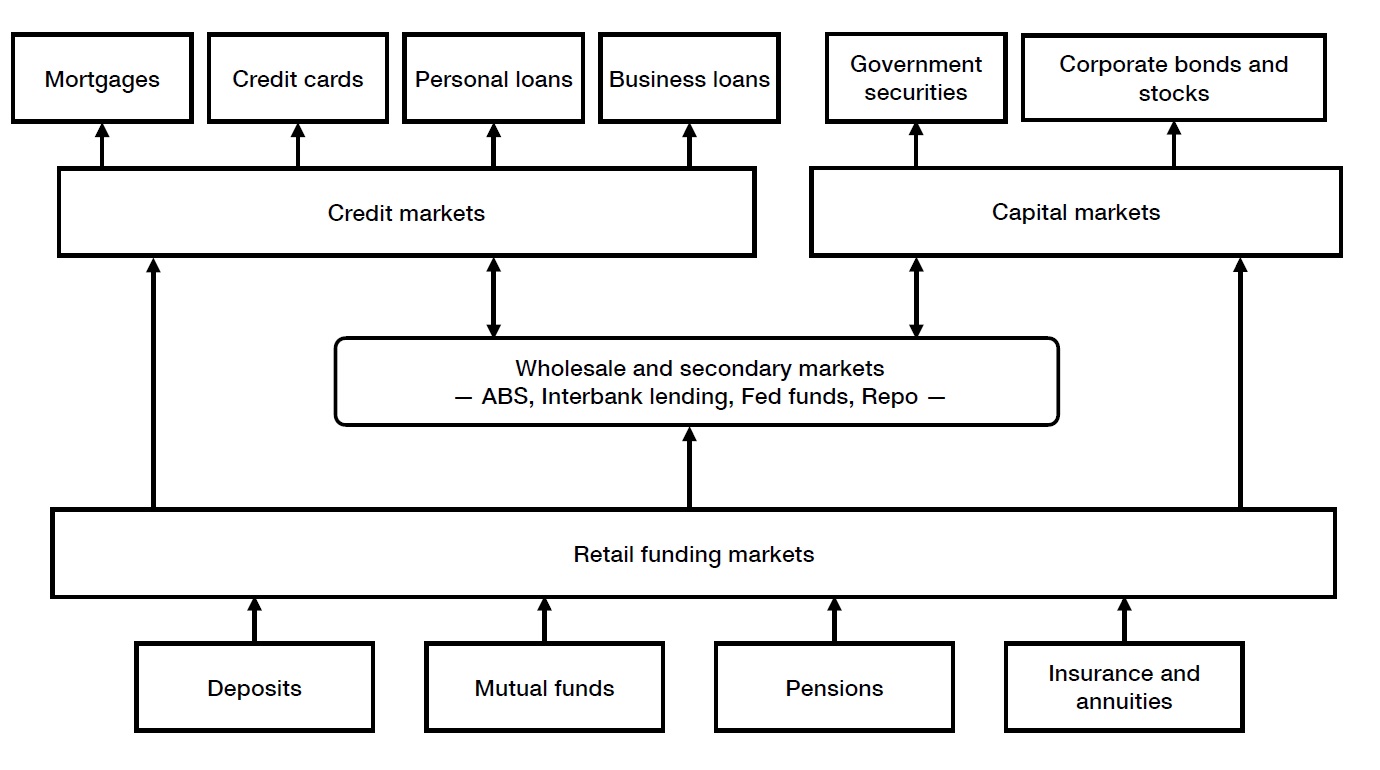}    
\end{center}
\caption{}\label{System fig:xxx et al}
\end{figure}

Figure \ref{System fig:xxx et al}, from \cite{Clarketal.21}, depicts where repos fit into the overall financial market vertical chain (or system). It's evident the essential role played by repos for the modern financial firm, e.g. banks and non-banks, since it's a core source of liquidity and collateral. For example, hedge funds could use the repo to fund long bond positions, or obtain collateral to implement short bond positions.  

\subsection{UK microstructure}

The UK repo market is dominated by bank dealers that intermediate transactions mostly backed by UK government bonds, or \textbf{gilts} \citep{KotidisvanHoren.18}. A wide range of other institutions also participate in the repo market including non-dealer banks, money market funds, central counterparties and corporates that invest cash through reverse repos. Hedge funds, pension funds, asset managers and insurance companies, on the other hand, borrow cash through repos to finance their investments \citep{Bicuetal.17, Huseretal.21}. The market includes transactions that occur between bank dealers (i.e. interdealer segment), which are mostly cleared through CCPs, and transactions that occur between dealers and clients that are mostly traded \textit{bilaterally} or OTC \citep{KotidisvanHoren.18}.

\begin{figure}
\begin{center}
    \includegraphics[width=0.6\linewidth]{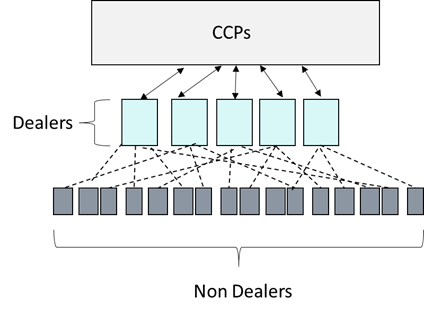}
\end{center}
\caption{Repo Market Structure}\label{Repo fig:Structure}
\end{figure}

Figure \ref{Repo fig:Structure} summarizes the stylized structure of the market in which dealers intervene.  

\begin{figure*}
        \begin{subfigure}{0.45\textwidth}
            \includegraphics[width=\linewidth]{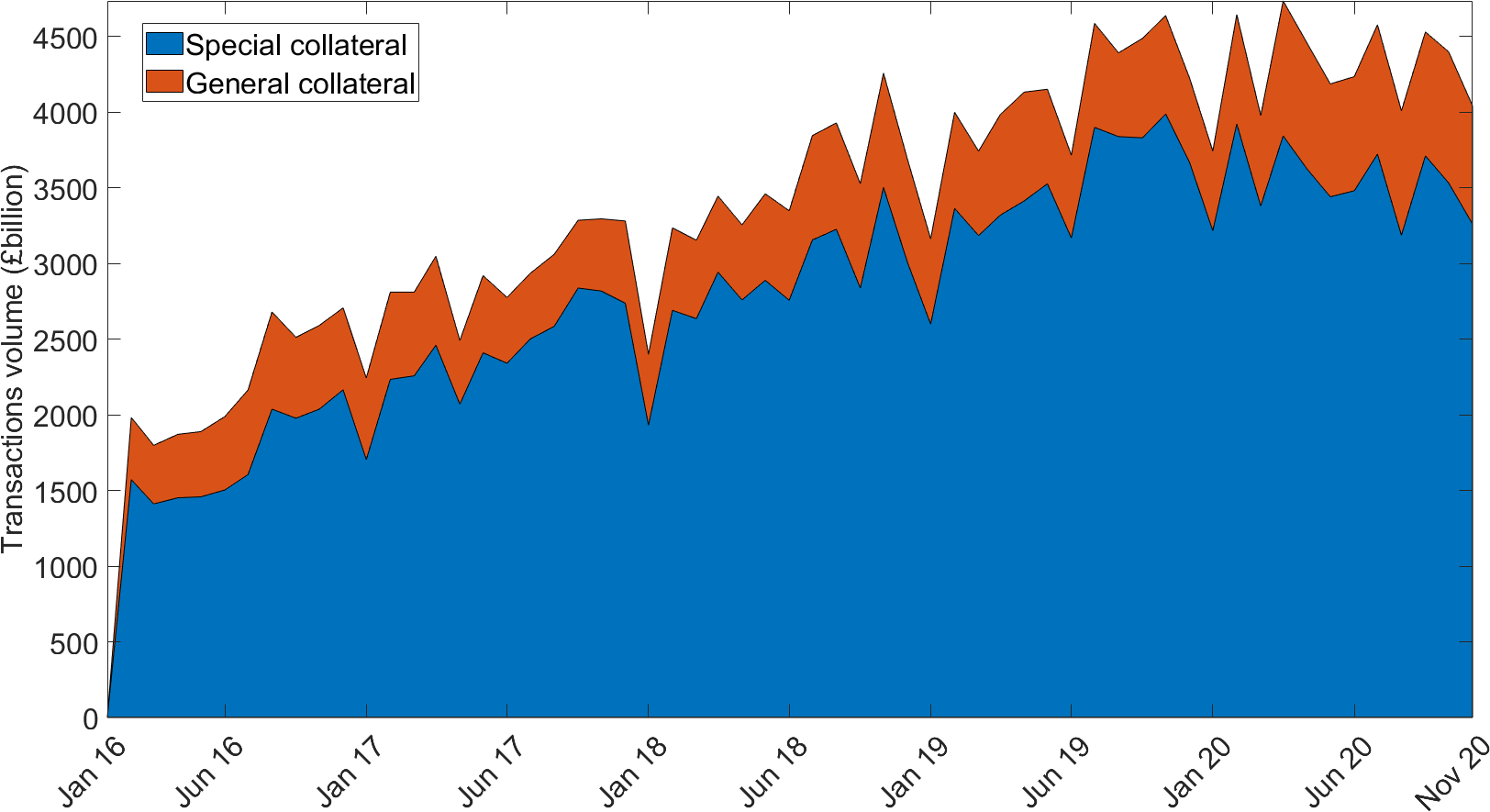}
            \caption{By collateral type (SMMD)}\label{fig:smmd_col}
        \end{subfigure}
        \hspace*{\fill}
        \begin{subfigure}{0.45\textwidth}
            \includegraphics[width=\linewidth]{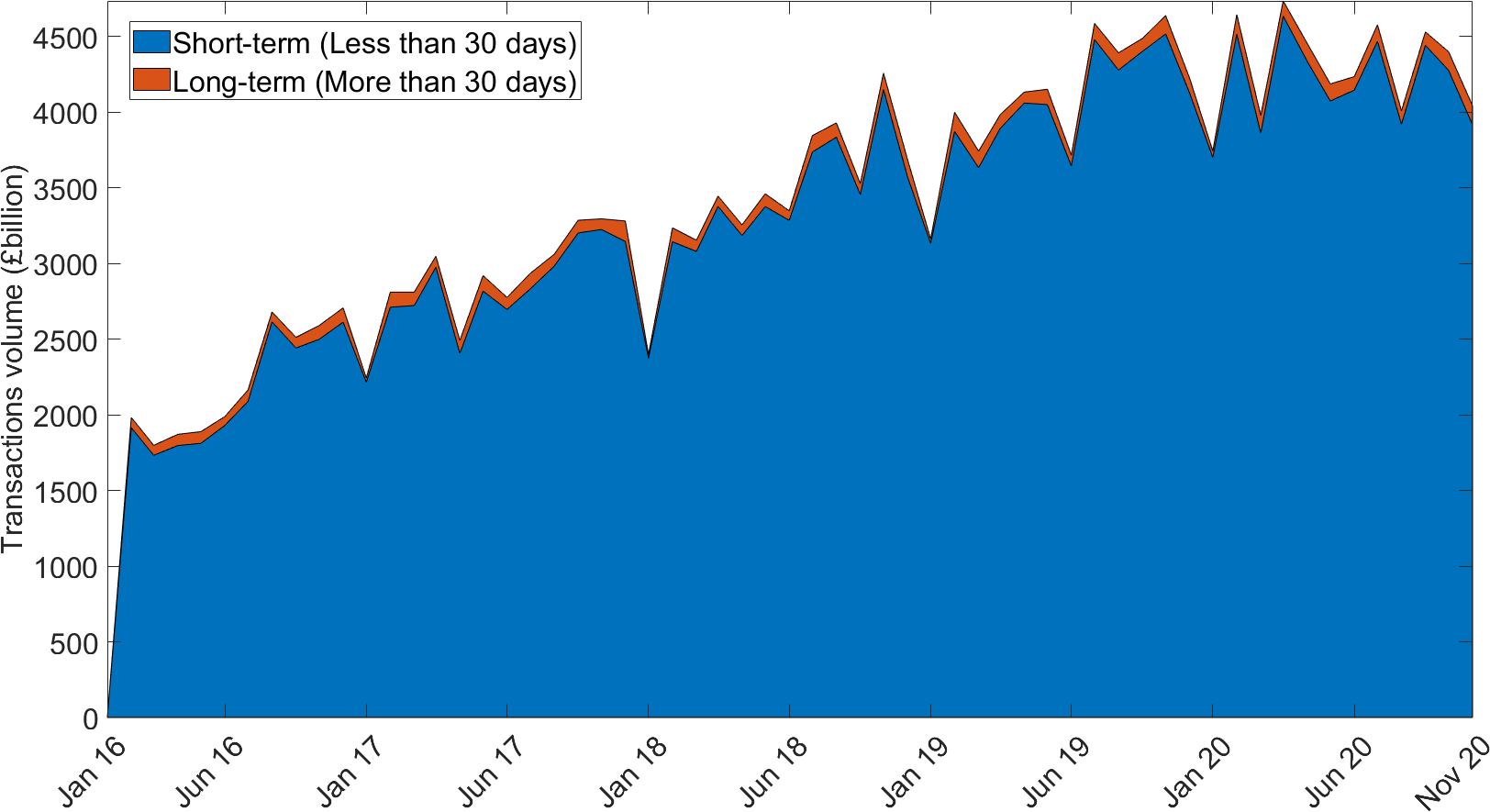}
            \caption{B bucket (SMMD)}\label{fig:smmd_mat}
        \end{subfigure}

        \caption{Monthly gilts repo and reverse repo transaction volumes}
\end{figure*}

\begin{figure}
\centering
\includegraphics[width=0.6\textwidth]{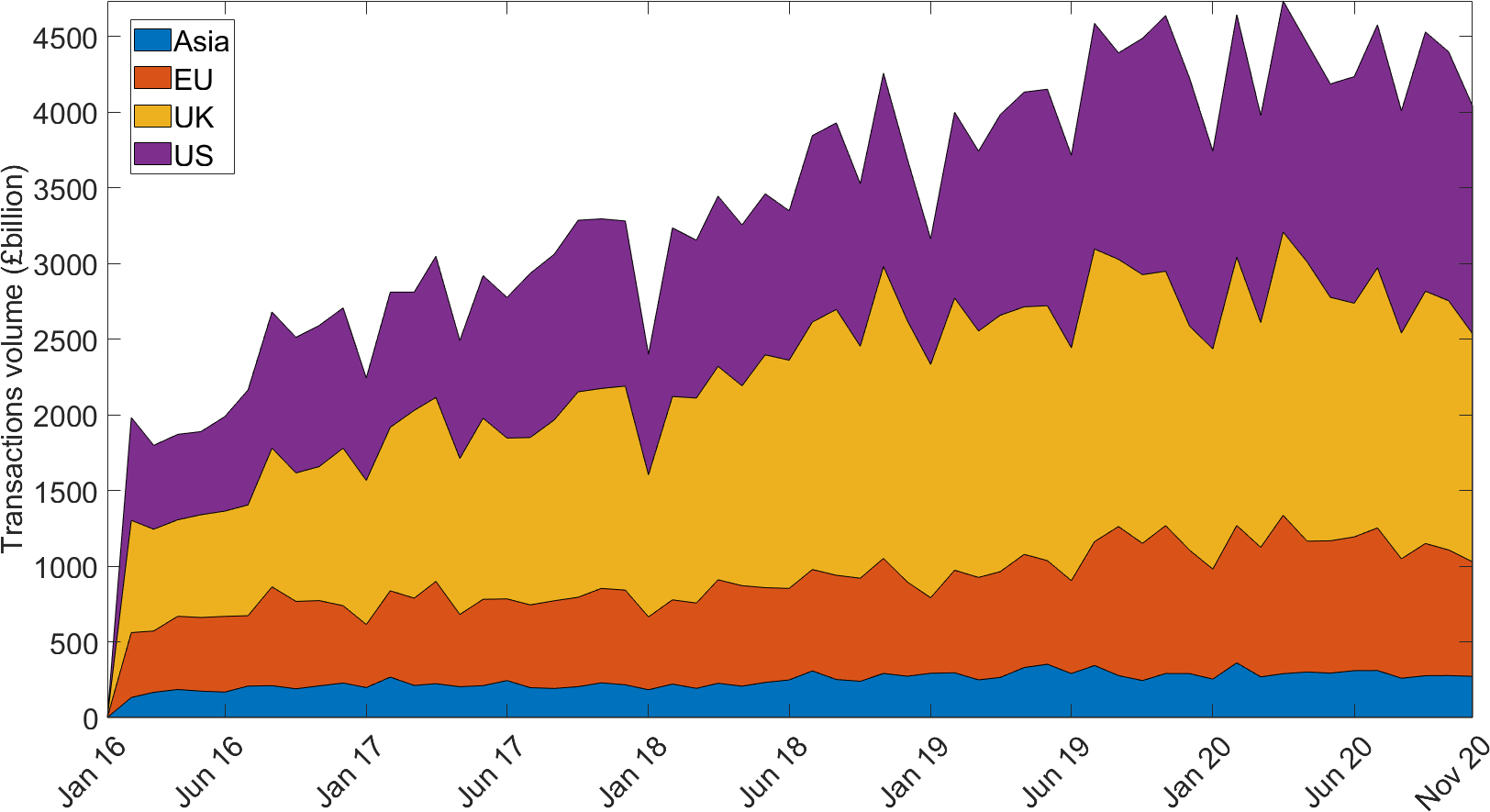}
\caption{Monthly gilts repo and reverse repo transaction volumes by region (SMMD)}
\label{fig:smmd_region}
\end{figure}

In Figure \ref{fig:smmd_col}, we plot the evolution of the UK gilts repo market using Bank of England's proprietary dataset SMMD (or Sterling Money Markets Data), splitting total monthly volume by general collaterial (GC) and special collateral (SC) trades. As can be seen, since 2016 there has been an increase in market activity, mainly SC trades, which constitute 95\% of transactions and 83\% of total transactions volume. In Figure \ref{fig:smmd_mat} we split total transaction volumes by maturity buckets, i.e. less than or equal to 30 days, and more than 30 days. The graph shows that the majority of trades have a maturity of less than 30 days (97\% of transactions and total volume). In Figure \ref{fig:smmd_region} we decompose total volume by participants' headquarters region. As is to be expected, UK-domiciled banks dominate the market in terms of traded volumes (32\% of transactions and 40\% of total volume), followed by US-domiciled banks (36\% of transactions and 33\% of total volume), EU-domiciled banks (24\% of transactions and 20\% of total volume), and Asia-domiciled banks (8\% of transactions and 7\% of total volume). Overall, on a monthly basis transaction volumes have doubled from \pounds2 trillion in 2016 to \pounds4 trillion in November 2020 which highlights the significance of repos as a key funding market in the UK.

To give perspective, we report gilts repo vis-a-vis reverse repo \textit{holdings} separated across maturity and region, as share of the total market. As can be seen, short-term holdings constitute 54\% of the total, indicating that while they represent the vast majority of transactions since they roll-over much more frequently, they are roughly equal to long-term repo and reverse repo holdings. Expressed differently, while in terms of flows (or volumes), short-term clearly dominates long-term, in terms of stocks (or holdings), the two are relatively equal. Furthermore, UK banks hold 45\% of the total volume, followed by US banks with 35\%, EU banks with 14\% and Asian banks with 6\%. 

\begin{figure}
\centering
\includegraphics[width=0.6\textwidth]{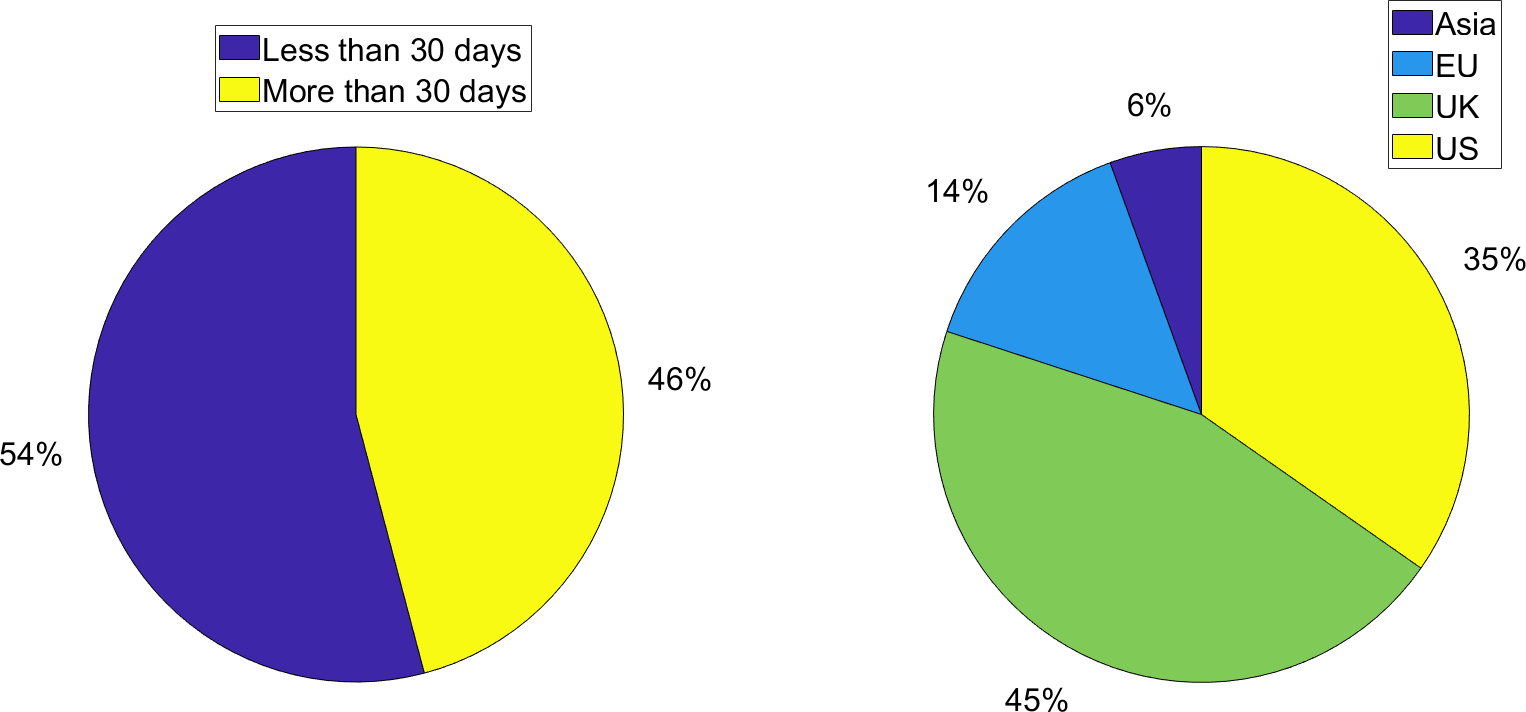}
\caption{gilts repo and reverse repo holdings volumes by maturity bucket and region (SMMD)}
\label{fig:pie_holdings}
\end{figure}

In Panel A of Table \ref{tab:descriptives}, descriptive statistics for daily and quarterly transaction volumes from SMMD are provided. On a daily frequency, the average general repo transaction volume across institutions is \pounds1,542.68 million, while the average general reverse repo transaction volume is \pounds469.48 million. On a quarterly frequency, the corresponding mean volumes are \pounds51,328.04 million and \pounds10,654.75 million. Special collateral transaction volumes follow a similar pattern, with daily average repo volumes being \pounds2,333.68 million and reverse repo volumes being \pounds1,728.02 million. On a quarterly frequency they average \pounds91,351.56 million and \pounds78,072.54 million. Overall, the higher volumes in repo compared to reverse repo transactions corroborates previous findings that banks and dealers are net borrowers in the UK repo market \citep{Huseretal.21}.

\section{Context}
\subsection{UK financial market}


As a global financial centre, London is home to some of the world's largest repo dealers, who also act as market makers in gilts. We analyse two over-the-counter (OTC) segments between 2016 and 2022: Repurchase agreements, where dealers provide cash in exchange for collateral from non-dealers, and reverse repurchase agreements, where dealers receive cash in exchange for collateral. Trading is concentrated within stable dealer–client relationships, which vary by sector. Reverse repo activity favours hedge funds and asset managers seeking on-the-run collateral, while repo activity favours pension funds and insurers looking to fund themselves with gilts. We combine transaction data with gilt-level measures of mispricing relative to a flexible yield curve benchmark and a market-wide liquidity proxy. The sample spans major events such as the US repo stress in September 2019, and the Dash-for-cash events of March 2020. This enables us to study both normal and stressed conditions \citep[e.g.][]{mancini2016euro,Huetal.13}.

We use two key terms throughout: (i) Market power refers to a dealer’s ability to set spreads and quantities differently in bilateral OTC trading than would prevail under perfect competition or full pass-through from interdealer markets, without conflating this with search, switching or transaction costs. Operationally, we estimate dealer-level markups (repo) and markdowns (reverse repo) from a structural supply–demand framework at the dealer–counterparty dyad level. This takes into account bond characteristics and rich time fixed effects.  Our short-term funding model, though stylized, reflects key market features. (ii) Persistent shocks refer to changes in dealers' intermediation capacity or willingness that endure beyond short-lived fluctuations (e.g. sustained balance sheet constraints or shifts in collateral availability). We identify persistence using a dynamic network approach based on time-varying VARs and generalised variance decompositions \citep{lutkepohl2005new,pesaran1998generalized,diebold:2012}. Additionally, we interpret dealer factors as market-wide influences arising when a subset of dealers transmits shocks of heterogeneous persistence through the trading network. These factors capture the extent to which repo market outcomes are shaped by a few influential dealers.

In line with standard IO literature \citep{Syverson.2024}) and using bilateral OTC segment data, we calculate dealer-level markup and markdown metrics. Additionally, in line with the literature on relationship lending \citep{Brauning.17,Jukartisetal.22}, we calculate dealer- and non-dealer-level metrics for relationship trading. Finally, based on recent literature on dynamic networks \citep{barunik2020dynamic,barunik2024}, we calculate market-wide dealer global factors. For the bond market, we construct bond-level mispricing metrics and a market-wide metric of market liquidity, following literature on mispricing \citep{LEWIS.21,pelizzon2025central,Huetal.13}.

Our contribution to the literature is characterised by several key findings. First of all, the market power of individual dealers is economically significant in bilateral over-the-counter (OTC) transactions. Higher market power metrics are associated with larger absolute spreads and quantities that deviate from competitive benchmarks. These distortions result in mispricing at the gilt level. Next, the dispersion of market power among dealers is important for aggregate liquidity. Holding the average level fixed, a more uneven distribution of bargaining power is associated with lower market-wide liquidity and greater deviations of gilt prices from benchmark values. Finally, the persistence of dealer shocks affects transmission: persistent shocks to influential reverse-repo dealers increase gilt mispricing, whereas persistent shocks to influential repo dealers reduce it. This is consistent with improved funding availability for pension funds and insurers against gilts. By contrast, transitory shocks have materially smaller effects.

We emphasise that market power has a stronger effect in the reverse repo segment. Here, the demand for scarce collateral from hedge funds and asset managers amplifies scarcity premia and pushes gilt prices away from the benchmark values. By contrast, a sustained increase in the influence of repo dealers tends to ease funding conditions against gilts and support the realignment of prices with fundamentals.

The second key finding, concerning the dispersion of market power, also shows that the impact operates at both the aggregate liquidity level and the individual bond level. The mispricing effects are in the same direction as the effects associated with illiquidity metrics, which highlights that dispersion in market power reinforces bond-specific pricing frictions.

\section{A stylized structural model for markups and markdowns}\label{appendix: Lerner}

We implement a simple structural model to retrieve dealer-level markups and markdown using data from bilateral OTC repo market segment. In the repo segment, we calculate the markup of dealers, as they are the only providers of liquidity to non-dealers. In the reverse repo segment, we calculate the markdown of dealers, as they are the only buyers of liquidity from non-dealers.

Below, we explain in detail the case for markups. Markdowns have an analogous specification and we comment briefly on the main differences.\footnote{We argue repo dealers have both product and factor market power. Our approach is an approximation for markup and markdown estimation as we do not jointly estimate them. Following \cite{Syverson.2024}, a profit maximizing dealer's first-order conditions show that a combination of both metrics is equal to the ratio of elasticity of output wrt liquidity, and dealer's expenditure on liquidity expressed as a share of revenue. Our approach responds to this , but with data limitation.} 

\textbf{Supply side}. Suppose that there are $D$ dealers in the repo market, indexed by $d=1,\dots,D$ each of which supplies liquidity or collateral through market making services. For the sake of simplicity, we assume that dealers are single-product firms; that is, each dealer supplies repo/reverse repo with fixed characteristics (e.g., interest rate, maturity and haircut) to non-dealers looking for liquidity/collateral in the bilateral OTC market. Accordingly, datawise, we aggregate all transactions made by a given dealer in a single product indexed by the same index of the dealer; the characteristics of that product correspond to averages of the observed lending terms across non-dealers that transacted from the same dealer in period $t$.\footnote{This is a restrictive assumption relative to how the market works in reality, in which a given dealer/non-dealer pair may agree on several transactions in the same day, each of them characterized by a particular volume, a given interest rate, a given maturity, and a given collateral. Products provided by the same dealer to the same non-dealer may, therefore, be heterogeneous; we, therefore, obtain an average measure of the true market power of each dealer, which can vary across non-dealers.} The variable profit of dealer $d$ derived from its transactions in the repo market is given by:
\begin{equation}\notag
\Pi_{dt} = (r_{dt}-c_{dt})M_ts_{d}(\mathbf{r}_d),
\end{equation}

\noindent where $r_{dt}$ is weighted average interest rate given by $d$ at time $t$, $c_{dt}$ is the dealer's marginal cost, $s_{dt}$ is the market share of dealer $d$ at time $t$, $\mathbf{r}_t$ is the $F\times 1$ vector of interest rates of all dealers in the market, and $M_t$ is the size of the repo market, which we take as all of the money borrowed by dealers and non-dealers in the repo market from any firm providing wholesale liquidity at time $t$.\footnote{For simplicity, we assume that dealers expect the full repayment of each loan from borrowers and that they actually do the full repayment of their loans; this implies that there is no loss of profits due to default. For a model that explicitly accounts for default, see \cite{Crawford.18}.} We assume that dealers compete in setting interest rates and that a pure-strategy Bertrand-Nash equilibrium in prices exist. Therefore, the interest rate of the lending supplied by dealer $d$ must satisfy the first order condition:
\begin{equation}\notag
s_{d}(\mathbf{r}_t)+(r_{dt}-c_{dt})\frac{\partial s_{d}(\mathbf{r}_t)}{\partial r_{dt}}=0.
\end{equation}

We have, therefore, a system of $D$ equations, one for each of the dealers (products) existing in the market. 
 Solving dealer $d$'s equation for its interest rate-cost margin yields, for $d=1,\dots,D$:
\begin{equation}
r_{dt}-c_{dt}=\frac{1}{-\frac{\partial s_{d}(\mathbf{r}_t)}{\partial r_{dt}}}s_d(\mathbf{r}_t).
\label{mkup}
\end{equation}

This optimal pricing rule allows us to back out a markup index for each dealer in each period $t$. Dividing the two sides of equation (\ref{mkup}) by $d$'s interest rate yields:
\begin{align}
&{Markup}_{dt}\equiv \frac{r_{dt}-c_{dt}}{r_{dt}} = \frac{1}{\eta_{dt}}, &\mbox{with}&
&\eta_{dt}=-\frac{\partial s_{d}(\mathbf{r}_t)}{\partial r_{dt}}\frac{r_{dt}}{s_d(\mathbf{r}_t)},
\label{LI}
\end{align}

\noindent being the positive own price elasticity of demand. The Lerner index being a function of the demand elasticity implies that we do not need to observe the marginal costs of dealers to estimate the Lerner index, but to have a good estimate of the own price elasticity of demand.

The case of markdown is similar.  Dealer $d$ maximize profit $\Pi^f_{dt} = (mrev_{dt}-r_{dt})M^f_ts^f_{d}(\mathbf{r}_d)$ , where $mrev_{dt}$ is the marginal revenue from aggregate funding obtained from non-dealers, and $r_{dt}$ is the average repo rate from such funding. The remaining variables are, just as with markup but superscript $f$  refers to the factor market (e.g. reverse repo segment), where dealer $d$ has market power.  Following a similar first order condition approach, the markdown is inversely related to the elasticity of supply ($\nu_{dt}$).
\begin{align}
&{Markdown}_{dt}\equiv \frac{mrev_{dt}-r_{dt}}{r_{dt}} = \frac{1}{\nu_{dt}}, &\mbox{with}&
&\nu_{dt}=\frac{\partial s^f_{d}(\mathbf{r}_t)}{\partial r_{dt}}\frac{r_{dt}}{s^f_d(\mathbf{r}_t)},
\label{LI_Markdown}
\end{align}

\textbf{Demand model.} The demand model presented in this section is in the spirit of the empirical industrial organization literature (in particular, \cite{Berry.94} and \cite{Nevo.00}). Non-dealers, indexed by $l=1,2,\dots,ND$ face a multiple-choice decision among $d$ dealers in each period. Assume that the conditional indirect utility of non-dealer $l$ from choosing to borrow money from dealer $d$ at time $t$ is given by:
\begin{equation}
u_{ldt}=\mathbf{x}_{d}\bm{\beta}-\alpha r_{dt}+ \gamma I_{dt}+\phi_t+\xi_{d}+\Delta\xi_{dt}+\varepsilon_{ldt}
\end{equation}

\noindent where $\mathbf{x}_{d}$ is a (row) vector of observable product (dealer) characteristics that do not vary with time; $r_{dt}$ is the mean interest rate of the lending granted by dealer $d$ to non-dealers at $t$; $I_{dt}$ is the mean intensity, across non-dealers, of dealer $d$'s lending relationships at $t$; $\phi_t$ accounts for time shocks that are common to all of the transactions observed at $t$ in the market; $\xi_{d}$ captures the mean valuation of the unobserved dealer characteristics that do not vary with time; $\Delta\xi_{dt}$ are unobserved dealer characteristics that vary with time; and $\varepsilon_{ldt}$ is an additively separable mean-zero random shock that captures idiosyncratic non-dealer preferences.\footnote{There is no practical distinction between markup and markdown specification. Noticeably, the main difference is the expected sign of $\alpha$, e.g. negative for markup and positive for markdown. } 

We assume that non-dealers' choice set includes an ``outside good'', which may capture all other liquidity sources not considered in this analysis. Normalizing its mean utility to zero, the indirect utility derived by non-dealer $l$ from the outside option writes as $u_{l0t}=\varepsilon_{l0t}$. Another key assumption of this model is that non-dealers choose at most one product (i.e., dealer) at each period $t$. The product (dealer) chosen is the one giving the highest utility. For given unobserved demand shocks, $\bm{\varepsilon}_{ldt}$, bank $l$ will choose product $d$ if:
\begin{equation}\notag
u_{ldt}\geqslant u_{lkt}, \; \forall \, k=0,1,\dots,D.
\end{equation}

Assuming that the shocks to utility $\varepsilon_{ldt}$ are independent of the product characteristics and of each other (i.i.d.), and drawn from a Type 1 Extreme Value distribution, the market share of dealer $d$ at time $t$ is given by:
\begin{equation}
s_{d}(\mathbf{X},\mathbf{r}_{t})=\frac{\exp(\mathbf{x}_{d}\bm{\beta}-\alpha r_{dt}+\gamma I_{dt}+\phi_t+\xi_{d}+\Delta\xi_{dt})}{1+\sum_{k}\exp(\mathbf{x}_{k}\bm{\beta}-\alpha r_{kt}+\gamma I_{kt}+\phi_t+\xi_{k}+\Delta\xi_{kt})},
\label{choiceprob}
\end{equation}

where $\mathbf{X}$ is the matrix of observed characteristics of all of the included dealers, that do not vary over time. 

\textbf{Elasticities.}
The non-dealer-level, own- and cross-price demand elasticities are given by:
\begin{align}
\eta_{ldkt}=\frac{\partial s_{dt}}{\partial r_{kt}}\frac{r_{kt}}{s_{dt}}=\left\{\begin{array}{rcl} -\alpha(1-s_{dt})r_{dt} & \mbox{id}\ \ d=k,\\[5pt]
\alpha  	s_{kt}r_{kt} & \mbox{if}\ \ d\neq k.\end{array}\right.
\label{elasts}
\end{align}

The non-dealer-level, own- and cross-price supply elasticities are given by:
\begin{align}
\nu_{ldkt}=\frac{\partial s^f_{dt}}{\partial r_{kt}}\frac{r_{kt}}{s^f_{dt}}=\left\{\begin{array}{rcl} \alpha(1-s^f_{dt})r_{dt} & \mbox{id}\ \ d=k,\\[5pt]
\alpha  	s^f_{kt}r_{kt} & \mbox{if}\ \ d\neq k.\end{array}\right.
\label{elasts_Markdown}
\end{align}

\textbf{Estimation and results.} We follow \cite{Berry.94} and use the equality between predicted shares, given by equation (\ref{choiceprob}), and observed market shares $S_{dt}$ to transform our non-linear model to a linear one. Formally, the model we obtain is given by:
\begin{equation}
   \ln S_{dt}-\ln S_{0t} =  \mathbf{x}_{d}\bm{\beta}-\alpha r_{dt}+\gamma SI_{dt}+\phi_t+\xi_{d}+\Delta\xi_{dt}.
   \label{berry}
\end{equation}

Notice that our demand model allows for unobserved factors at time $\phi_t$ and fund levels $\xi_{d}$. We account for those unobservables by including time and fund dummies, respectively. The latter capture also all of the observed fund attributes that do not vary over time  $\mathbf{x}_{d}\bm{\beta}$. We do not account for dealer-time unobserved factors, $\Delta\xi_{dt}$; thus, we leave it as the error term of the model.



\section{Global Dealer Factors: Estimation}
\label{appendix: global_factor}

\subsection{Overview of the time-varying parameter VAR model}

Since the behaviour of dealers and non-dealers is highly dynamic, we approximate it with time-varying parameter vector autoregression (TVP-VAR) model: 
\begin{equation}
x_{t,T}=\phi_0\left(\frac{t}{T}\right)+\phi_1\left(\frac{t}{T}\right)x_{t-1,T}+\ldots+\phi_p \left(\frac{t}{T}\right)x_{t-p,T} + \epsilon_t,
\label{eq:VAR2}
\end{equation}
estimated from ($N^D+N^{ND}$) dimensional vector of volume of the repo or reverse repo dealers and non-dealers $x_{t,T}=\left(Vol^D_{1,t,T},\ldots,Vol^{D}_{N^D,t,T},Vol^{ND}_{1,t,T}\ldots,Vol^{ND}_{N^{ND},t,T}\right)'$ in one system, or from ($N^D+N^{ND}$) dimensional vector of rate spread of the repo or reverse repo dealers and non-dealers $x_{t,T}=\left(Rate^D_{1,t,T},\ldots,Rate^{D}_{N^D,t,T},Rate^{ND}_{1,t,T}\ldots,Rate^{ND}_{N^{ND},t,T}\right)'$ in day $t=1,\ldots,T$ following the methodology of \cite{barunik2024}. Note the variable $x_{t,T}$ is triangular array of observations $(x_{t,T};t=1,\ldots,T)$ with sample size $T$ that can be interpreted as a locally stationary approximation around a fixed point $t/T$. As a consequence, the process can change its properties smoothly over time. 

An important feature of such local approximation is the possibility to represent a locally stationary process $x_{t,T}$ as a time-varying vector moving average:
\begin{equation}
\label{eq:wold}
x_{t,T} = \sum_{h=-\infty}^{+\infty} \alpha_{t,T}\left(h\right) \epsilon_{t-h},
\end{equation}
where coefficients $\alpha_{t,T}\left(h\right)$ can be approximated under certain smoothness assumptions a $\alpha_{t,T}\left(h\right) \approx \alpha(t/T,h)$, see \cite{dahlhaus1996kullback}. The innovations $\epsilon_t$ are independent random variables with zero mean and finite variation. The original process $x_{t,T}$ can be represented as a linear combination of uncorrelated innovations with time-varying impulse response (TV-IRF) functions $\alpha_{t,T}\left(h\right)$. The information contained in the impulse response functions allows the contribution of shocks in the system to be measured. Thus, their transformations over time determine the interconnectedness of dealers and non-dealers. 

We further transform local impulse responses in the system into local impulse transfer functions using Fourier transformations. This allows us to identify the persistence dynamics of interconnectedness based on the heterogeneous persistence of shocks in the system. A dynamic representation of the variance decomposition of shocks from dealer (or non-dealers) $j$ to dealer (or non-dealers) $k$ then establishes a dynamic horizon-specific adjacency matrix, which is central to our measures. 

Specifically, the element of such matrix that captures how shocks propagate from a dealer (or non-dealers) $j$ to a dealer (or non-dealers) $k$ at a given time and horizon\footnote{In the empirical investigation, 20 business days divides the transitory and persistent horizons.} $d_i \in \mathcal{H} = \{\text{Tr},\text{Per}\},$ is formally defined as:
\begin{equation} \label{eq:dynamicadjmatrix}
\Big[ \theta_{t,T}^d \Big]_{j,k} = \frac{\widehat{\sigma}_{kk}^{-1} \displaystyle \sum_{\omega \in d_i} \left( 
    \bigg[ \widehat\alpha_{t,T}(\omega) \widehat \Sigma_{t,T} \bigg]_{j,k} \right)^2 }{ \displaystyle \sum_{\omega \in \mathcal{H}} \Bigg[ \widehat\alpha_{t,T}(\omega) \widehat \Sigma_{t,T} \widehat\alpha^{\top}_{t,T}(\omega)  \Bigg]_{j,j} },\end{equation}
where $\widehat\alpha_{t,T}(\omega) = \sum_{h=0}^{H-1} \sum_h \widehat\alpha_{t,T}(h) e^{-i\omega h}$ is an impulse transfer function estimated from Fourier frequencies $\omega$ of impulse responses covering a given horizon $d_i$ frequencies.\footnote{Note that $i=\sqrt{-1}$.} Since the rows of the dynamic adjacency matrix do not necessarily sum up to one, we normalise the element in each row by the corresponding row sum: $\Big[ \widetilde \theta_{t,T}^d \Big]_{j,k} = \Big[ \theta_{t,T}^d \Big]_{j,k}\Big/ \sum\limits_{k=1}^N\Big[ \theta_{t,T}^d \Big]_{j,k}$. Equation (\ref{eq:dynamicadjmatrix}) fully defines a dynamic horizon-specific link between banks. Note that in the paper we denote $\Big[ \widehat \theta_{t,d} \Big]_{j,k} = \Big[ \widetilde \theta_{t,T}^d \Big]_{j,k}$ to ease the notation burden.

To obtain the time-varying coefficient estimates
$\widehat{\phi}_{1,t,T},...,\widehat{\phi}_{p,t,T}$
and the time-varying covariance matrix
$\widehat{\Sigma}_{t,T}$
we estimate the approximating model in Equation (\ref{eq:VAR2}) using Quasi-Bayesian Local-Likelihood (QBLL) methods \citep{petrova2019quasi}. We provide a detailed discussion of the estimation algorithm in Appendix \ref{app:estimate_var}.

For the estimation we use a kernel weighting function, which gives larger weights to those observations surrounding the period whose coefficient and covariance matrices are of interest. Using conjugate priors, the (quasi) posterior distribution of the parameters of the model are available analytically. This alleviates the need to use a Markov Chain Monte Carlo (MCMC) simulation algorithm and permits the use of parallel computing. 

Finally, the variance decompositions of the forecast errors from the VMA($\infty$) representation require a truncation of the infinite horizon with a $H$ horizon approximation. As $H\rightarrow \infty$ the error disappears \citep{lutkepohl2005new}. We note here that $H$ serves as an approximating factor and has no interpretation in the time-domain. We obtain horizon specific measures using Fourier transforms and set our truncation horizon $H$=100; the results are qualitatively similar for $H\in \{50,100,200\}$. In computing our measures, we also diagonalise the covariance matrix because our objective is to focus on the connections controlled for possible contemporaneous correlation in residuals of the system. The $\alpha(u,d)$ matrix embeds the causal nature of connections, and the covariance matrix $\Sigma(u)$ contains contemporaneous covariances within the off-diagonal elements.

\subsection{Estimation of the time-varying parameter VAR model}\label{app:estimate_var}

Let $x_{t}$ be an $N \times 1$ vector generated by a stable time-varying parameter (TVP) heteroskedastic VAR model with $p$ lags:
\begin{equation}\label{eq:VAR}
x_{t,T}=\phi_{1}(t/T)x_{t-1,T}+\ldots+\phi_{p}(t/T)x_{t-p,T} + \epsilon_{t,T},
\end{equation}
where 
$\epsilon_{t,T}=\Sigma^{-1/2}(t/T)\beta_{t,T}, \beta_{t,T}\sim NID(0,I_M)$ 
and 
$\phi(t/T)=(\phi_{1}(t/T),\ldots,\phi_{p}(t/T))^{\top}$ 
are the time varying autoregressive coefficients.

Note that all roots of the polynomial $\chi(z)=\text{det}\left(I_{N}-\sum^{L}_{p=1}z^{p}x_{p,t}\right)$ lie outside the unit circle, and $\Sigma^{-1}_{t}$ is a positive definite time-varying covariance matrix. Stacking the time-varying intercepts and autoregressive matrices in the vector $\phi_{t,T}$ with $\overline{x}^{\top}_{t} = \left(\text{I}_{N} \otimes x_{t}\right),\: x_{t}=\left(1,x^{\top}_{t-1},\dots,x^{\top}_{t-p}\right)$ and denoting the Kronecker product by $\otimes,$ the model can be written as:
\begin{eqnarray}
x_{t,T} = \overline{x}^{\top}_{t,T}\phi_{t,T} + \Sigma^{-\frac{1}{2}}_{t/T}\beta_{t,T}
\end{eqnarray}

We obtain the time-varying parameters of the model by employing the Quasi-Bayesian Local-Likelihood (QBLL) approach of \cite{petrova2019quasi}. The estimation of Equation (\ref{eq:VAR}) requires re-weighting the likelihood function. The weighting function gives higher proportions to observations surrounding the time period whose parameter values are of interest. The local likelihood function at period $k$ is given by:
\begin{gather}
\text{L}_{k}\left(x|\theta_{k},\Sigma_{k},\overline{x} \right) \propto\\ 
\nonumber
|\Sigma_{k}|^{\text{trace}(D_{k})/2}\exp\left\{-\frac{1}{2}(x-\overline{x}^{\top}\phi_{k})^{\top}\left(\Sigma_{k}\otimes D_{k}\right)(x-\overline{x}^{\top}\phi_{k})\right\}
\end{gather}
The $D_{k}$ is a diagonal matrix whose elements hold the weights:
\begin{eqnarray}
D_{k} &=& \text{diag}(\varrho_{k1},\dots,\varrho_{kT})\\
\varrho_{kt} &=& \phi_{T,k}w_{kt}/\sum^{T}_{t=1}w_{kt}\\
\label{eq:weight}
w_{kt} &=& (1/\sqrt{2\pi})\exp((-1/2)((k-t)/H)^{2}),\quad\text{for}\: k,t\in\{1,\dots,T\}\\
\zeta_{Tk} &=& \left(\left(\sum^{T}_{t=1}w_{kt}\right)^{2}\right)^{-1}
\end{eqnarray}
where $\varrho_{kt}$ is a normalised kernel function. $w_{kt}$ uses a Normal kernel weighting function. $\zeta_{Tk}$ gives the rate of convergence and behaves like the bandwidth parameter $H$ in (\ref{eq:weight}). The kernel function puts a greater weight on the observations surrounding the parameter estimates at time $k$ relative to more distant observations.

We use a Normal-Wishart prior distribution for $\phi_{k}|\:\Sigma_{k}$ for $k\in\{1,\dots,T\}$:
\begin{eqnarray}
\phi_{k}|\Sigma_{k} \backsim \mathcal{N}\left(\phi_{0k},(\Sigma_{k} \otimes \Xi_{0k})^{-1}\right)\\
\Sigma_{k} \backsim \mathcal{W}\left(\alpha_{0k},\Gamma_{0k}\right)
\end{eqnarray}
where $\phi_{0k}$ is a vector of prior means, $\Xi_{0k}$ is a positive definite matrix, $\alpha_{0k}$ is a scale parameter of the Wishart distribution ($\mathcal{W}$), and $\Gamma_{0k}$ is a positive definite matrix. 

The prior and weighted likelihood function implies a Normal-Wishart quasi posterior distribution for $\phi_{k}|\:\Sigma_{k}$ for $k=\{1,\dots,T\}$. Formally, let $A = (\overline{x}^{\top}_{1},\dots,\overline{x}^{\top}_{T})^{\top}$ and $Y=(x_{1},\dots,x_{T})^{\top}$, then:
\begin{eqnarray}
\phi_{k}|\Sigma_{k},A,Y &\backsim & \mathcal{N}\left(\widetilde{\theta}_{k},\left(\Sigma_{k}\otimes\widetilde{\Xi}_{k}\right)^{-1}\right)\\
\Sigma_{k} &\backsim & \mathcal{W}\left(\widetilde{\alpha}_{k},{\widetilde{\Gamma}}^{-1}_{k} \right)
\end{eqnarray}
with quasi posterior parameters:
\begin{eqnarray}
\widetilde{\phi}_{k} &=& \left(I_{N}\otimes {\widetilde{\Xi}}^{-1}_{k}\right)\left[\left({I}_{N}\otimes {A}^{\top}{D}_{k}{A}\right)\hat{\phi}_{k}+ \left({I}_{N}\otimes {\Xi}_{0k}\right)\phi_{0k} \right]\\
{\widetilde{\Xi}}_{k} &=& {\widetilde{\Xi}}_{0k} + {A}^{\top}{D}_{k}{A}\\
\widetilde{\alpha}_{k} &=& \alpha_{0k}+\sum^{T}_{t=1}\varrho_{kt}\\
{\widetilde{\Gamma}}_{k} &=& {\Gamma}_{0k} + {Y}'{D}_{k}{Y} + {\Phi}_{0k}{\Gamma}_{0k}{\Phi}^{\top}_{0k} - {\widetilde{\Phi}}_{k}{\widetilde{\Gamma}}_{k}{\widetilde{\Phi}}^{\top}_{k}
\end{eqnarray}
where $\widehat{\phi}_{k} = \left({I}_{N}\otimes {A}^{\top}{D}_{k}{A}\right)^{-1}\left({I}_{N} \otimes {A}^{\top}{D}_{k}\right)y$ is the local likelihood estimator for $\phi_{k}$. The matrices ${\Phi}_{0k},\:{\widetilde{\Phi}}_{k}$ are conformable matrices from the vector of prior means, $\phi_{0k}$, and a draw from the quasi posterior distribution, $\widetilde{\phi}_{k}$, respectively.

The motivation for employing these methods are threefold. First, we are able to estimate large systems that conventional Bayesian estimation methods do not permit. This is typically because the state-space representation of an $N$-dimensional TVP VAR ($p$) requires an additional $N(3/2 + N(p+1/2))$ state equations for every additional variable. Conventional Markov Chain Monte Carlo (MCMC) methods fail to estimate larger models, which in general confine one to (usually) fewer than 6 variables in the system. Second, the standard approach is fully parametric and requires a law of motion. This can distort inference if the true law of motion is misspecified. Third, the methods used here permit direct estimation of the VAR's time-varying covariance matrix, which has an inverse-Wishart density and is symmetric positive definite at every point in time. 

In estimating the model, we use $p$=2 and a Minnesota Normal-Wishart prior with a shrinkage value $\varphi=0.05$ and centre the coefficient on the first lag of each variable to 0.1 in each respective equation. The prior for the Wishart parameters are set following \cite{kadiyala1997numerical}. For each point in time, we run 100 simulations of the model to generate the (quasi) posterior distribution of parameter estimates. Note we experiment with various lag lengths, $p=\{2,3,4,5\}$; shrinkage values, $\varphi=\{0.01, 0.25, 0.5\}$; and values to centre the coefficient on the first lag of each variable, $\{0, 0.05, 0.2, 0.5\}$. Network measures from these experiments are qualitatively similar. Notably, adding lags to the VAR  and increasing the persistence in the prior value of the first lagged dependent variable in each equation increases computation time.


\section{Additional results}\label{appendix: results}
\subsection{Non-Dealer Sectors and Repo Market Segmentation}

We observe qualitative differences between non-dealers from different sectors. Evidence from the UK and elsewhere, e.g. \cite{Czechetal.21, Schrimpf.2020}, suggests that hedge funds and asset managers behave more procyclically than pension funds and insurance companies.\footnote{In the ``Dash-for-Cash'' period, at the Gilt repo, while hedge funds and pension funds increased the demands to dealers, money market funds reduced dealers' funding provision.  In the Gilt market, \cite{Czechetal.21} document that hedge funds in aggregate did not behave as procyclically as reported by \cite{Schrimpf.2020}.}\footnote{ Procyclicality here is viewed in terms of market volumes. In business cycle terms, they would be considered countercylical.}  

Table \ref{tab:volume and spread OTC by sector by frequency} presents our benchmark specification in equation \ref{estimation: non dealer} for two sectors, namely hedge funds \& asset managers (HFs \& AMs) on one hand, and pension funds \& insurance companies (PFs \& ICs) on the other. The left panel shows estimates for the repo segment and the right panel estimates for the reverse repo segment. Finally, each panel presents estimates for each sector as well as dealer factors, constituted by transitory and persistent shocks. 

Focusing on the volume, we highlight two features of market power. First, the market power of dealers does not affect HFs \& AMs, but it does affect PFs \& ICs.\footnote{In previous versions, we show that the volume of \textit{non-dealer banks} is not affected by the market power of repo dealers. They are available on request.} This is consistent with the fact that these institutions are responsible for most daily transactions. In our estimation of market power in Table \ref{tab:demand} we show that market power decreases with a reduction in variable frequency, i.e. monthly interactions. Second, the impact of market power is higher (in absolute terms) in the reverse repo segment than in the repo segment. In terms of magnitude, the reduction in volume is small, as in the worst case it is equivalent to a reduction of $10^{1.72} = \pounds 52.5m$. 

Table \ref{tab:volume and spread OTC by sector by frequency} highlights that HFs \& AMs and PFs \& ICs highly value relationships with dealers. Estimates of both proxies, i.e. depth and frequency, are statistically significant for HFs \& AMs. PFs \& ICs tend to operate less frequently and that's why Depth is the only relationship metric that is statistically significant. In terms of magnitude, just as with market power, the impact is rather small.

We now proceed to discuss market segmentation (or specialization). According to \cite{Sambalaibat.23}, there is evidence of segmentation if, as main dealers become more interconnected, specific non-dealer sectors also increase their funding from/to repo dealers. The common characteristic of these sectors is that they trade more frequently relative to other sectors and account for a large share of the total volume.  In terms of pricing, as main dealer are more interconnected, repo rates should decrease for all sectors reflecting a centrality discount.\footnote{The effect on pricing is also consistent with \cite{Duffieetal.2005} as most sophisticated dealers, i.e. with higher intensity-adjusted bargaining power, will trade at lower prices relative to less sophisticated dealers.  See \cite{Weill.2020, Feldhutter.2012}}

We posit that persistent and transitory shocks operate differently across non-dealers sectors, and that dealers segment the repo and the reverse repo segments when dealer factors have a heterogeneous impact across non-dealers. 

Table \ref{tab:volume and spread OTC by sector by frequency} shows how persistent and transitory shocks operate on different sectors. On the one hand, persistent shocks to dealer factors greatly increase PFs \& ICs funding. For the repo segment, while HFs \& AMs reduce slightly their funding in $10^{2.3} = \pounds 199m$, PFs \& ICs increases their funding by large orders of magnitude. In the reverse repo segment, we observe qualitatively similar results. On the other hand, dealer-originated transitory shocks affect HFs \& AMs at the reverse repo segment only. We argue that this evidence supports the segmentation (or specialization) hypothesis, as HFs \& AMs are the most relevant institutions in the reverse repo segment, meanwhile PFs \& ICs mostly implement long-term strategies.

\begin{table}[htbp]
  \centering
  \caption{OTC impact by sector and Persistence of shocks: Volume and Spread}
  \scriptsize
    \begin{tabular}{lccccccccc}
    \toprule
    & \multicolumn{9}{c}{Volume} \\
    \cmidrule{2-10}
    & \multicolumn{4}{c}{Repo segment}  &  & \multicolumn{4}{c}{Reverse Repo segment} \\
    \cmidrule{2-5}  \cmidrule{7-10}
    & \multicolumn{2}{c}{HFs \& AMs} & \multicolumn{2}{c}{PFs \& Ics} &  &  \multicolumn{2}{c}{HFs \& AMs} & \multicolumn{2}{c}{PFs \& Ics} \\
         \cmidrule{2-5}  \cmidrule{7-10}
    Market Power & {0.053} & {0.032} & {-0.518*} & {-0.373} &  & {-0.128} & {-0.176} & {-1.717***} & {-1.468***} \\
                 & {[0.116]} & {[0.114]} & {[0.266]} & {[0.234]}  &  & {[0.258]} & {[0.261]} & {[0.556]} & {[0.553]} \\
    Depth        & {0.623**} & {0.623**} & {0.963***} & {0.964***}  &  & {0.514} & {0.521} & {2.154**} & {2.051**} \\
                 & {[0.246]} & {[0.246]} & {[0.349]} & {[0.352]}  &  & {[0.327]} & {[0.326]} & {[0.864]} & {[0.845]} \\
    Frequency    & {0.009***} & {0.009***} & {0.002} & {0.001}  &  & {0.014***} & {0.014***} & {-0.002} & {-0.001} \\
                 & {[0.001]} & {[0.001]} & {[0.002]} & {[0.002]}  &  & {[0.002]} & {[0.002]} & {[0.003]} & {[0.003]} \\
    DF tran repo & {-0.046} & {} & {0.353} & {}  &  & {4.021***} & {} & {-4.707} & {} \\
                 & {[0.693]} & {} & {[1.908]} & {}  &  & {[0.726]} & {} & {[3.290]} & {} \\
    DF tran reverse & {-0.221} & {} & {-4.697} & {}  &  & {2.038***} & {} & {-5.110*} & {} \\
                & {[0.693]} & {} & {[4.115]} & {}  &  & {[0.653]} & {} & {[2.788]} & {} \\
    DF per repo & {} & {1.455} & {} & {17.607***}  &  & {} & {-1.172} & {} & {12.144*} \\
                & {} & {[1.251]} & {} & {[6.582]}  &  & {} & {[1.378]} & {} & {[7.272]} \\
    DF per reverse & {} & {-2.284**} & {} & {-0.264}  &  & {} & {3.705***} & {} & {1.996} \\
                & {} & {[1.045]} & {} & {[3.844]}  &  & {} & {[1.215]} & {} & {[7.428]} \\
     \cmidrule{2-5}  \cmidrule{7-10}
    R-squared & {0.632} & {0.632} & {0.504} & {0.507}  &  & {0.455} & {0.454} & {0.364} & {0.367} \\
     \cmidrule{2-10}
    & \multicolumn{9}{c}{Spread} \\
    \cmidrule{2-10}\textbf{}
    & \multicolumn{2}{c}{HFs \& AMs} & \multicolumn{2}{c}{PFs \& Ics} &  &  \multicolumn{2}{c}{HFs \& AMs} & \multicolumn{2}{c}{PFs \& Ics} \\
    \cmidrule{2-5}  \cmidrule{7-10}
Market Power & {0.115***} & {0.118***} & {-0.205***} & {-0.205***}  &  & {0.564***} & {0.559***} & {0.182***} & {0.180***} \\
          & {[0.008]} & {[0.008]} & {[0.027]} & {[0.029]}  &  & {[0.022]} & {[0.022]} & {[0.023]} & {[0.023]} \\
    Depth & {0.001} & {0.002} & {0.005} & {0.001}  &  & {0.007} & {0.007} & {0.023} & {0.024} \\
          & {[0.010]} & {[0.010]} & {[0.015]} & {[0.015]}  &  & {[0.012]} & {[0.012]} & {[0.017]} & {[0.017]} \\
    Frequency & {-0.000} & {-0.000} & {-0.000} & {-0.000}  &  & {0.000} & {0.000} & {-0.000} & {-0.000} \\
          & {[0.000]} & {[0.000]} & {[0.000]} & {[0.000]}  &  & {[0.000]} & {[0.000]} & {[0.000]} & {[0.000]} \\
    DF tran repo & {-0.147***} & {} & {-0.921***} & {}  &  & {-0.139***} & {} & {-0.208***} & {} \\
          & {[0.026]} & {} & {[0.164]} & {}  &  & {[0.030]} & {} & {[0.061]} & {} \\
    DF tran reverse & {-0.119***} & {} & {0.218**} & {}  &  & {-0.082***} & {} & {0.000} & {} \\
          & {[0.031]} & {} & {[0.097]} & {}  &  & {[0.029]} & {} & {[0.043]} & {} \\
    DF per repo & {} & {0.024} & {} & {-0.323*}  &  & {} & {-0.045} & {} & {-0.131} \\
          & {} & {[0.053]} & {} & {[0.189]}  &  & {} & {[0.051]} & {} & {[0.090]} \\
    DF per reverse & {} & {-0.025} & {} & {0.298}  &  & {} & {-0.282***} & {} & {-0.130} \\
          & {} & {[0.052]} & {} & {[0.236]}  &  & {} & {[0.056]} & {} & {[0.121]} \\
     \cmidrule{2-5}  \cmidrule{7-10}
     R-squared & {0.055} & {0.053} & {0.314} & {0.298}  &  & {0.142} & {0.142} & {0.158} & {0.156} \\
     \cmidrule{2-5}  \cmidrule{7-10}
     D/ND FE & {Yes} & {Yes} & {Yes} & {Yes}  &  & {Yes} & {Yes} & {Yes} & {Yes} \\
    ND/D*Week FE & {Yes} & {Yes} & {Yes} & {Yes}  &  & {Yes} & {Yes} & {Yes} & {Yes} \\
    Year FE & {Yes} & {Yes} & {Yes} & {Yes}  &  & {Yes} & {Yes} & {Yes} & {Yes} \\
    Controls & {Yes} & {Yes} & {Yes} & {Yes}  &  & {Yes} & {Yes} & {Yes} & {Yes} \\
    Observations & {68,781} & {68,781} & {11,955} & {11,955}  &  &  {85,643} & {85,643} & {9,867} & {9,867} \\
    \bottomrule
    \multicolumn{10}{l}{$^{*}$p$<$0.1; $^{**}$p$<$0.05; $^{***}$p$<$0.01} \\
\multicolumn{10}{@{}l}{\parbox[t]{.9\linewidth}{Note: This table reports the results of the regressions for repo market volume and rate spread (in absolute value) impact by non-dealer sector and persistence of shocks to dealers, as discussed in Section \ref{Sect Results}, for the period 2016:M1 to 2022:M1.  Definitions, sources and frequency of all independent variables are presented in Section \ref{Sect Data Identification}. Panels report the results for the dependent variables as follows: the left panel uses repo segment transactions, and the right panel uses reverse repo segment transactions. Clustered standard errors on dealer / non-dealer dyads.}} \\
    \end{tabular}%
  \label{tab:volume and spread OTC by sector by frequency}%
\end{table}%

Table \ref{tab:volume and spread OTC by sector by frequency} reports the sector-specific impact on repo spread by non-dealer sector. The impact of market power on HFs \& AMs has the expected sign, and in terms of magnitude, dealers with $1\%$ higher market power increase the spread in the reverse segment by 0.55 pp, while it's only 0.11 pp for the repo segment. The impact on PFs \& ICs is roughly 0.2 pp on both sides.  Noticeably, impact on the repo segment has the opposite sign, i.e. $-0.208$ pp, probably explained by the size of transaction.

\section{Tables}\label{appendix: tables}

\begin{table}[htbp]
\footnotesize
  \centering
  \caption{Summary statistics for the repo market by type of transaction (repo vs. reverse repo) between dealers and CCPs. Our sample only uses gilts as collateral.  Volumes are  expressed in $10^7$ of  sterling and the interest rate spread is relative to the BoE reference rate.}
    \begin{tabular}{lcccccc}
    \toprule
          & Median   & Mean  & SD    & Min   & Max   & N \\
    \midrule
          & \multicolumn{6}{c}{Repo} \\
          \cmidrule{2-7}
    Volume   & 2.34 & 3.47 & 3.85 & 0.10 & 33.00 & 1149310 \\
    Rate spread & -0.06 & -0.08 & 0.08  & -0.95 & 0.65  & 1149310 \\
          &       &       &       &       &       &  \\
    \cmidrule{2-7}
          & \multicolumn{6}{c}{Reverse} \\
          \cmidrule{2-7}
    Volume   & 2.27 & 3.44 & 4.11 & 0.10 & 50.00 & 1103752 \\
    Rate spread & -0.06 & -0.07 & 0.08  & -0.95 & 0.56  & 1103752 \\
    \bottomrule
    \end{tabular}%
  \label{SumStats: Repo CCP}%
\end{table}%

\begin{table}[htbp]
\footnotesize
  \centering
  \caption{\% Overnight Funding}
   \small
    \begin{tabular}{lllllll}
    \hline
          &       & \multicolumn{2}{c}{OTC} & & \multicolumn{2}{c}{CCP} \\
          \cmidrule{3-4} \cmidrule{6-7}
          &       & Repo  & Reverse & & Repo  & Reverse \\
    \cmidrule{3-4} \cmidrule{6-7}
    Overnight Funding &       & 31.59\% &  46.61\% & & 99.06\% & 98.64\% \\
    Unique ISINs &       &  2.42  &  2.11  & &  25.19 & 22.25 \\
    \hline
    \end{tabular}%
  \label{fact: repo maturities, ISIN}%
\end{table}%


\begin{table}[htbp]
\footnotesize
  \centering
  \scriptsize
  \caption{Demand Estimation with Jointly Repo and Reverse Repo Segments}
    \begin{tabular}{lccccc}
    \toprule
          & \multicolumn{5}{c}{Repo and Reverse Repo Data} \\
    \cmidrule{2-6}
    Repo Rate & -0.094 & 0.086 & -0.209*** & -0.118 & 0.129 \\
          & 0.067 & 0.075 & 0.089 & 0.100 & 0.096 \\
    Residual Maturity & 0.003*** & 0.010*** & 0.010*** & 0.013*** & 0.014*** \\
          & 0.000 & 0.001 & 0.000 & 0.002 & 0.002 \\
    Frequency & 0.045*** & -0.752*** & 0.0914*** & -0.369*** &  \\
          & 0.001 & 0.022 & 0.001 & 0.033 &  \\
    Collateral Market Value & -0.000*** & -0.000*** & -0.000*** & -0.000*** & -0.000*** \\
          & 0.000 & 0.000 & 0.000 & 0.000 & 0.000 \\
          &       &       &       &       &  \\
    Sample & All   & All   & Restricted & Restricted & Restricted \\
    \cmidrule{2-6}
    Control function & no    & yes   & no    & yes   & yes \\
    Month FE & yes   & yes   & yes   & yes   & yes \\
    Year FE & yes   & yes   & yes   & yes   & yes \\
    Obs   & 12,024,349 & 12,024,349 & 10,882,728 & 10,882,728 & 10,882,728 \\
    R2    & 0.0847 & 0.0931 & 0.0881 & 0.0901 & 0.0501 \\
    \bottomrule
    \multicolumn{6}{l}{$^{*}$p$<$0.1; $^{**}$p$<$0.05; $^{***}$p$<$0.01} \\
\multicolumn{6}{@{}l}{\parbox[t]{.75\linewidth}{Note: This table reports the results of the regressions for demand estimation, as discussed in Section \ref{Sect Data Identification}, for the period 2016:M1 to 2022:M1.  Definitions, sources and frequency of all independent variables are presented in Section \ref{Sect Data Identification}. This tables uses all repo transactions without distinguishing between repo and reverse repo. }} \\
    \end{tabular}%
  \label{tab: demand_both_segments}%
\end{table}%


\begin{table}[htbp]
\footnotesize
  \centering
  \caption{OTC impact by persistence of shocks: simultaneous dyads}
  \scriptsize
    \begin{tabular}{lccccc}
    \toprule
          & \multicolumn{5}{c}{Log volume} \\
          \cmidrule{2-6}
         & \multicolumn{2}{c}{Repo} && \multicolumn{2}{c}{Reverse} \\
          \cmidrule{2-3} \cmidrule{5-6}
    Market Power & -0.119 & -0.121 && -0.188 & -0.301 \\
          & [0.132] & [0.131] && [0.259] & [0.249] \\
    Depth & 0.512** & 0.508** && 0.583*** & 0.585*** \\
          & [0.247] & [0.247] & &[0.191] & [0.194] \\
    Frequency & 0.009*** & 0.009*** && 0.009*** & 0.009*** \\
          & [0.001] & [0.001] && [0.001] & [0.001] \\
    Log CB Reserves & -0.193 & -0.044& & -0.328* & -0.148 \\
          & [0.214] & [0.243] && [0.170] & [0.188] \\
    DF tran repo & 0.466 &     &  & 2.065*** &  \\
          & [0.627] &      & & [0.648] &  \\
    DF tran reverse & 0.796 &   &    & 1.452** &  \\
          & [0.719] &      & & [0.727] &  \\
    DF per repo &       & 1.128 &   &    & -1.062 \\
          &       & [1.196] &    &   & [1.390] \\
    DF per reverse &       & 0.396 &  &     & 0.583 \\
          &       & [1.267] &    &   & [1.053] \\
          &       &       &     &  &  \\
    Dealer FE & Yes   & Yes  & & No    & No \\
    NonDealer FE & No    & No   & & Yes   & Yes \\
    NonDealer*Week FE & Yes   & Yes  & & No    & No \\
    Dealer*Week FE & No    & No  &  & Yes   & Yes \\
    Year FE & Yes   & Yes &  & Yes   & Yes  \\
    Controls & Yes   & Yes &  & Yes   & Yes \\
    Observations & 52,396 & 52,396 && 52,597 & 52,597 \\
    R-squared & 0.624 & 0.624 && 0.596 & 0.596 \\
    \cmidrule{2-6}
          & \multicolumn{5}{c}{Repo rate spread} \\
          \cmidrule{2-6}
         & \multicolumn{2}{c}{Repo} && \multicolumn{2}{c}{Reverse} \\
          \cmidrule{2-3} \cmidrule{5-6}
    Market Power & 0.103*** & 0.107***& & 0.581*** & 0.577*** \\
          & [0.007] & [0.007] && [0.023] & [0.022] \\
    Depth & 0.014 & 0.015& & 0.025* & 0.026* \\
          & [0.014] & [0.014]& & [0.014] & [0.014] \\
    Frequency & -0.000 & -0.000 && -0.000 & -0.000 \\
          & [0.000] & [0.000]& & [0.000] & [0.000] \\
    Log CB Reserves & 0.029** & 0.009 & &0.065*** & 0.046*** \\
          & [0.012] & [0.013] && [0.013] & [0.011] \\
    DF tran repo & -0.158*** &    &   & -0.134*** &  \\
          & [0.030] &       && [0.037] &  \\
    DF tran reverse & -0.119*** &  &     & -0.126*** &  \\
          & [0.037] &     &  & [0.041] &  \\
    DF per repo &       & 0.047 &  &     & 0.019 \\
          &       & [0.059] &     &  & [0.062] \\
    DF per reverse &       & 0.032 & &      & -0.328*** \\
          &       & [0.076] &     &  & [0.073] \\
          &       &       &     &  &  \\
    Dealer FE & Yes   & Yes   && No    & No \\
    NonDealer FE & No    & No   & & Yes   & Yes \\
    NonDealer*Week FE & Yes   & Yes &  & No    & No \\
    Dealer*Week FE & No    & No   & & Yes   & Yes \\
    Year FE & Yes   & Yes   && Yes   & Yes  \\
    Controls & Yes   & Yes  & & Yes   & Yes \\
    Observations & 52,396 & 52,396& & 52,597 & 52,597 \\
    R-squared & 0.112 & 0.110 && 0.142 & 0.142 \\
    \bottomrule
    \multicolumn{6}{l}{$^{*}$p$<$0.1; $^{**}$p$<$0.05; $^{***}$p$<$0.01} \\
\multicolumn{6}{@{}l}{\parbox[t]{.7\linewidth}{Note: This table reports the results of the regressions for repo market impact for simultaneous dyads by frequency of shock to dealers, as discussed in Section \ref{Sect Results}, for the period 2016:M1 to 2022:M1. Definitions, sources and frequency of all independent variables are presented in Section \ref{Sect Data Identification}. Panels report the results for the dependent variables as follows: the top panel uses log of volume, and the bottom panel uses the absolute value of repo spread to the reference rate determined by Bank of England. Clustered standard errors on dealer / non-dealer dyads.}} \\
    \end{tabular}%
  \label{tab:OTC repo simultaneous by frequency}%
\end{table}%

 
\begin{table}[htbp]
\footnotesize
  \centering
  \caption{OTC impact by persistence: High Client vs Low Client}
  \scriptsize
    \begin{tabular}{l|cccc}
    \toprule
          & \multicolumn{4}{c}{Log volume} \\
         & \multicolumn{2}{c}{Repo} & \multicolumn{2}{c}{Reverse} \\
          &       &       &       &  \\
    Market Power & -0.224* & -0.206* & -0.776*** & -0.792*** \\
          & [0.121] & [0.116] & [0.272] & [0.273] \\
    Client High & 0.785*** & 0.781*** & 1.559*** & 1.564*** \\
          & [0.252] & [0.252] & [0.193] & [0.194] \\
    Client Low & -1.091*** & -1.096*** & -0.013 & -0.005 \\
          & [0.284] & [0.284] & [0.317] & [0.320] \\
    Log CB Reserves & -0.608*** & -0.326* & 1.487*** & 1.843*** \\
          & [0.182] & [0.189] & [0.245] & [0.248] \\
    DF tran repo & 2.045*** &       & 3.854*** &  \\
          & [0.630] &       & [0.748] &  \\
    DF tran reverse & -0.161 &       & 2.004*** &  \\
          & [0.978] &       & [0.701] &  \\
    DF per repo &       & 3.412** &       & -1.482 \\
          &       & [1.445] &       & [1.330] \\
    DF per reverse &       & 0.420 &       & 4.741*** \\
          &       & [1.210] &       & [1.321] \\
          &       &       &       &  \\
    Dealer FE & Yes   & Yes   & No    & No \\
    NonDealer FE & No    & No    & Yes   & Yes \\
    NonDealer*Week FE & Yes   & Yes   & No    & No \\
    Dealer*Week FE & No    & No    & Yes   & Yes \\
    Year FE & Yes   & Yes   & Yes   & Yes  \\
    Controls & Yes   & Yes   & Yes   & Yes \\
    Observations & 92,314 & 92,314 & 115,555 & 115,555 \\
    R-squared & 0.568 & 0.568 & 0.420 & 0.419 \\
    \hline
          & \multicolumn{4}{c}{Repo rate spread} \\
         & \multicolumn{2}{c}{Repo} & \multicolumn{2}{c}{Reverse} \\
          &       &       &       &  \\
    Market Power & 0.035*** & 0.040*** & 0.530*** & 0.527*** \\
          & [0.013] & [0.013] & [0.020] & [0.020] \\
    Client High & -0.016 & -0.015 & 0.007 & 0.006 \\
          & [0.011] & [0.011] & [0.006] & [0.006] \\
    Client Low & -0.012 & -0.012 & 0.001 & 0.001 \\
          & [0.011] & [0.011] & [0.006] & [0.006] \\
    Log CB Reserves & 0.046*** & 0.025*** & 0.070*** & 0.053*** \\
          & [0.008] & [0.009] & [0.009] & [0.008] \\
    DF tran repo & -0.263*** &       & -0.136*** &  \\
          & [0.039] &       & [0.025] &  \\
    DF tran reverse & -0.086** &       & -0.066*** &  \\
          & [0.038] &       & [0.025] &  \\
    DF per repo &       & 0.022 &       & -0.031 \\
          &       & [0.067] &       & [0.041] \\
    DF per reverse &       & 0.058 &       & -0.236*** \\
          &       & [0.082] &       & [0.047] \\
          &       &       &       &  \\
    Dealer FE & Yes   & Yes   & No    & No \\
    NonDealer FE & No    & No    & Yes   & Yes \\
    NonDealer*Week FE & Yes   & Yes   & No    & No \\
    Dealer*Week FE & No    & No    & Yes   & Yes \\
    Year FE & Yes   & Yes   & Yes   & Yes  \\
    Controls & Yes   & Yes   & Yes   & Yes \\
    Observations & 92,314 & 92,314 & 115,555 & 115,555 \\
    R-squared & 0.075 & 0.072 & 0.129 & 0.129 \\
    \bottomrule
    \multicolumn{5}{l}{$^{*}$p$<$0.1; $^{**}$p$<$0.05; $^{***}$p$<$0.01} \\
\multicolumn{5}{@{}l}{\parbox[t]{.7\linewidth}{Note: This table reports the results of the regressions for repo market impact by frequency of shocks to repo dealers, replacing relationship trading metrics with dummy variables ``High Client'' and ``Low Client'', as discussed in Section \ref{Sect Results}, for the period 2016:M1 to 2022:M1.  Definitions, sources and frequency of all independent variables are presented in Section \ref{Sect Data Identification}. Panels report the results for the dependent variables as follows: the top panel uses log of volume, and the bottom panel uses the absolute value of repo spread to the reference rate determined by Bank of England. Clustered standard errors on dealer / non-dealer dyads.}} \\
    \end{tabular}%
  \label{tab: impact on repo Client non-Client}%
\end{table}%


\begin{table}[htbp]
\footnotesize
  \centering
  \caption{OTC impact by persistence, Simultaneous Dyads: High Clients vs Low Clients}
  \scriptsize
    \begin{tabular}{l|cccc}
    \toprule
          & \multicolumn{4}{c}{Log volume} \\
        & \multicolumn{2}{c}{Repo} & \multicolumn{2}{c}{Reverse} \\
          &       &       &       &  \\
    Market Power & -0.342** & -0.356** & -0.488* & -0.606** \\
          & [0.151] & [0.149] & [0.286] & [0.271] \\
    Client High & 0.996*** & 0.992*** & 0.919*** & 0.915*** \\
          & [0.147] & [0.147] & [0.087] & [0.087] \\
    Client Low & -0.796*** & -0.799*** & -0.359*** & -0.360*** \\
          & [0.185] & [0.186] & [0.116] & [0.116] \\
    Log CB Reserves & 0.232 & 0.434* & -0.103 & 0.021 \\
          & [0.202] & [0.234] & [0.188] & [0.204] \\
    DF tran repo & 0.591 &       & 2.018*** &  \\
          & [0.668] &       & [0.686] &  \\
    DF tran reverse & 1.905** &       & 0.842 &  \\
          & [0.812] &       & [0.735] &  \\
    DF per repo &       & 0.603 &       & -1.155 \\
          &       & [1.184] &       & [1.470] \\
    DF per reverse &       & 0.925 &       & -0.033 \\
          &       & [1.535] &       & [1.194] \\
          &       &       &       &  \\
    Dealer FE & Yes   & Yes   & No    & No \\
    NonDealer FE & No    & No    & Yes   & Yes \\
    NonDealer*Week FE & Yes   & Yes   & No    & No \\
    Dealer*Week FE & No    & No    & Yes   & Yes \\
    Year FE & Yes   & Yes   & Yes   & Yes  \\
    Controls & Yes   & Yes   & Yes   & Yes \\
    Observations & 52,396 & 52,396 & 52,597 & 52,597 \\
    R-squared & 0.585 & 0.585 & 0.574 & 0.575 \\
    \hline
          & \multicolumn{4}{c}{Repo rate spread} \\
         & \multicolumn{2}{c}{Repo} & \multicolumn{2}{c}{Reverse} \\
          &       &       &       &  \\
    Market Power & 0.106*** & 0.110*** & 0.581*** & 0.577*** \\
          & [0.007] & [0.007] & [0.023] & [0.022] \\
    Client High & -0.001 & -0.001 & 0.005 & 0.005 \\
          & [0.004] & [0.004] & [0.008] & [0.008] \\
    Client Low & 0.011 & 0.011 & 0.002 & 0.002 \\
          & [0.008] & [0.008] & [0.006] & [0.006] \\
    Log CB Reserves & 0.026** & 0.006 & 0.067*** & 0.048*** \\
          & [0.011] & [0.012] & [0.014] & [0.012] \\
    DF tran repo & -0.159*** &       & -0.134*** &  \\
          & [0.031] &       & [0.037] &  \\
    DF tran reverse & -0.125*** &       & -0.126*** &  \\
          & [0.040] &       & [0.042] &  \\
    DF per repo &       & 0.053 &       & 0.024 \\
          &       & [0.061] &       & [0.063] \\
    DF per reverse &       & 0.020 &       & -0.328*** \\
          &       & [0.084] &       & [0.072] \\
          &       &       &       &  \\
    Dealer FE & Yes   & Yes   & No    & No \\
    NonDealer FE & No    & No    & Yes   & Yes \\
    NonDealer*Week FE & Yes   & Yes   & No    & No \\
    Dealer*Week FE & No    & No    & Yes   & Yes \\
    Year FE & Yes   & Yes   & Yes   & Yes  \\
    Controls & Yes   & Yes   & Yes   & Yes \\
    Observations & 52,396 & 52,396 & 52,597 & 52,597 \\
    R-squared & 0.104 & 0.101 & 0.141 & 0.141 \\
    \bottomrule
    \multicolumn{5}{l}{$^{*}$p$<$0.1; $^{**}$p$<$0.05; $^{***}$p$<$0.01} \\
\multicolumn{5}{@{}l}{\parbox[t]{.7\linewidth}{Note: This table reports the results of the regressions for repo market impact by frequency of shocks to repo dealers and simultaneous dyads, replacing relationship trading metrics with dummy variables ``High Client'' and ``Low Client'', as discussed in Section \ref{Sect Results}, for the period 2016:M1 to 2022:M1. Definitions, sources and frequency of all independent variables are presented in Section \ref{Sect Data Identification}. Panels report the results for the dependent variables as follows: the top panel uses log of volume, and the bottom panel uses the absolute value of repo spread to the reference rate determined by Bank of England. Clustered standard errors on dealer / non-dealer dyads.}} \\
    \end{tabular}%
  \label{tab: impact on repo Client non-Client by frequency}%
\end{table}%


\begin{landscape}
\begin{table}[htbp]
\footnotesize
  \centering
  \caption{OTC impact by sector, by persistence: High Client and Low Client}
  \tiny
    \begin{tabular}{l|cccccccc}
    \toprule
          & \multicolumn{4}{c}{Log Volume} & \multicolumn{4}{c}{Rate spread} \\
    Repo segment & \multicolumn{2}{c}{HFs \& AMs} & \multicolumn{2}{c}{PFs \& Ics} & \multicolumn{2}{c}{HFs \& AMs} & \multicolumn{2}{c}{PFs \& Ics} \\
          &       &       &       &       &       &       &       &  \\
    Market Power & -0.165 & -0.186 & -0.312 & -0.150 & 0.117*** & 0.120*** & -0.206*** & -0.206*** \\
          & [0.137] & [0.134] & [0.333] & [0.307] & [0.008] & [0.008] & [0.029] & [0.030] \\
    Client High & 0.948*** & 0.944*** & -1.458** & -1.546** & 0.004 & 0.004 & -0.034 & -0.039 \\
          & [0.187] & [0.188] & [0.622] & [0.662] & [0.005] & [0.005] & [0.040] & [0.037] \\
    Client Low & -0.995*** & -1.000*** & -2.547*** & -2.591*** & 0.016** & 0.016** & -0.024 & -0.028 \\
          & [0.254] & [0.255] & [0.668] & [0.690] & [0.007] & [0.007] & [0.027] & [0.027] \\
    Log CB Reserves & -0.968*** & -0.843*** & -0.801 & -0.285 & 0.032*** & 0.015 & -0.047 & -0.085*** \\
          & [0.198] & [0.205] & [0.575] & [0.538] & [0.008] & [0.009] & [0.029] & [0.030] \\
    DF tran repo & 0.357 &       & 2.727 &       & -0.148*** &       & -0.885*** &  \\
          & [0.712] &       & [2.394] &       & [0.027] &       & [0.186] &  \\
    DF tran reverse & 0.724 &       & -6.533 &       & -0.118*** &       & 0.172 &  \\
          & [0.784] &       & [4.355] &       & [0.032] &       & [0.115] &  \\
    DF per repo &       & 1.183 &       & 17.654** &       & 0.029 &       & -0.292 \\
          &       & [1.212] &       & [7.800] &       & [0.054] &       & [0.194] \\
    DF per reverse &       & -1.091 &       & -0.129 &       & -0.029 &       & 0.292 \\
          &       & [1.216] &       & [3.644] &       & [0.053] &       & [0.235] \\
          &       &       &       &       &       &       &       &  \\
    Dealer FE & Yes   & Yes   & Yes   & Yes   & Yes   & Yes   & Yes   & Yes \\
    NonDealer*Week FE & Yes   & Yes   & Yes   & Yes   & Yes   & Yes   & Yes   & Yes \\
    Year FE & Yes   & Yes   & Yes   & Yes   & Yes   & Yes   & Yes   & Yes \\
    Controls & Yes   & Yes   & Yes   & Yes   & Yes   & Yes   & Yes   & Yes \\
    Observations & 68,781 & 68,781 & 11,955 & 11,955 & 68,781 & 68,781 & 11,955 & 11,955 \\
    R-squared & 0.588 & 0.588 & 0.336 & 0.326 & 0.043 & 0.040 & 0.289 & 0.265 \\
    \hline
          & \multicolumn{4}{c}{Log Volume} & \multicolumn{4}{c}{Rate spread} \\
    Reverse segment & \multicolumn{2}{c}{HFs \& AMs} & \multicolumn{2}{c}{PFs \& Ics} & \multicolumn{2}{c}{HFs \& AMs} & \multicolumn{2}{c}{PFs \& Ics} \\
          &       &       &       &       &       &       &       &  \\
    Market Power & -0.382 & -0.393 & -1.746*** & -1.503** & 0.563*** & 0.558*** & 0.187*** & 0.185*** \\
          & [0.279] & [0.279] & [0.643] & [0.651] & [0.022] & [0.022] & [0.025] & [0.024] \\
    Client High & 1.350*** & 1.351*** & 0.418 & 0.405 & 0.009 & 0.008 & -0.009 & -0.009 \\
          & [0.136] & [0.136] & [0.705] & [0.693] & [0.007] & [0.007] & [0.011] & [0.011] \\
    Client Low & -0.315 & -0.314 & -0.398 & -0.498 & 0.008 & 0.008 & -0.012 & -0.013 \\
          & [0.262] & [0.262] & [0.857] & [0.811] & [0.006] & [0.006] & [0.010] & [0.010] \\
    Log CB Reserves & 0.853*** & 1.156*** & 0.424 & 0.476 & 0.089*** & 0.070*** & 0.057*** & 0.037** \\
          & [0.233] & [0.241] & [0.910] & [0.915] & [0.010] & [0.009] & [0.017] & [0.017] \\
    DF tran repo & 3.747*** &       & -4.909 &       & -0.137*** &       & -0.190*** &  \\
          & [0.748] &       & [3.757] &       & [0.030] &       & [0.061] &  \\
    DF tran reverse & 1.475** &       & -4.587* &       & -0.088*** &       & 0.006 &  \\
          & [0.677] &       & [2.651] &       & [0.030] &       & [0.044] &  \\
    DF per repo &       & -1.400 &       & 12.365* &       & -0.040 &       & -0.129 \\
          &       & [1.452] &       & [7.278] &       & [0.051] &       & [0.093] \\
    DF per reverse &       & 4.203*** &       & 1.966 &       & -0.282*** &       & -0.124 \\
          &       & [1.249] &       & [7.071] &       & [0.056] &       & [0.126] \\
          &       &       &       &       &       &       &       &  \\
    NonDealer FE & Yes   & Yes   & Yes   & Yes   & Yes   & Yes   & Yes   & Yes \\
    Dealer*Week FE & Yes   & Yes   & Yes   & Yes   & Yes   & Yes   & Yes   & Yes \\
    Year FE & Yes   & Yes   & Yes   & Yes   & Yes   & Yes   & Yes   & Yes \\
    Controls & Yes   & Yes   & Yes   & Yes   & Yes   & Yes   & Yes   & Yes \\
    Observations & 85,643 & 85,643 & 9,867 & 9,867 & 85,643 & 85,643 & 9,867 & 9,867 \\
    R-squared & 0.402 & 0.401 & 0.362 & 0.364 & 0.138 & 0.139 & 0.138 & 0.135 \\
    \bottomrule
    \multicolumn{9}{l}{$^{*}$p$<$0.1; $^{**}$p$<$0.05; $^{***}$p$<$0.01} \\
\multicolumn{9}{@{}l}{\parbox[t]{.7\linewidth}{Note: This table reports the results of the regressions for repo market impact by frequency of shocks to repo dealers and non-dealer sectors, replacing relationship trading metrics with dummy variables ``High Client'' and ``Low Client'', as discussed in Section \ref{Sect Results}, for the period 2016:M1 to 2022:M1. Definitions, sources and frequency of all independent variables are presented in Section \ref{Sect Data Identification}. Panels report the results for the dependent variables as follows: the top panel uses log of volume, and the bottom panel uses the absolute value of repo spread to the reference rate determined by Bank of England. Clustered standard errors on dealer / non-dealer dyads.}} \\
    \end{tabular}%
  \label{tab: impact on repo Client non-Client by sector by frequency}%
\end{table}%

\end{landscape}


\begin{table}[htbp]
\footnotesize
  \centering
  \caption{Mispricing: One-day mispricing lag}
  \scriptsize
    \begin{tabular}{lcccccccc}
    \toprule
       & \multicolumn{8}{c}{} \\
          & \multicolumn{2}{c}{All} & \multicolumn{2}{c}{Short} & \multicolumn{2}{c}{Medium} & \multicolumn{2}{c}{Long} \\
         \cmidrule{2-9}
    Markup & \multicolumn{1}{c}{0.248*} & \multicolumn{1}{c}{0.270*} & \multicolumn{1}{c}{0.385} & \multicolumn{1}{c}{0.393} & \multicolumn{1}{c}{0.194} & \multicolumn{1}{c}{0.219} & \multicolumn{1}{c}{0.100} & \multicolumn{1}{c}{0.127} \\
          & \multicolumn{1}{c}{[0.143]} & \multicolumn{1}{c}{[0.144]} & \multicolumn{1}{c}{[0.267]} & \multicolumn{1}{c}{[0.264]} & \multicolumn{1}{c}{[0.256]} & \multicolumn{1}{c}{[0.254]} & \multicolumn{1}{c}{[0.158]} & \multicolumn{1}{c}{[0.158]} \\
    Markdown & \multicolumn{1}{c}{-0.175} & \multicolumn{1}{c}{-0.147} & \multicolumn{1}{c}{0.542} & \multicolumn{1}{c}{0.558} & \multicolumn{1}{c}{-0.168} & \multicolumn{1}{c}{-0.149} & \multicolumn{1}{c}{-0.436*} & \multicolumn{1}{c}{-0.400} \\
          & \multicolumn{1}{c}{[0.200]} & \multicolumn{1}{c}{[0.193]} & \multicolumn{1}{c}{[0.353]} & \multicolumn{1}{c}{[0.351]} & \multicolumn{1}{c}{[0.244]} & \multicolumn{1}{c}{[0.238]} & \multicolumn{1}{c}{[0.255]} & \multicolumn{1}{c}{[0.245]} \\
    DF tran repo & \multicolumn{1}{c}{-0.134} & \multicolumn{1}{c}{} & \multicolumn{1}{c}{-0.500} & \multicolumn{1}{c}{} & \multicolumn{1}{c}{-0.170} & \multicolumn{1}{c}{} & \multicolumn{1}{c}{0.051} & \multicolumn{1}{c}{} \\
          & \multicolumn{1}{c}{[1.304]} & \multicolumn{1}{c}{} & \multicolumn{1}{c}{[1.169]} & \multicolumn{1}{c}{} & \multicolumn{1}{c}{[1.484]} & \multicolumn{1}{c}{} & \multicolumn{1}{c}{[1.547]} & \multicolumn{1}{c}{} \\
    DF tran reverse & \multicolumn{1}{c}{0.122} & \multicolumn{1}{c}{} & \multicolumn{1}{c}{0.517} & \multicolumn{1}{c}{} & \multicolumn{1}{c}{0.389} & \multicolumn{1}{c}{} & \multicolumn{1}{c}{0.090} & \multicolumn{1}{c}{} \\
          & \multicolumn{1}{c}{[1.382]} & \multicolumn{1}{c}{} & \multicolumn{1}{c}{[1.355]} & \multicolumn{1}{c}{} & \multicolumn{1}{c}{[1.515]} & \multicolumn{1}{c}{} & \multicolumn{1}{c}{[1.752]} & \multicolumn{1}{c}{} \\
    DF per repo & \multicolumn{1}{c}{} & \multicolumn{1}{c}{-1.782} & \multicolumn{1}{c}{} & \multicolumn{1}{c}{-0.400} & \multicolumn{1}{c}{} & \multicolumn{1}{c}{-1.595} & \multicolumn{1}{c}{} & \multicolumn{1}{c}{-3.073} \\
          & \multicolumn{1}{c}{} & \multicolumn{1}{c}{[1.567]} & \multicolumn{1}{c}{} & \multicolumn{1}{c}{[1.674]} & \multicolumn{1}{c}{} & \multicolumn{1}{c}{[1.630]} & \multicolumn{1}{c}{} & \multicolumn{1}{c}{[1.949]} \\
    DF per reverse & \multicolumn{1}{c}{} & \multicolumn{1}{c}{0.091} & \multicolumn{1}{c}{} & \multicolumn{1}{c}{-0.504} & \multicolumn{1}{c}{} & \multicolumn{1}{c}{-0.533} & \multicolumn{1}{c}{} & \multicolumn{1}{c}{0.803} \\
          & \multicolumn{1}{c}{} & \multicolumn{1}{c}{[1.865]} & \multicolumn{1}{c}{} & \multicolumn{1}{c}{[2.082]} & \multicolumn{1}{c}{} & \multicolumn{1}{c}{[2.089]} & \multicolumn{1}{c}{} & \multicolumn{1}{c}{[2.133]} \\
    CB Market Share & \multicolumn{1}{c}{-0.045} & \multicolumn{1}{c}{-0.044} & \multicolumn{1}{c}{0.231*} & \multicolumn{1}{c}{0.225*} & \multicolumn{1}{c}{-0.046} & \multicolumn{1}{c}{-0.053} & \multicolumn{1}{c}{-0.031} & \multicolumn{1}{c}{-0.020} \\
          & \multicolumn{1}{c}{[0.048]} & \multicolumn{1}{c}{[0.049]} & \multicolumn{1}{c}{[0.126]} & \multicolumn{1}{c}{[0.127]} & \multicolumn{1}{c}{[0.067]} & \multicolumn{1}{c}{[0.068]} & \multicolumn{1}{c}{[0.083]} & \multicolumn{1}{c}{[0.086]} \\
    Log CB Reserves & \multicolumn{1}{c}{0.010} & \multicolumn{1}{c}{-0.157} & \multicolumn{1}{c}{-0.581**} & \multicolumn{1}{c}{-0.626***} & \multicolumn{1}{c}{0.041} & \multicolumn{1}{c}{-0.115} & \multicolumn{1}{c}{0.364} & \multicolumn{1}{c}{0.101} \\
          & \multicolumn{1}{c}{[0.318]} & \multicolumn{1}{c}{[0.242]} & \multicolumn{1}{c}{[0.254]} & \multicolumn{1}{c}{[0.212]} & \multicolumn{1}{c}{[0.349]} & \multicolumn{1}{c}{[0.264]} & \multicolumn{1}{c}{[0.378]} & \multicolumn{1}{c}{[0.305]} \\
          &       &       &       &       &       &       &       &  \\
    Bond*Month FE & \multicolumn{1}{c}{Yes} & \multicolumn{1}{c}{Yes} & \multicolumn{1}{c}{Yes} & \multicolumn{1}{c}{Yes} & \multicolumn{1}{c}{Yes} & \multicolumn{1}{c}{Yes} & \multicolumn{1}{c}{Yes} & \multicolumn{1}{c}{Yes} \\
    Year FE & \multicolumn{1}{c}{Yes} & \multicolumn{1}{c}{Yes} & \multicolumn{1}{c}{Yes} & \multicolumn{1}{c}{Yes} & \multicolumn{1}{c}{Yes} & \multicolumn{1}{c}{Yes} & \multicolumn{1}{c}{Yes} & \multicolumn{1}{c}{Yes} \\
    Controls & \multicolumn{1}{c}{Yes} & \multicolumn{1}{c}{Yes} & \multicolumn{1}{c}{Yes} & \multicolumn{1}{c}{Yes} & \multicolumn{1}{c}{Yes} & \multicolumn{1}{c}{Yes} & \multicolumn{1}{c}{Yes} & \multicolumn{1}{c}{Yes} \\
    Observations & \multicolumn{1}{c}{21,532} & \multicolumn{1}{c}{21,532} & \multicolumn{1}{c}{5,222} & \multicolumn{1}{c}{5,222} & \multicolumn{1}{c}{7,507} & \multicolumn{1}{c}{7,507} & \multicolumn{1}{c}{8,799} & \multicolumn{1}{c}{8,799} \\
    R-squared & \multicolumn{1}{c}{0.041} & \multicolumn{1}{c}{0.042} & \multicolumn{1}{c}{0.024} & \multicolumn{1}{c}{0.024} & \multicolumn{1}{c}{0.031} & \multicolumn{1}{c}{0.032} & \multicolumn{1}{c}{0.056} & \multicolumn{1}{c}{0.059} \\
    \bottomrule
    \multicolumn{9}{l}{$^{*}$p$<$0.1; $^{**}$p$<$0.05; $^{***}$p$<$0.01} \\
\multicolumn{9}{@{}l}{\parbox[t]{.9\linewidth}{Note: This table reports the results of the regressions for bond-level one-day-ahead mispricing, as discussed in Section \ref{Sect Results}, for the period 2016:M1 to 2022:M1.  Definitions, sources and frequency of all independent variables are presented in Section \ref{Sect Data Identification}. The dependent variable is the absolute value of the spread between the bond-level yield and the predicted yield based on a spline. We use Driscoll-Kraay standard errors with 20 working days lag.}} \\
    \end{tabular}%
  \label{tab:benchmark 1 day lag}%
\end{table}%


\begin{table}[htbp]
\footnotesize
  \centering
  \caption{Mispricing benchmark: Three-day mispricing lag}
  \scriptsize
    \begin{tabular}{l|cccccccc}
    \toprule
       & \multicolumn{8}{c}{} \\
          & \multicolumn{2}{c}{All} & \multicolumn{2}{c}{Short} & \multicolumn{2}{c}{Medium} & \multicolumn{2}{c}{Long} \\
          & \multicolumn{1}{c}{} & \multicolumn{1}{c}{} & \multicolumn{1}{c}{} & \multicolumn{1}{c}{} & \multicolumn{1}{c}{} & \multicolumn{1}{c}{} & \multicolumn{1}{c}{} & \multicolumn{1}{c}{} \\
    Markup & \multicolumn{1}{c}{0.441**} & \multicolumn{1}{c}{0.486***} & \multicolumn{1}{c}{0.180} & \multicolumn{1}{c}{0.242} & \multicolumn{1}{c}{0.595*} & \multicolumn{1}{c}{0.625*} & \multicolumn{1}{c}{0.221} & \multicolumn{1}{c}{0.242} \\
          & \multicolumn{1}{c}{[0.191]} & \multicolumn{1}{c}{[0.187]} & \multicolumn{1}{c}{[0.370]} & \multicolumn{1}{c}{[0.363]} & \multicolumn{1}{c}{[0.337]} & \multicolumn{1}{c}{[0.331]} & \multicolumn{1}{c}{[0.183]} & \multicolumn{1}{c}{[0.179]} \\
    Markdown & \multicolumn{1}{c}{0.135} & \multicolumn{1}{c}{0.191} & \multicolumn{1}{c}{1.124*} & \multicolumn{1}{c}{1.222**} & \multicolumn{1}{c}{0.245} & \multicolumn{1}{c}{0.290} & \multicolumn{1}{c}{-0.257} & \multicolumn{1}{c}{-0.228} \\
          & \multicolumn{1}{c}{[0.300]} & \multicolumn{1}{c}{[0.294]} & \multicolumn{1}{c}{[0.573]} & \multicolumn{1}{c}{[0.571]} & \multicolumn{1}{c}{[0.339]} & \multicolumn{1}{c}{[0.334]} & \multicolumn{1}{c}{[0.323]} & \multicolumn{1}{c}{[0.314]} \\
    DF tran repo & \multicolumn{1}{c}{-1.543} & \multicolumn{1}{c}{} & \multicolumn{1}{c}{-3.857***} & \multicolumn{1}{c}{} & \multicolumn{1}{c}{-1.416} & \multicolumn{1}{c}{} & \multicolumn{1}{c}{-0.127} & \multicolumn{1}{c}{} \\
          & \multicolumn{1}{c}{[1.589]} & \multicolumn{1}{c}{} & \multicolumn{1}{c}{[1.372]} & \multicolumn{1}{c}{} & \multicolumn{1}{c}{[1.796]} & \multicolumn{1}{c}{} & \multicolumn{1}{c}{[1.781]} & \multicolumn{1}{c}{} \\
    DF tran reverse & \multicolumn{1}{c}{-0.834} & \multicolumn{1}{c}{} & \multicolumn{1}{c}{-1.804} & \multicolumn{1}{c}{} & \multicolumn{1}{c}{-0.008} & \multicolumn{1}{c}{} & \multicolumn{1}{c}{-0.598} & \multicolumn{1}{c}{} \\
          & \multicolumn{1}{c}{[1.777]} & \multicolumn{1}{c}{} & \multicolumn{1}{c}{[2.105]} & \multicolumn{1}{c}{} & \multicolumn{1}{c}{[1.870]} & \multicolumn{1}{c}{} & \multicolumn{1}{c}{[2.018]} & \multicolumn{1}{c}{} \\
    DF per repo & \multicolumn{1}{c}{} & \multicolumn{1}{c}{-1.868} & \multicolumn{1}{c}{} & \multicolumn{1}{c}{-2.630} & \multicolumn{1}{c}{} & \multicolumn{1}{c}{-1.698} & \multicolumn{1}{c}{} & \multicolumn{1}{c}{-1.796} \\
          & \multicolumn{1}{c}{} & \multicolumn{1}{c}{[2.317]} & \multicolumn{1}{c}{} & \multicolumn{1}{c}{[2.773]} & \multicolumn{1}{c}{} & \multicolumn{1}{c}{[2.337]} & \multicolumn{1}{c}{} & \multicolumn{1}{c}{[2.663]} \\
    DF per reverse & \multicolumn{1}{c}{} & \multicolumn{1}{c}{2.084} & \multicolumn{1}{c}{} & \multicolumn{1}{c}{4.308} & \multicolumn{1}{c}{} & \multicolumn{1}{c}{1.367} & \multicolumn{1}{c}{} & \multicolumn{1}{c}{1.199} \\
          & \multicolumn{1}{c}{} & \multicolumn{1}{c}{[2.524]} & \multicolumn{1}{c}{} & \multicolumn{1}{c}{[3.247]} & \multicolumn{1}{c}{} & \multicolumn{1}{c}{[2.841]} & \multicolumn{1}{c}{} & \multicolumn{1}{c}{[2.863]} \\
    CB Market Share & \multicolumn{1}{c}{-0.096*} & \multicolumn{1}{c}{-0.099*} & \multicolumn{1}{c}{-0.070} & \multicolumn{1}{c}{-0.094} & \multicolumn{1}{c}{-0.158**} & \multicolumn{1}{c}{-0.157**} & \multicolumn{1}{c}{-0.020} & \multicolumn{1}{c}{-0.019} \\
          & \multicolumn{1}{c}{[0.055]} & \multicolumn{1}{c}{[0.055]} & \multicolumn{1}{c}{[0.136]} & \multicolumn{1}{c}{[0.134]} & \multicolumn{1}{c}{[0.069]} & \multicolumn{1}{c}{[0.070]} & \multicolumn{1}{c}{[0.093]} & \multicolumn{1}{c}{[0.094]} \\
    Log CB Reserves & \multicolumn{1}{c}{-0.013} & \multicolumn{1}{c}{-0.293} & \multicolumn{1}{c}{-0.301} & \multicolumn{1}{c}{-0.852**} & \multicolumn{1}{c}{-0.076} & \multicolumn{1}{c}{-0.289} & \multicolumn{1}{c}{0.358} & \multicolumn{1}{c}{0.173} \\
          & \multicolumn{1}{c}{[0.356]} & \multicolumn{1}{c}{[0.339]} & \multicolumn{1}{c}{[0.398]} & \multicolumn{1}{c}{[0.383]} & \multicolumn{1}{c}{[0.402]} & \multicolumn{1}{c}{[0.366]} & \multicolumn{1}{c}{[0.364]} & \multicolumn{1}{c}{[0.353]} \\
    Bond*Month FE & Yes   & Yes   & Yes   & Yes   & Yes   & Yes   & Yes   & Yes \\
    Year FE & Yes   & Yes   & Yes   & Yes   & Yes   & Yes   & Yes   & Yes \\
    Controls & Yes   & Yes   & Yes   & Yes   & Yes   & Yes   & Yes   & Yes \\
    Observations & \multicolumn{1}{c}{15,089} & \multicolumn{1}{c}{15,089} & \multicolumn{1}{c}{3,639} & \multicolumn{1}{c}{3,639} & \multicolumn{1}{c}{5,269} & \multicolumn{1}{c}{5,269} & \multicolumn{1}{c}{6,174} & \multicolumn{1}{c}{6,174} \\
    R-squared & \multicolumn{1}{c}{0.038} & \multicolumn{1}{c}{0.037} & \multicolumn{1}{c}{0.057} & \multicolumn{1}{c}{0.052} & \multicolumn{1}{c}{0.032} & \multicolumn{1}{c}{0.032} & \multicolumn{1}{c}{0.042} & \multicolumn{1}{c}{0.043} \\
    \bottomrule
    \multicolumn{9}{l}{$^{*}$p$<$0.1; $^{**}$p$<$0.05; $^{***}$p$<$0.01} \\
\multicolumn{9}{@{}l}{\parbox[t]{.9\linewidth}{Note: This table reports the results of the regressions for bond-level three-days-ahead mispricing, as discussed in Section \ref{Sect Results}, for the period 2016:M1 to 2022:M1. Definitions, sources and frequency of all independent variables are presented in Section \ref{Sect Data Identification}. The dependent variable is the absolute value of the spread between the bond-level yield and the predicted yield based on a spline. We use Driscoll-Kraay standard errors with 20 working days lag.}} \\
    \end{tabular}%
  \label{tab:benchmark 3 days lag}%
\end{table}%


\begin{table}[htbp]
\footnotesize
  \centering
  \caption{Mispricing benchmark: Five-day mispricing lag}
  \scriptsize
    \begin{tabular}{l|cccccccc}
    \toprule
          & \multicolumn{2}{c}{All} & \multicolumn{2}{c}{Short} & \multicolumn{2}{c}{Medium} & \multicolumn{2}{c}{Long} \\
          & \multicolumn{1}{c}{} & \multicolumn{1}{c}{} & \multicolumn{1}{c}{} & \multicolumn{1}{c}{} & \multicolumn{1}{c}{} & \multicolumn{1}{c}{} & \multicolumn{1}{c}{} & \multicolumn{1}{c}{} \\
    Markup & \multicolumn{1}{c}{0.159} & \multicolumn{1}{c}{0.230} & \multicolumn{1}{c}{0.084} & \multicolumn{1}{c}{0.110} & \multicolumn{1}{c}{0.228} & \multicolumn{1}{c}{0.282} & \multicolumn{1}{c}{0.084} & \multicolumn{1}{c}{0.154} \\
          & \multicolumn{1}{c}{[0.179]} & \multicolumn{1}{c}{[0.180]} & \multicolumn{1}{c}{[0.362]} & \multicolumn{1}{c}{[0.369]} & \multicolumn{1}{c}{[0.299]} & \multicolumn{1}{c}{[0.307]} & \multicolumn{1}{c}{[0.214]} & \multicolumn{1}{c}{[0.211]} \\
    Markdown & \multicolumn{1}{c}{0.113} & \multicolumn{1}{c}{0.198} & \multicolumn{1}{c}{0.459} & \multicolumn{1}{c}{0.508} & \multicolumn{1}{c}{0.419} & \multicolumn{1}{c}{0.524*} & \multicolumn{1}{c}{-0.245} & \multicolumn{1}{c}{-0.155} \\
          & \multicolumn{1}{c}{[0.249]} & \multicolumn{1}{c}{[0.251]} & \multicolumn{1}{c}{[0.493]} & \multicolumn{1}{c}{[0.489]} & \multicolumn{1}{c}{[0.313]} & \multicolumn{1}{c}{[0.314]} & \multicolumn{1}{c}{[0.265]} & \multicolumn{1}{c}{[0.267]} \\
    DF tran repo & \multicolumn{1}{c}{-2.613*} & \multicolumn{1}{c}{} & \multicolumn{1}{c}{-2.406} & \multicolumn{1}{c}{} & \multicolumn{1}{c}{-3.527**} & \multicolumn{1}{c}{} & \multicolumn{1}{c}{-2.022} & \multicolumn{1}{c}{} \\
          & \multicolumn{1}{c}{[1.559]} & \multicolumn{1}{c}{} & \multicolumn{1}{c}{[1.732]} & \multicolumn{1}{c}{} & \multicolumn{1}{c}{[1.796]} & \multicolumn{1}{c}{} & \multicolumn{1}{c}{[1.713]} & \multicolumn{1}{c}{} \\
    DF tran reverse & \multicolumn{1}{c}{-1.711} & \multicolumn{1}{c}{} & \multicolumn{1}{c}{-0.096} & \multicolumn{1}{c}{} & \multicolumn{1}{c}{-0.749} & \multicolumn{1}{c}{} & \multicolumn{1}{c}{-2.504} & \multicolumn{1}{c}{} \\
          & \multicolumn{1}{c}{[1.909]} & \multicolumn{1}{c}{} & \multicolumn{1}{c}{[2.431]} & \multicolumn{1}{c}{} & \multicolumn{1}{c}{[2.076]} & \multicolumn{1}{c}{} & \multicolumn{1}{c}{[2.097]} & \multicolumn{1}{c}{} \\
    DF per repo & \multicolumn{1}{c}{} & \multicolumn{1}{c}{-1.396} & \multicolumn{1}{c}{} & \multicolumn{1}{c}{-0.125} & \multicolumn{1}{c}{} & \multicolumn{1}{c}{-1.551} & \multicolumn{1}{c}{} & \multicolumn{1}{c}{-1.836} \\
          & \multicolumn{1}{c}{} & \multicolumn{1}{c}{[2.609]} & \multicolumn{1}{c}{} & \multicolumn{1}{c}{[3.449]} & \multicolumn{1}{c}{} & \multicolumn{1}{c}{[2.782]} & \multicolumn{1}{c}{} & \multicolumn{1}{c}{[2.551]} \\
    DF per reverse & \multicolumn{1}{c}{} & \multicolumn{1}{c}{1.916} & \multicolumn{1}{c}{} & \multicolumn{1}{c}{0.996} & \multicolumn{1}{c}{} & \multicolumn{1}{c}{1.958} & \multicolumn{1}{c}{} & \multicolumn{1}{c}{1.412} \\
          & \multicolumn{1}{c}{} & \multicolumn{1}{c}{[3.164]} & \multicolumn{1}{c}{} & \multicolumn{1}{c}{[4.410]} & \multicolumn{1}{c}{} & \multicolumn{1}{c}{[3.495]} & \multicolumn{1}{c}{} & \multicolumn{1}{c}{[2.923]} \\
    CB Market Share & \multicolumn{1}{c}{-0.052} & \multicolumn{1}{c}{-0.065} & \multicolumn{1}{c}{-0.043} & \multicolumn{1}{c}{-0.059} & \multicolumn{1}{c}{-0.038} & \multicolumn{1}{c}{-0.036} & \multicolumn{1}{c}{0.022} & \multicolumn{1}{c}{0.011} \\
          & \multicolumn{1}{c}{[0.069]} & \multicolumn{1}{c}{[0.070]} & \multicolumn{1}{c}{[0.133]} & \multicolumn{1}{c}{[0.137]} & \multicolumn{1}{c}{[0.082]} & \multicolumn{1}{c}{[0.084]} & \multicolumn{1}{c}{[0.116]} & \multicolumn{1}{c}{[0.114]} \\
    Log CB Reserves & \multicolumn{1}{c}{0.084} & \multicolumn{1}{c}{-0.276} & \multicolumn{1}{c}{-0.576} & \multicolumn{1}{c}{-0.733*} & \multicolumn{1}{c}{-0.028} & \multicolumn{1}{c}{-0.396} & \multicolumn{1}{c}{0.532} & \multicolumn{1}{c}{0.104} \\
          & \multicolumn{1}{c}{[0.387]} & \multicolumn{1}{c}{[0.363]} & \multicolumn{1}{c}{[0.443]} & \multicolumn{1}{c}{[0.428]} & \multicolumn{1}{c}{[0.428]} & \multicolumn{1}{c}{[0.386]} & \multicolumn{1}{c}{[0.403]} & \multicolumn{1}{c}{[0.380]} \\
          &       &       &       &       &       &       &       &  \\
    Bond*Month FE & Yes   & Yes   & Yes   & Yes   & Yes   & Yes   & Yes   & Yes \\
    Year FE & Yes   & Yes   & Yes   & Yes   & Yes   & Yes   & Yes   & Yes \\
    Controls & Yes   & Yes   & Yes   & Yes   & Yes   & Yes   & Yes   & Yes \\
    Observations & \multicolumn{1}{c}{15,547} & \multicolumn{1}{c}{15,547} & \multicolumn{1}{c}{3,759} & \multicolumn{1}{c}{3,759} & \multicolumn{1}{c}{5,457} & \multicolumn{1}{c}{5,457} & \multicolumn{1}{c}{6,326} & \multicolumn{1}{c}{6,326} \\
    R-squared & \multicolumn{1}{c}{0.161} & \multicolumn{1}{c}{0.158} & \multicolumn{1}{c}{0.078} & \multicolumn{1}{c}{0.075} & \multicolumn{1}{c}{0.107} & \multicolumn{1}{c}{0.103} & \multicolumn{1}{c}{0.261} & \multicolumn{1}{c}{0.259} \\
    \bottomrule
    \multicolumn{9}{l}{$^{*}$p$<$0.1; $^{**}$p$<$0.05; $^{***}$p$<$0.01} \\
\multicolumn{9}{@{}l}{\parbox[t]{.9\linewidth}{Note: This table reports the results of the regressions for bond-level five-days-ahead mispricing, as discussed in Section \ref{Sect Results}, for the period 2016:M1 to 2022:M1. Definitions, sources and frequency of all independent variables are presented in Section \ref{Sect Data Identification}. The dependent variable is the absolute value of the spread between the bond-level yield and the predicted yield based on a spline. We use Driscoll-Kraay standard errors with 20 working days lag.}} \\
    \end{tabular}%
  \label{tab: benchmark 5 days lags}%
\end{table}%

\end{document}